\documentclass[prx,twocolumn,superscriptaddress,longbibliography]{revtex4-2}
\pdfoutput=1

\usepackage{amsmath,braket,amssymb}
\usepackage{amsthm}
\usepackage[final]{graphicx}
\usepackage{subfigure}
\usepackage{xcolor}

\usepackage[english]{babel}
\usepackage{color}
\usepackage{mathtools}
\usepackage{physics}
\usepackage{times}
\usepackage{comment}

\newcommand{\dket}[1]{\ket{#1} \rangle}
\newcommand{\dbra}[1]{\langle \bra{#1}}
\usepackage{empheq}
\usepackage{tensor}

\usepackage{tikz}
\usetikzlibrary{positioning,calc,shapes.geometric}

\tikzset{
    tensorbox/.style={
        draw,
        thick,
        rectangle,
        minimum width=0.85cm,
        minimum height=0.75cm,
        inner sep=1pt
    },
    scalarnode/.style={
        draw,
        thick,
        diamond,
        aspect=1.4,
        minimum width=1cm,
        minimum height=0.8cm,
        inner sep=1pt
    },
    ketnode/.style={
        draw,
        thick,
        regular polygon,
        regular polygon sides=3,
        shape border rotate=-90,
        minimum size=9mm,
        inner sep=1pt
    },
    branode/.style={
        draw,
        thick,
        regular polygon,
        regular polygon sides=3,
        shape border rotate=90,
        minimum size=9mm,
        inner sep=1pt
    },
    leg/.style={thick},
    every node/.style={font=\small}
}

\usepackage{hhline}

\usepackage[colorlinks,citecolor=teal,linkcolor=teal,
urlcolor=teal,hyperindex]{hyperref}

\usepackage[normalem]{ulem}
\normalfont

\usepackage[linesnumbered, ruled,vlined]{algorithm2e}

\graphicspath{{./images/},{./imagesAppendix/}}

\definecolor{forestgreen}{rgb}{0.08, 0.4, 0.13}
\definecolor{darkBlue}{rgb}{0.08, 0.13, 0.4}

\definecolor{THc}{rgb}{0.9,0.3,0.2}

\makeatletter
\newif\ifappendixtocrecord
\appendixtocrecordfalse

\AtBeginDocument{%
  \let\appendixtoc@addtocontents\addtocontents
  \long\def\addtocontents#1#2{%
    \appendixtoc@addtocontents{#1}{#2}%
    \ifappendixtocrecord
      \def\appendixtoc@extension{#1}%
      \def\appendixtoc@toc{toc}%
      \ifx\appendixtoc@extension\appendixtoc@toc
        \appendixtoc@addtocontents{atoc}{#2}%
      \fi
    \fi
  }%
}

\newcommand{\appendixtableofcontents}{%
  \begingroup
    \appendixtocrecordfalse
    \section*{Contents of Appendices}%
    \setcounter{tocdepth}{3}%
    \let\appendix\appendix@toc
    \@starttoc{atoc}%
  \endgroup
  \appendixtocrecordtrue
}
\makeatother

\newtheorem{theorem}{Theorem}

\newtheorem{observation}{Observation}

\newtheorem{lemma}{Lemma}

\renewcommand{\eqref}[1]{Eq.~(\ref{#1})}

\newcommand{\OTrXDiag}{
\vcenter{\hbox{
\begin{tikzpicture}[baseline=(current bounding box.center)]
    \node[tensorbox, minimum height=0.95cm] (X) at (0,0) {$X$};
    \coordinate (L1) at ($(X.west)+(-0.42,0.25)$);
    \coordinate (L2) at ($(X.west)+(-0.42,-0.25)$);
    \coordinate (R1) at ($(X.east)+(0.42,0.25)$);
    \coordinate (R2) at ($(X.east)+(0.42,-0.25)$);

    \draw[leg] (X.west |- L1) -- (L1);
    \draw[leg] (X.west |- L2) -- (L2);
    \draw[leg] (X.east |- R1) -- (R1);
    \draw[leg] (X.east |- R2) -- (R2);

    \draw[gray, dashed, thick] ($(L1 |- X.north)$) -- ($(L1 |- X.south)$);
    \draw[gray, dashed, thick] ($(R1 |- X.north)$) -- ($(R1 |- X.south)$);
\end{tikzpicture}
}}
}

\newcommand{\OTrSXDiag}{
\vcenter{\hbox{
\begin{tikzpicture}[baseline=(current bounding box.center)]
    \node[tensorbox, minimum height=0.95cm] (X) at (0,0) {$X$};

    \coordinate (LN1) at ($(X.west)+(-0.18,0.25)$);
    \coordinate (LN2) at ($(X.west)+(-0.18,-0.25)$);
    \coordinate (LF1) at ($(X.west)+(-0.80,0.25)$);
    \coordinate (LF2) at ($(X.west)+(-0.80,-0.25)$);

    \coordinate (R1) at ($(X.east)+(0.42,0.25)$);
    \coordinate (R2) at ($(X.east)+(0.42,-0.25)$);

    \draw[leg] (X.west |- LN1) -- (LN1);
    \draw[leg] (X.west |- LN2) -- (LN2);

    \draw[leg] (LN1) -- (LF2);
    \draw[leg] (LN2) -- (LF1);

    \draw[leg] (X.east |- R1) -- (R1);
    \draw[leg] (X.east |- R2) -- (R2);

    \draw[gray, dashed, thick] ($(LF1 |- X.north)$) -- ($(LF1 |- X.south)$);
    \draw[gray, dashed, thick] ($(R1 |- X.north)$) -- ($(R1 |- X.south)$);
\end{tikzpicture}
}}
}

\newcommand{\OTrPhiXDiag}{
\vcenter{\hbox{
\begin{tikzpicture}[baseline=(current bounding box.center)]
    \node[tensorbox, minimum height=0.95cm] (X) at (0,0) {$X$};

    \coordinate (L1) at ($(X.west)+(-0.28,0.25)$);
    \coordinate (L2) at ($(X.west)+(-0.28,-0.25)$);
    \coordinate (R1) at ($(X.east)+(0.28,0.25)$);
    \coordinate (R2) at ($(X.east)+(0.28,-0.25)$);

    \draw[leg] (X.west |- L1) -- (L1);
    \draw[leg] (X.west |- L2) -- (L2);
    \draw[leg] (X.east |- R1) -- (R1);
    \draw[leg] (X.east |- R2) -- (R2);

    \draw[leg] (L1) .. controls ($(L1)+(-0.24,0)$) and ($(L2)+(-0.24,0)$) .. (L2);

    \draw[leg] (R1) .. controls ($(R1)+(0.24,0)$) and ($(R2)+(0.24,0)$) .. (R2);
\end{tikzpicture}
}}
}

\newcommand{\OIdDiag}{
\vcenter{\hbox{
\begin{tikzpicture}[baseline=(current bounding box.center)]
    \draw[leg] (-0.38,0.18) -- (0.38,0.18);
    \draw[leg] (-0.38,-0.18) -- (0.38,-0.18);
\end{tikzpicture}
}}
}

\newcommand{\OSwapDiag}{
\vcenter{\hbox{
\begin{tikzpicture}[baseline=(current bounding box.center)]
    \draw[leg] (-0.38,0.18) -- (0.38,-0.18);
    \draw[leg] (-0.38,-0.18) -- (0.38,0.18);
\end{tikzpicture}
}}
}

\newcommand{\OPhiDiag}{
\vcenter{\hbox{
\begin{tikzpicture}[baseline=(current bounding box.center)]
    \draw[leg] (-0.22,0.20) arc[start angle=90,end angle=-90,radius=0.20];
    \draw[leg] (0.22,0.20) arc[start angle=90,end angle=270,radius=0.20];
\end{tikzpicture}
}}
}

\newcommand{\SpTrPhiXDiag}{
\vcenter{\hbox{
\begin{tikzpicture}[baseline=(current bounding box.center)]
    \node[tensorbox, minimum height=0.95cm] (X) at (0,0) {$X$};

    \coordinate (L1) at ($(X.west)+(-0.28,0.25)$);
    \coordinate (L2) at ($(X.west)+(-0.28,-0.25)$);
    \coordinate (R1) at ($(X.east)+(0.28,0.25)$);
    \coordinate (R2) at ($(X.east)+(0.28,-0.25)$);

    \draw[leg] (X.west |- L1) -- (L1);
    \draw[leg] (X.west |- L2) -- (L2);
    \draw[leg] (X.east |- R1) -- (R1);
    \draw[leg] (X.east |- R2) -- (R2);

    \draw[leg] (L1) .. controls ($(L1)+(-0.24,0)$) and ($(L2)+(-0.24,0)$) .. (L2);

    \draw[leg] (R1) .. controls ($(R1)+(0.24,0)$) and ($(R2)+(0.24,0)$) .. (R2);

    \fill[black] ($(L1)!0.5!(L2)+(-0.18,0)$) circle (1.4pt);
    \fill[black] ($(R1)!0.5!(R2)+(0.18,0)$) circle (1.4pt);
\end{tikzpicture}
}}
}

\newcommand{\SpPhiDiag}{
\vcenter{\hbox{
\begin{tikzpicture}[baseline=(current bounding box.center)]
    \draw[leg] (-0.30,0.20) arc[start angle=90,end angle=-90,radius=0.20];
    \draw[leg] (0.30,0.20) arc[start angle=90,end angle=270,radius=0.20];

    \fill[black] (-0.10,0) circle (1.4pt);
    \fill[black] (0.10,0) circle (1.4pt);
\end{tikzpicture}
}}
}

\newcommand{\COETraceDiag}{
\vcenter{\hbox{
\begin{tikzpicture}[baseline=(current bounding box.center)]
    \node[tensorbox] (X) at (0,0) {$X$};
    \coordinate (L) at ($(X.west)+(-0.42,0)$);
    \coordinate (R) at ($(X.east)+(0.42,0)$);

    \draw[leg] (X.west) -- (L);
    \draw[leg] (X.east) -- (R);

    \draw[gray, dashed, thick] ($(L |- X.north)$) -- ($(L |- X.south)$);
    \draw[gray, dashed, thick] ($(R |- X.north)$) -- ($(R |- X.south)$);
\end{tikzpicture}
}}
}

\newcommand{\COEIdDiag}{
\vcenter{\hbox{
\begin{tikzpicture}[baseline=(current bounding box.center)]
    \draw[leg] (-0.38,0) -- (0.38,0);
\end{tikzpicture}
}}
}

\newcommand{\COETransposeDiag}{
\vcenter{\hbox{
\begin{tikzpicture}[baseline=(current bounding box.center)]
    \node[tensorbox] (A) at (0,0) {$X$};

    \coordinate (L) at ($(A.west)+(-0.25,0)$);
    \coordinate (R) at ($(A.east)+(0.25,0)$);
    \draw[leg] (A.west) -- (L);
    \draw[leg] (A.east) -- (R);

    \draw[leg] (R) .. controls ($(R)+(0.35,0)$) and ($(R)+(0.35,0.55)$) .. ($(R)+(-0.25,0.55)$);
    \draw[leg] ($(R)+(-0.25,0.55)$) -- ($(R)+(-0.90,0.55)$);

    \draw[leg] (L) .. controls ($(L)+(-0.35,0)$) and ($(L)+(-0.35,-0.55)$) .. ($(L)+(0.25,-0.55)$);
    \draw[leg] ($(L)+(0.25,-0.55)$) -- ($(L)+(0.90,-0.55)$);
\end{tikzpicture}
}}
}

\newcommand{\CSEDualDiag}{
\vcenter{\hbox{
\begin{tikzpicture}[baseline=(current bounding box.center)]
    \node[tensorbox] (A) at (0,0) {$X$};

    \coordinate (L) at ($(A.west)+(-0.25,0)$);
    \coordinate (R) at ($(A.east)+(0.25,0)$);
    \draw[leg] (A.west) -- (L);
    \draw[leg] (A.east) -- (R);

    \draw[leg]
        (R)
        .. controls ($(R)+(0.35,0)$) and ($(R)+(0.35,0.55)$)
        .. node[pos=0.55,circle,fill=black,inner sep=1.4pt] {}
        ($(R)+(-0.25,0.55)$);
    \draw[leg] ($(R)+(-0.25,0.55)$) -- ($(R)+(-0.90,0.55)$);

    \draw[leg]
        (L)
        .. controls ($(L)+(-0.35,0)$) and ($(L)+(-0.35,-0.55)$)
        .. node[pos=0.55,circle,fill=black,inner sep=1.4pt] {}
        ($(L)+(0.25,-0.55)$);
    \draw[leg] ($(L)+(0.25,-0.55)$) -- ($(L)+(0.90,-0.55)$);
\end{tikzpicture}
}}
}

\begin{document}

\title{Efficient certification of time-reversal symmetry requires entanglement}

\author{Zhenhuan Liu}
\thanks{ZL and ZD contributed equally to this work.\\\href{qubithuan@gmail.com}{qubithuan@gmail.com}\\ \href{mailto:du-zy23@mails.tsinghua.edu.cn}{du-zy23@mails.tsinghua.edu.cn}}
\affiliation{Quantum Research Center, Technology Innovation Institute (TII), Abu Dhabi, United Arab Emirates}

\author{Zhenyu Du}
\thanks{ZL and ZD contributed equally to this work.\\\href{qubithuan@gmail.com}{qubithuan@gmail.com}\\ \href{mailto:du-zy23@mails.tsinghua.edu.cn}{du-zy23@mails.tsinghua.edu.cn}}
\affiliation{Center for Quantum Information, Institute for Interdisciplinary Information Sciences, Tsinghua University, Beijing 100084, China}

\author{Yifan Tang}
\affiliation{Dahlem Center for Complex Quantum Systems, Freie Universität Berlin, 14195 Berlin, Germany}

\author{Zi-Wen Liu}
\affiliation{Yau Mathematical Sciences Center, Tsinghua University, Beijing 100084, China} 

\author{Jens Eisert}
\affiliation{Dahlem Center for Complex Quantum Systems, Freie Universität Berlin, 14195 Berlin, Germany}
\affiliation{Helmholtz-Zentrum Berlin für Materialien und Energie, 14109 Berlin, Germany}

\author{Ingo Roth}
\affiliation{Quantum Research Center, Technology Innovation Institute (TII), Abu Dhabi, United Arab Emirates}

\begin{abstract}
Time-reversal symmetry is a fundamental principle of physics describing the invariance of physical laws under reversal of the direction of time.
We formulate a Bell-inequality-like test of this antiunitary symmetry using only forward access and trusted quantum operations: entanglement converts temporal input--output relations into measurable spatial exchange symmetry.
For $n$-qubit unitary dynamics, we prove that reliably distinguishing the time-reversal-symmetric circular ensembles from Haar-random dynamics requires $\Omega(\min\{2^{n/2},2^{n-e}\})$ queries for any classically adaptive protocol. Here, $e=\min\{e_{\mathrm s},e_{\mathrm m}\}$ with $e_{\mathrm s}$ and $e_{\mathrm m}$ representing the probe and measurement logarithmic entanglement negativities, respectively.
Maximally entangled probes and SWAP measurements reduce this cost to a constant number of queries.
Furthermore, we develop a time-reversal symmetry test for arbitrary fixed, compatible probes and measurements, relate its query complexity to their logarithmic negativities, and match the lower-bound scaling in the high-entanglement regime by optimizing the probe and measurement.
Our results establish a quantitative connection between entanglement and time-reversal symmetry, bridging two central concepts in quantum information science and fundamental physics.
\end{abstract}
	
\maketitle

\section{Introduction}

\emph{Time-reversal symmetry} (TRS) expresses the 
invariance of physical laws under reversal of the direction of time.
The study of TRS and other discrete symmetries has played a central role in the development of modern physics.
CPT symmetry combines \emph{charge conjugation} (C), \emph{parity} (P), and \emph{time reversal} (T), and tests of these discrete symmetries have profoundly shaped modern physics~\cite{Lehnert2016CPT}.
The discovery of parity violation in weak interactions~\cite{LeeYang1956Parity,Wu1957Parity} was followed by the observation of CP violation in neutral kaons~\cite{Christenson1964CPViolation}, implying time-reversal violation if CPT symmetry is preserved.
Time-reversal violation was subsequently observed directly by comparing transitions between neutral kaons and their antiparticles~\cite{Angelopoulos1998TimeReversal}.
In quantum many-body systems, TRS enforces Kramers degeneracy for half-integer total spin~\cite{Roberts2012Kramers} and protects the edge states of quantum spin Hall insulators~\cite{KaneMele2005Z2}.
It also governs interference between time-reversed paths in disordered transport~\cite{HikamiLarkinNagaoka1980} and helps determine the universal spectral statistics of quantum-chaotic systems~\cite{Dyson1962Threefold,BohigasGiannoniSchmit1984}.
Reliable certification of TRS is therefore essential for characterizing physical dynamics and testing predictions based on TRS.

Despite its fundamental importance, certifying TRS presents a distinctive operational challenge: the symmetry relates forward transitions to their time-reversed counterparts, whereas experiments access inputs and outputs in a fixed causal order.
Various measurement schemes have been proposed or implemented in specific experimental settings, including neutral-meson systems~\cite{Angelopoulos1998TimeReversal,Banuls1999EntangledSymmetries}, electric-dipole-moment searches~\cite{Baker2006NeutronEDM}, magnetic and optical measurements~\cite{Stolle1997TimeReversal,Luke1998TimeReversal,Xia2006TimeReversal,Cho2016Kerr}, microwave cavities, and superconducting artificial atoms~\cite{Dietz2007TimeReversal,Brito2015TimeReversal}.
Research in quantum information and quantum learning has addressed both the design of symmetry-testing protocols and fundamental query-complexity limits under resource constraints~\cite{Chen2021LearningWithoutQuantumMemory,Aharonov2022AlgorithmicMeasurement,Chen2025HypothesisTestingSymmetry,Hayashi2026OptimalUnitaryTesting}.
These works, however, concern antiunitary operator invariance, a symmetry condition distinct from the dynamical TRS considered here.
Understanding the complexity limits and resource requirements of dynamical TRS certification is therefore important for assessing the feasibility of physical symmetry tests and clarifying the role of quantum resources in learning unknown dynamics~\cite{chen2022pauli,chen2023unitarity,chen2024displacement,king2024conjugate,chen2024pauli,liu2024exponential,kang2026bound}.

\begin{figure}
\centering
\includegraphics[width=0.8\linewidth]{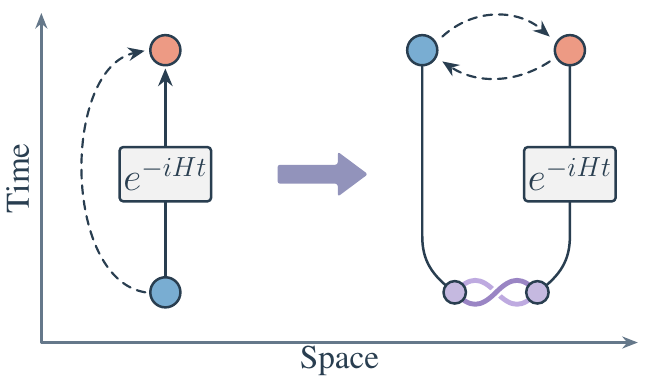}
\caption{Quantum entanglement places the input and output of an evolution, which play asymmetric roles in time, on equal footing as two spatial subsystems.
This translates dynamical time-reversal symmetry into a spatial exchange constraint on the associated bipartite state, with symmetry breaking witnessed by an otherwise forbidden measurement outcome.}
\label{fig:overview}
\end{figure}

Bell's theorem reveals a fundamental distinction between quantum theory and local classical descriptions through experimentally testable constraints on measurement probabilities~\cite{Bell1964JE}.
Entanglement is essential to violating these constraints: suitable local measurements on entangled states can generate correlations that no local hidden-variable model can reproduce.
Inspired by this perspective, we develop a Bell-inequality-like test of dynamical TRS and establish entanglement as a key resource for its efficient certification.
Here, entanglement provides a coherent reference for the unknown evolution.
As illustrated in Fig.~\ref{fig:overview}, it converts the temporal input--output relation underlying TRS into a measurable spatial exchange constraint, without implementing an antiunitary map or querying inverse dynamics.
The analogy concerns the conversion of physical assumptions into constraints on observable statistics.

Specifically, we consider equal-prior discrimination among the two time-reversal-symmetric unitary ensembles and the Haar ensemble for $n$-qubit unitary dynamics, using forward queries and retaining only classical information between rounds.
We prove that any classically adaptive protocol whose probe and measurement logarithmic negativities~\cite{EisertPhD2001,VidalWerner2002Negativity} are bounded by $e_{\mathrm s}$ and $e_{\mathrm m}$, respectively, requires $\Omega(\min\{2^{n/2},2^{n-e}\})$ queries to succeed with probability at least $2/3$, where $e=\min\{e_{\mathrm s},e_{\mathrm m}\}$.
This bound allows arbitrary auxiliary dimensions and pure or mixed probes.
In particular, separable or positive-partial-transpose (PPT) bound-entangled probes, as well as measurements with only PPT effects, still entail an exponential query cost.
Conversely, maximally entangled probes and SWAP measurements enable constant-query certification at fixed accuracy and confidence using only one call to the unknown unitary per run, with applications to probing TRS-breaking interactions relevant to symmetry-protected phenomena.
More generally, we prove that TRS can be certified using an arbitrary fixed, known probe state and a compatible joint positive operator-valued measure (POVM).
We relate its query complexity to the logarithmic negativities of the probe state and measurement and show that optimizing this pair achieves the optimal scaling $\Theta(2^{n-e})$ for $n/2\leq e\leq n$ at fixed accuracy and confidence.

\section{Time-reversal symmetry}

Time reversal is represented by a physically specified antiunitary operator $\Xi$ and can be written as $\Xi=U_\Xi K$, where $K$ denotes complex conjugation and $U_\Xi$ is unitary.
Although this representation depends on the basis, the relevant basis is fixed by the physical degrees of freedom and is therefore regarded as known in the certification problem.
For example, for a spinless particle in the position basis, one may take $\Xi=K$, which leaves position invariant and reverses momentum.
Throughout this work, the physical basis and the corresponding time-reversal operator are fixed, while the target dynamics is unknown.

For a time-independent Hamiltonian, TRS requires $\Xi H\Xi^{-1}=H$.
Physically, this means that forward and time-reversed motion are governed by the same Hamiltonian, with time-odd observables such as momentum and angular momentum changing sign.
External parameters are held fixed in the symmetry test, so an applied magnetic field can break TRS.
Since $\Xi$ is antiunitary, the corresponding evolution $U(t)=e^{-iHt}$ satisfies
\begin{equation}
    \Xi U(t)\Xi^{-1}
    =
    U(t)^\dagger,
    \label{eq:main-dynamical-trs}
\end{equation}
for every evolution time.
Thus time reversal reverses the direction of evolution while leaving the underlying dynamical law unchanged.

Dyson's threefold classification distinguishes TRS with $\Xi^2=+I$, TRS with $\Xi^2=-I$, and the absence of TRS, conventionally labeled $+1$, $-1$, and $0$, respectively~\cite{Dyson1962Threefold}.
The $+1$ class includes spinless orbital systems and integer-spin systems, whereas the $-1$ class is associated with half-integer total spin and gives rise to Kramers degeneracy~\cite{Roberts2012Kramers}.
For example, for $n$ spin-$1/2$ degrees of freedom,
$\Xi=(i\sigma_y)^{\otimes n}K$ and hence $\Xi^2=(-1)^n I$.
Independently of the microscopic realization, one can choose canonical bases in which
\begin{equation}
    \begin{aligned}
        \Xi^2=+I:\quad&
        \Xi=K,
        & U&=U^{\mathrm T},\\
        \Xi^2=-I:\quad&
        \Xi=JK,
        & U&=JU^{\mathrm T}J^\dagger,
    \end{aligned}
    \label{eq:main-trs-classes}
\end{equation}
Here $J=i\sigma_y\otimes I_{d/2}$ is the canonical symplectic form pairing the $d$ basis state vectors into $d/2$ Kramers pairs, so the second class requires even dimension.

The circular orthogonal and symplectic ensembles (COE, CSE) provide invariant random-matrix models for these two classes.
Let $W$ be Haar distributed on $\mathrm{U}(d)$ and define the symplectic dual by $X^{\mathrm D}:=JX^{\mathrm T}J^\dagger$.
Then
\begin{equation}
    U_{\mathrm{COE}}=W^{\mathrm T}W,\qquad
    U_{\mathrm{CSE}}=W^{\mathrm D}W,
    \label{eq:coe-cse-definitions}
\end{equation}
so that $U_{\mathrm{COE}}=U_{\mathrm{COE}}^{\mathrm T}$ and
$U_{\mathrm{CSE}}=U_{\mathrm{CSE}}^{\mathrm D}$~\cite{Matsumoto2013SymmetricSpaces}.
In the absence of TRS, we use the unrestricted Haar unitary ensemble as the reference.
Accordingly, COE, CSE, and Haar-random unitaries model the three symmetry classes considered in this work.
We refer to Appendix~\ref{app:pre} for a detailed introduction to TRS and the mathematical structures of the COE and CSE.

Some quantum information studies consider the operator symmetry $\Xi U\Xi^{-1}=U$~\cite{Chen2021LearningWithoutQuantumMemory,Aharonov2022AlgorithmicMeasurement,Chen2025HypothesisTestingSymmetry,Hayashi2026OptimalUnitaryTesting}.
This condition requires the evolution operator itself to remain unchanged under the antiunitary transformation.
In contrast, dynamical TRS in \eqref{eq:main-dynamical-trs} requires forward and backward evolution to obey the same physical laws, rather than to be identical.
Our work concerns this dynamical notion of TRS.

\section{TRS certification with limited entanglement}
\label{sec:limited-entanglement}

\begin{figure}[htbp]
\centering
\includegraphics[width=1\linewidth]{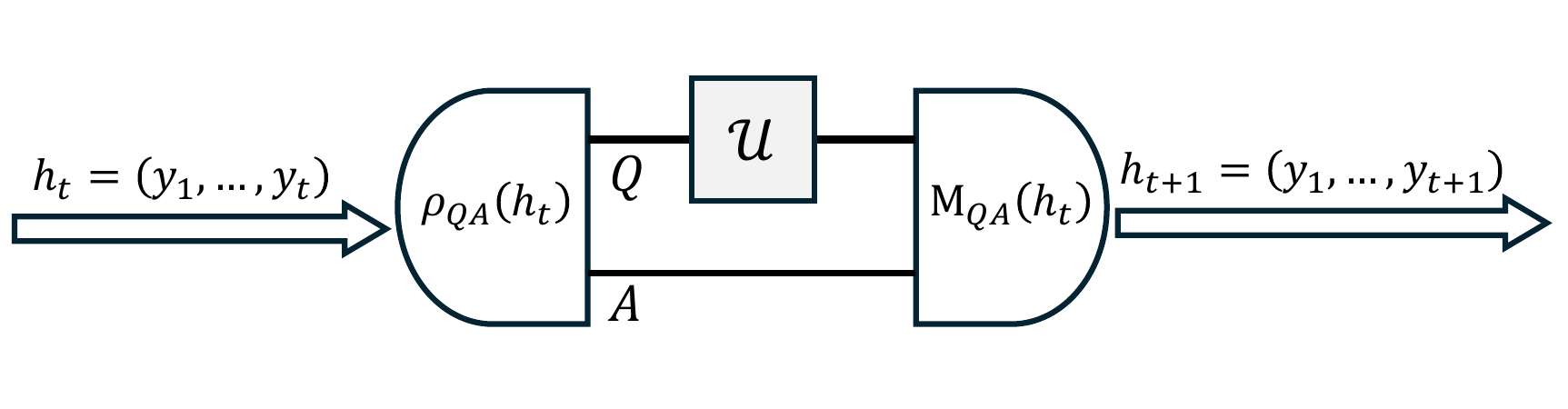}
\caption{Adaptive protocol for certifying TRS with entangled input states and joint measurements.}
\label{fig:learning_protocol}
\end{figure}

We consider adaptive protocols that access the same unknown $n$-qubit unitary channel $\mathcal U$ in $T$ rounds, as illustrated in Fig.~\ref{fig:learning_protocol}.
Let $Q$ denote the $n$-qubit query register, with $\dim Q=d=2^n$, and let $A$ be an auxiliary register of arbitrary finite dimension.
In round $t+1$, conditioned on the preceding transcript $h_{t}=(y_1,\ldots,y_{t})$, the protocol prepares a state $\rho_{Q,A}(h_{t})$ on $Q\otimes A$, applies $\mathcal U\otimes\operatorname{id}_A$, and performs an arbitrary joint POVM $\mathsf M_{Q,A}(h_{t})$ on the output.
The outcome $y_{t+1}$ updates the transcript to $h_{t+1}=(y_1,\ldots,y_{t+1})$, which may determine both the input state and the measurement used in the next round.
All quantum registers are discarded after each round, so only the classical transcript is retained between channel uses.
The complete measurement record $h_T$ is then processed to infer the time-reversal symmetry properties of $\mathcal U$.

To emphasize the role of entanglement in TRS certification, we consider entanglement resources in both the input probes and the joint measurements across the bipartition $Q:A$, without restricting the auxiliary dimension.
For a bipartite state $\rho_{Q,A}$, define its logarithmic negativity by
\begin{equation}
    E_N^{\mathrm{s}}(\rho_{Q,A})
    :=
    \log_2\|\rho_{Q,A}^{\mathrm T_Q}\|_1,
    \label{eq:main-negativity}
\end{equation}
where $\mathrm T_Q$ denotes partial transpose over $Q$.
Logarithmic negativity is a computable entanglement monotone under completely PPT-preserving operations and provides an upper bound on distillable entanglement~\cite{VidalWerner2002Negativity,Plenio2005LogarithmicNegativity}.
Negativity-based methods also play an important role in entanglement detection~\cite{Elben2020MixedStateEntanglement,Zhou2020Negativity} and quantum device benchmarking~\cite{Shaw2024Benchmarking}.
For a joint POVM $\mathsf M_{Q,A}=\{M_y\}_y$, we introduce the corresponding measurement negativity
\begin{equation}
    E_N^{\mathrm{m}}(\mathsf M_{Q,A})
    :=
    \log_2
    \sup_{\substack{\sigma_{Q,A}\geq0\\
                    \operatorname{Tr}\sigma_{Q,A}=1}}
    \sum_y
    \left|
        \operatorname{Tr} \left(
            M_y^{\mathrm T_Q}\sigma_{Q,A}
        \right)
    \right|.
    \label{eq:main-measurement-negativity}
\end{equation}
This quantity vanishes precisely when every POVM effect is PPT.
It lower-bounds the number of preshared EPR pairs required to implement $\mathsf M$ exactly on arbitrary inputs using \emph{local operations and classical communication} (LOCC), 
providing an operational interpretation in terms of nonlocal measurement costs~\cite{Bandyopadhyay2009MeasurementCost,GourScandolo2021BipartiteChannel} that also applies to the more restrictive setting of measurement localization~\cite{Pauwels2025MeasurementLocalization}.

\begin{theorem}[Bounded probe and measurement entanglement]
\label{thm:main-negativity-lower-bound}
\label{thm:main-ancilla-lower-bound}
Let $d=2^n$ and $0\leq e_{\mathrm{s}},e_{\mathrm{m}}\leq n$.
With equal prior probability, an unknown unitary $U$ is sampled from the COE, CSE, or Haar ensemble.
The associated channel $\mathcal U(X)=UXU^\dagger$ is fixed and reused in every round.
Suppose that every conditional probe and POVM satisfy
\begin{equation}
E_N^{\mathrm{s}}\bigl(\rho_{Q,A}(h_{t})\bigr)
\leq e_{\mathrm{s}}, \ \ 
E_N^{\mathrm{m}}\bigl(
    \mathsf M_{Q,A}(h_{t})
\bigr)
\leq e_{\mathrm{m}}
\label{eq:main-negativity-constraint}
\end{equation}
on every adaptive branch.
The probes may be pure or mixed, and the auxiliary system may have arbitrary dimension.
Then any classically adaptive protocol in the learning model of Fig.~\ref{fig:learning_protocol}, retaining only classical information between rounds, that identifies the correct hypothesis with average success probability at least $2/3$ requires
\begin{equation}
    T
    =
    \Omega \left(
        \min\left\{
            2^{n/2},
            2^{n-\min\{e_{\mathrm{s}},e_{\mathrm{m}}\}}
        \right\}
    \right)
    \label{eq:main-negativity-query-lower-bound}
\end{equation}
uses of the unknown channel.
\end{theorem}

The proof, as detailed in Appendix~\ref{app:proof}, is based on the learning tree formalism and Le Cam-type argument~\cite{LeCam1986AsymptoticMethods,Chen2021LearningWithoutQuantumMemory}.
Theorem~\ref{thm:main-negativity-lower-bound} establishes non-PPT entanglement in both probe states and measurements as the necessary resource for efficient TRS certification in this access model.
Protocols restricted to PPT probes, including separable states and PPT bound-entangled states, or to POVMs with only PPT effects, still require $\Omega(2^{n/2})$ queries, even when the other resource and the auxiliary dimension are unrestricted.
Constant-query certification therefore requires at least one conditional probe with $E_N^{\mathrm{s}}\geq n-\mathcal{O}(1)$ and at least one conditional POVM with $E_N^{\mathrm m}\geq n-\mathcal{O}(1)$.
The near-maximal probe entanglement requirement also makes low-query certification difficult to achieve under practical noise~\cite{Cotler2026NoisyQuantumLearning}.

\section{Quantifying TRS and physical interpretation}
\label{sec:quantifying-trs}

Entanglement makes dynamical TRS accessible by encoding it as an exchange symmetry of an enlarged bipartite state~\cite{Liu2021GaugingAnomalousUnitaries}.
This representation allows us to quantify symmetry breaking as well as certify the symmetry.
Specifically, prepare the maximally entangled state vector
$\ket{\Phi_d}_{Q,A}=d^{-1/2}\sum_{j=0}^{d-1}\ket{j}_Q\ket{j}_A$
and apply $\mathcal U$ to $Q$ once, obtaining the normalized Choi state $\rho_{\mathcal U}$.
Define the swap operator $\mathbb S$ which exchanges $Q$ and $A$, and define its symplectic counterpart by
$\mathbb S_J=\mathbb S(J\otimes J)$, with $J$ as in \eqref{eq:main-trs-classes}.
These observables define two bounded linear channel signatures:
\begin{equation}
    f_+(\mathcal U)=\operatorname{Tr}(\mathbb S\rho_{\mathcal U}),\qquad
    f_-(\mathcal U)=\operatorname{Tr}(\mathbb S_J\rho_{\mathcal U}).
    \label{eq:main-trs-signatures}
\end{equation}
Their deviations from unity measure the discrepancy between the evolution and its time-reversed counterpart (see proof in Appendix~\ref{app:trs-dynamical-interpretation}).

\begin{observation}[Quantifying dynamical TRS breaking]
\label{obs:main-trs-exchange}
In the canonical bases with $\Xi_+=K$ and $\Xi_-=JK$, we have
\begin{equation}
    1-f_\pm(\mathcal U)
    =\frac{\|\Xi_\pm U\Xi_\pm^{-1}-U^\dagger\|_{\mathrm F}^2}{2d},
    \label{eq:main-trs-defect}
\end{equation}
where $\|\cdot\|_{\mathrm F}$ is the 
Frobenius norm.
Consequently, $\Xi_\pm U\Xi_\pm^{-1}=U^\dagger$ if and only if $f_\pm(\mathcal U)=1$.
\end{observation}

These signatures admit a Bell-inequality-like interpretation: a physical assumption is translated into a constraint on observable probabilities that can be falsified experimentally.
Here, entanglement converts the temporal TRS relation into a spatial exchange constraint: for a specified time-reversal operator and evolution time, exact TRS forbids the $-1$ outcome of the corresponding exchange measurement.
Observing this outcome therefore rules out the symmetry in the ideal trusted model.
The analogy concerns the testability of physical assumptions rather than Bell nonlocality: our protocol relies on trusted preparations, controls, and measurements and is not device independent.

Beyond certification, the signatures quantify the strength of TRS-breaking interactions when the evolution time can be varied.
Their short-time expansions, specializations of operator fidelity susceptibility~\cite{Lu2008OperatorFidelity}, isolate the TRS-odd part of $H$.

\begin{theorem}[Hamiltonian TRS-breaking strength]
\label{thm:main-trs-curvature}
Let $H$ be a time-independent Hamiltonian on a $d$-dimensional Hilbert space and let $\mathcal U_t$ denote the channel generated by $U(t)=e^{-iHt}$. 
Define $H_\pm:=(H-\Xi_\pm H\Xi_\pm^{-1})/2$, then
\begin{equation}
-\frac14\left.\frac{d^2}{dt^2}f_\pm(\mathcal U_t)\right|_{t=0}=d^{-1}\operatorname{Tr}(H_\pm^2).
    \label{eq:main-trs-short-time}
\end{equation}
This curvature vanishes if and only if $\Xi_\pm H\Xi_\pm^{-1}=H$.
\end{theorem}

The proof is provided in Appendix~\ref{app:trs-dynamical-interpretation}. These signatures provide dynamical probes of TRS-breaking interactions relevant to symmetry-protected physical phenomena.
Applications include magnetic perturbations affecting Kramers degeneracy and helical-edge transport~\cite{Roberts2012Kramers,Wu2018QuantumSpinHall,Wu2006HelicalLiquid}, and scalar-spin-chirality couplings in quantum magnets~\cite{Motrunich2006SpinChirality,Kitamura2017SpinChirality}.

\section{Certifying TRS with general probe states and measurements}
\label{sec:trs-certification-protocol}

We now turn to the certification task considered throughout this work. We use a fixed, known input state and a fixed joint POVM.
Its query complexity depends on a joint parameter of the probe and measurement.
For a fixed sign $\tau\in\{+1,-1\}$, with $f_{+1}=f_+$ and
$f_{-1}=f_-$, the goal is to distinguish
\begin{equation}
    H_0^{(\tau)}:\ f_\tau(\mathcal U)=1
    \quad\text{from}\quad
    H_1^{(\tau)}:\ f_\tau(\mathcal U)\leq 1-\epsilon ,
    \label{eq:main-trs-testing-promise}
\end{equation}
for an unknown but fixed unitary channel $\mathcal U$.
Let $\rho_{Q,A}$ be a probe,
$\mathsf M=\{M_y\}_y$ be a POVM, and write $\rho_A=\operatorname{Tr}_Q\rho_{Q,A}$.
Define
\begin{equation}
    a_y:=\operatorname{Tr}
    \left[M_y(I_Q\otimes\rho_A)\right], \ \
    b_y:=\operatorname{Tr}
    \left(M_y\rho_{Q,A}^{\mathrm{T}_Q}\right),
\label{eq:gfp-ab}
\end{equation}
and
\begin{equation}
    \lambda_\tau(\rho,\mathsf M)
    := \sum_{\substack{y:\,a_y>0\\ b_y=-\tau a_y}} a_y .
\label{eq:gfp-lambda}
\end{equation}
We obtain the following upper bound for the query complexity.

\begin{theorem}[One-sided TRS testing with fixed resources]
\label{thm:gfp-testing}
Let $d=2^n$ with $n\geq1$, and let $\rho_{Q,A}$ and
$\mathsf M$ be fixed and known, with arbitrary finite auxiliary dimension.
Fix $\tau\in\{+1,-1\}$, $0<\epsilon\leq1$, and
$0<\delta\leq1/4$.
If $\lambda_\tau(\rho,\mathsf M)>0$, then the promise problem
in~\eqref{eq:main-trs-testing-promise} can be solved for every unknown
unitary channel $\mathcal U$ satisfying either promise, accepting with probability one under $H_0^{(\tau)}$ and rejecting with probability at least $1-\delta$ under $H_1^{(\tau)}$,
using
\begin{equation}
    T_\tau
    =
    \left\lceil
    \frac{d-\tau}
         {\lambda_\tau(\rho,\mathsf M)\epsilon}
    \log\frac1\delta
    \right\rceil
    \label{eq:gfp-query}
\end{equation}
queries.
Each query uses the same probe and POVM, known local controls on $Q$,
and no quantum memory between queries.
Conversely, any classically adaptive protocol that solves the same promise problem with worst-case error at most $\delta$ requires
\begin{equation}
    T=\Omega\left(
        \epsilon^{-1}\log(1/\delta)
    \right)
\end{equation}
queries, even with unrestricted auxiliary dimension and unrestricted
probe and measurement entanglement.
\end{theorem}

Under the constraints $E_N^{\mathrm s}(\rho_{Q,A})\leq e_{\mathrm s}$ and $E_N^{\mathrm m}(\mathsf M)\leq e_{\mathrm m}$, let $e_*:=\min\{e_{\mathrm s},e_{\mathrm m}\}$.
It holds that
\begin{equation}
    0\leq
    \lambda_\tau(\rho,\mathsf M)
    \leq
    \frac{2^{e_*}-\tau}{2}.
    \label{eq:gfp-resource}
\end{equation}
The upper bound is attainable whenever $2^{e_*}$ is an integer.
Optimizing the probe state and measurement yields
\begin{equation}
    T_\tau
    =
    \mathcal O\left(
        2^{n-e_*}
        \epsilon^{-1}
        \log(1/\delta)
    \right)
    \label{eq:gfp-scaling}
\end{equation}
for both signs when $e_*\geq1$.
Combining this upper bound with Theorem~\ref{thm:main-negativity-lower-bound}, the optimal query complexity at fixed accuracy and confidence satisfies
\begin{equation}
    T^\star
    =
    \Theta(2^{n-e_*}),
    \qquad
    n/2\leq e_*\leq n.
    \label{eq:gfp-optimal-scaling}
\end{equation}
Thus, throughout this high-entanglement regime, the smaller one of the probe and measurement logarithmic negativities determines the optimal dependence on the system dimension $d=2^n$.
This optimality is achieved by optimizing the probe--measurement pair; large negativities alone do not guarantee the compatibility condition $\lambda_\tau>0$ for a specified pair.
Proofs are provided in Appendix~\ref{app:gfp-testing}.

Theorem~\ref{thm:gfp-testing} also recovers the exchange-measurement protocol of the preceding section.
For a maximally entangled input state and the SWAP POVM with the prescribed local controls, the probe and measurement logarithmic negativities are both exactly $n$, and the resource bound is saturated with $\lambda_\tau=(d-\tau)/2$.
Substituting this value into the theorem gives $T_\tau=\lceil 2\epsilon^{-1}\log(1/\delta)\rceil=\Theta(\epsilon^{-1}\log(1/\delta))$, independently of $d$.
Each run requires only one query to the unknown unitary, with no quantum memory retained between runs.
The lower bound in Theorem~\ref{thm:gfp-testing} shows that this dependence on both the accuracy gap $\epsilon$ and failure probability $\delta$ is optimal, even among classically adaptive protocols with unrestricted auxiliary dimension and probe and measurement entanglement.

While the protocol of Theorem~\ref{thm:gfp-testing} is nonadaptive, classical adaptation can improve its dimension dependence when no auxiliary system is available ($e_*=0$).
In Appendix~\ref{app:ancilla-free-protocol}, we construct and analyze an adaptive protocol that uses $O(2^n\log n/n)$ queries at fixed accuracy and confidence.
This improves the $\mathcal{O}(2^n)$ scaling of the nonadaptive construction for $\tau=-1$ and also enables certification for $\tau=+1$, for which $\lambda_+=0$ makes Theorem~\ref{thm:gfp-testing} inapplicable.

\section{Discussion}

Our results suggest that the empirical content of a symmetry depends on the correlations an observer can prepare and measure. An experiment proceeds forward in time, yet can test a relation involving reversed motion.
Unlike continuous unitary symmetries, TRS is discrete and antiunitary, so it 
is not directly certified by measuring a conserved Hermitian generator supplied by Noether's theorem~\cite{Noether1971InvariantVariation}.
Entangled probes instead map dynamical TRS to a unitary exchange symmetry on an enlarged bipartite state, converting a temporal input--output relation into accessible spatial correlations.
The exponential advantage established here makes this quantitative: access to these correlations can determine whether a symmetry is practically empirically testable.

This insight also helps separate microscopic time-reversal symmetry from the thermodynamic arrow. The former concerns the dynamical law, while the latter can arise from boundary conditions and coarse graining even when 
that law is actually time-reversal symmetric. This distinction is central to quantum cosmology~\cite{GellMannHartle1994TimeJE}. Our protocol tests an unknown evolution relative to a specified physical time reversal, using an entangled reference to access input-output 
correlations without reversing the experiment's temporal order. For an open subsystem, extending this approach would require distinguishing genuine symmetry breaking from loss of access to the environment.

A more concrete connection to quantum gravity comes through random-matrix universality. 
Antiunitary symmetries help determine the spectral statistics of chaotic systems, including models of black-hole dynamics~\cite{Cotler2017BlackHolesJE}. In Jackiw-Teitelboim gravity, 
boundary time-reversal symmetry calls for unorientable geometries in the bulk path integral~\cite{StanfordWitten2020JTJE}. Could entangled probes diagnose boundary 
symmetry data with a bulk interpretation without resolving individual energy levels? Our signatures diagnose dynamical 
TRS directly in finite dimensions. Establishing the cost for a specific holographic theory would require going beyond the circular ensembles used in our lower bounds and accounting for accessible energies, evolution times, and boundary operations.

The entangled reference also invites comparison with thermofield doubles in holography~\cite{Maldacena2003EternalJE,Chapman2019ThermofieldJE}. For a 
finite-dimensional system and a conjugate reference copy, the maximally entangled reference vector is the infinite-temperature limit of a thermofield-double state vector. In the black-hole regime of a 
holographic theory, a finite-temperature thermofield double describes an eternal two-sided anti-de Sitter black hole~\cite{Maldacena2003EternalJE}. A thermally weighted exchange test would raise questions about how its sensitivity depends on temperature and accessible correlations between the boundaries. Such a test would bring energy weighting and state preparation into the problem; a bulk interpretation would 
need to be established for its observables explicitly.

These questions bring the physical cost of the test into focus. Few uses of an unknown evolution can still demand substantial preparation and measurement resources. In many-body or gravitational settings, locality, energy, and control over the reference may be as consequential as query complexity. Determining which symmetry signatures survive these restrictions would connect our bounds to an observer's capabilities; robustness would also be needed to distinguish weak symmetry breaking from imperfect control. More broadly, entanglement gives empirical access to a relation between forward and time-reversed dynamics. In our trusted setting, the Bell-inequality-like step expresses that relation as a probability constraint and quantifies the resources needed to test it. Whether similar reasoning can make other assumptions about time and spacetime operationally precise remains an intriguing question for fundamental physics.

\begin{acknowledgements}
We thank Qi Ye, Yingfei Gu, and Hong-Ye Hu for insightful discussions. 
Y.~T. and J.~E. have been supported by the BMFTR (QuSol, MUNIQC-Atoms, PasQuops, PrakiQom), the DFG (CRC 183, BoLaCo, and SPP 2514), the Quantum Flagship (Millenion, PasQuanS2), the Munich Quantum Valley, Berlin Quantum, QuantERA (SDPCode), and the European Research Council (DebuQC).
Z.-W.~L.~is supported in part by NSFC under Grant No.~12475023, Dushi Program, and startup funding from YMSC.
Z.~D. is supported by the National Natural Science Foundation of China (Grant No. 12575023), the Quantum Science and Technology-National Science and Technology Major Project (Grants No.~2021ZD0300804 and No.~2021ZD0300702), the CCF-QuantumCtek Superconducting Quantum Computing Special Cooperation Program (Grant No.~CCF-QC2025005), and the Turing AI Institute of Nanjing.
\end{acknowledgements}


%

\onecolumngrid
\appendix
\clearpage

\makeatletter
\def\l@section{\@dottedtocline{1}{0em}{2em}}
\def\l@subsection{\@dottedtocline{2}{2em}{1.8em}}
\def\l@subsubsection{\@dottedtocline{3}{3.8em}{1.8em}}
\def\numberline#1{\hb@xt@\@tempdima{#1.\hfil}}
\def\@pnumwidth{2em}
\def\@tocrmarg{3em}
\makeatother
\appendixtableofcontents

\section{Preliminaries}\label{app:pre}

Throughout these appendices, $d$ denotes the total Hilbert-space dimension of the queried system; for an $n$-qubit system, $d=2^n$.
The symplectic cases require even $d$, with the canonical basis organized into $d/2$ Kramers pairs.

\subsection{Time-reversal symmetry}

We first fix the time-reversal convention used throughout the paper.
Physical time reversal is represented by an antiunitary operator $\Xi$.
Antiunitarity ensures the correct transformation of the Schr\"odinger equation, with $\hbar=1$,
\begin{equation}
    i\frac{d}{dt}\ket{\psi(t)}=H\ket{\psi(t)}.
\end{equation}
Reversing time changes the sign of the time derivative, while antiunitarity supplies the accompanying complex conjugation, $i\mapsto-i$.
Consequently, the reversed state vector $\Xi\ket{\psi(-t)}$ evolves under the Hamiltonian $\Xi H\Xi^{-1}$.

In any fixed orthonormal basis, every antiunitary operator can be written as a unitary operator followed by complex conjugation.
We write the time-reversal operator as
\begin{equation}
    \Xi=U_\Xi K,
\end{equation}
where $K$ denotes complex conjugation in the chosen basis and $U_\Xi$ is unitary.
Thus $U_\Xi$ specifies the physical transformation once the basis is fixed.

The microscopic degrees of freedom determine the action of $\Xi$; the Hilbert-space dimension alone does not.
In the certification problem, this action and its matrix representation in the physical basis are known independently of the unknown dynamics.
For a spinless system, one can often choose a basis invariant under time reversal, such as the position basis or a real orbital basis.
In this basis,
\begin{equation}
    \Xi=K,
     U_\Xi=I.
\end{equation}
For example, in the position representation, $p=-i\partial_x$, so
\begin{equation}
    KxK^{-1}=x,
     KpK^{-1}=-p.
\end{equation}
Thus, complex conjugation implements the physical reversal of momentum while leaving position unchanged.
For a spin-$1/2$ degree of freedom, time reversal must flip all spin components.
In the standard $\sigma_z$ eigenbasis, we may choose
\begin{equation}
    \Xi=i\sigma_yK,
\end{equation}
which satisfies
\begin{equation}
    \Xi\sigma_\alpha\Xi^{-1}=-\sigma_\alpha,
     \alpha=x,y,z.
\end{equation}
For $n$ spin-$1/2$ degrees of freedom in the product basis, this gives
\begin{equation}
    \Xi=J_{\mathrm{spin}}K,
     J_{\mathrm{spin}}=(i\sigma_y)^{\otimes n}.
\end{equation}
Since $J_{\mathrm{spin}}$ is real and $(i\sigma_y)^2=-I$, we obtain
\begin{equation}
    \Xi^2=J_{\mathrm{spin}}J_{\mathrm{spin}}^*=(-1)^nI,
     J_{\mathrm{spin}}^{\mathrm T}=(-1)^nJ_{\mathrm{spin}}.
\end{equation}
Consequently, an even number of spin-$1/2$ degrees of freedom gives $\Xi^2=I$, whereas an odd number gives $\Xi^2=-I$.
When orbital and spin degrees of freedom are both present, $U_\Xi$ is the tensor product of their unitary parts, with $K$ acting in the joint product basis.
An abstract $n$-qubit Hilbert space need not represent $n$ physical spins, so its physical time-reversal operator must be specified separately.

Under a change of orthonormal basis described by a unitary matrix $W$, the Hamiltonian transforms as
\begin{equation}
    H'=W^\dagger HW.
\end{equation}
The same physical time-reversal operator is represented in the new basis by
\begin{equation}
    \Xi=U_\Xi'K',
     U_\Xi'=W^\dagger U_\Xi W^*,
\end{equation}
where $K'$ denotes complex conjugation in the new basis.
The appearance of $W^*$ reflects the antiunitary nature of $\Xi$.
Transforming both $H$ and the representation of $\Xi$ ensures that the symmetry condition is basis independent.

For a time-independent Hamiltonian, time-reversal symmetry means
\begin{equation}
    \Xi H\Xi^{-1}=H,
    \text{equivalently}
    U_\Xi H^*U_\Xi^\dagger=H.
\end{equation}
This condition states that the forward and time-reversed state vectors obey the same dynamical law.
If $U(t)=e^{-iHt}$ is generated by such a Hamiltonian, antiunitarity gives
\begin{equation}
    \Xi U(t)\Xi^{-1}
    =e^{+iHt}
    =U(-t)
    =U(t)^\dagger.
\end{equation}
Accordingly, for a finite-time unitary $U$, we consider the dynamical time-reversal condition
\begin{equation}
    \Xi U\Xi^{-1}=U^\dagger,
    \text{equivalently}
    U_\Xi U^*U_\Xi^\dagger=U^\dagger.
\end{equation}
For evolution generated by a fixed Hamiltonian, this condition is equivalent to Hamiltonian time-reversal symmetry when required for all times.
Its validity at a single time, however, need not imply that the original Hamiltonian is time-reversal symmetric.
The finite-time condition is unchanged by a global phase of $U$, so it is well defined for the unitary channel $\mathcal U(X)=UXU^\dagger$.

Taking the Hermitian conjugate gives the equivalent transpose condition
\begin{equation}
    U=U_\Xi U^{\mathrm T}U_\Xi^\dagger.
\end{equation}
For a spinless system in a time-reversal-invariant basis, $U_\Xi=I$, and hence
\begin{equation}
    U=U^{\mathrm T}.
\end{equation}
For $n$ physical spin-$1/2$ degrees of freedom in the product basis, the condition instead reads
\begin{equation}
    U=J_{\mathrm{spin}}U^{\mathrm T}J_{\mathrm{spin}}^\dagger,
     J_{\mathrm{spin}}=(i\sigma_y)^{\otimes n}.
\end{equation}

More generally, the distinction relevant to the circular ensembles is between $\Xi^2=I$ and $\Xi^2=-I$.
When $\Xi^2=I$, an orthonormal basis can be chosen in which $\Xi=K$, and the allowed unitaries are symmetric.
When $\Xi^2=-I$, the Hilbert-space dimension is even, and a basis can be chosen in which
\begin{equation}
    \Xi=JK,
\qquad    
    J=i\sigma_y\otimes I_{d/2}=
    \begin{bmatrix}
        0&I_{d/2}\\
        -I_{d/2}&0
    \end{bmatrix},
\end{equation}
Here $J$ is the canonical form used in \eqref{eq:main-trs-classes}, rather than the product-basis matrix $J_{\mathrm{spin}}$.
The allowed unitaries satisfy the self-duality condition $U=JU^{\mathrm T}J^\dagger$.
These are the respective symmetry constraints underlying COE and CSE; the ensembles additionally specify the corresponding invariant probability measures.

The operator-invariance condition $\Xi U\Xi^{-1}=U$ is different from dynamical TRS.
For $\Xi=K$, it requires $U$ to be real orthogonal, whereas dynamical TRS requires $U$ to be symmetric.
Imposed on $U(t)=e^{-iHt}$ for every $t$, operator invariance gives $\Xi H\Xi^{-1}=-H$, rather than Hamiltonian TRS.
All these conditions refer to a specified physical $\Xi$; choosing a different antiunitary transformation changes the symmetry being tested.

\subsection{Tensor graph}

We use tensor graphs to display the contractions appearing in the ensemble moment calculations.
A tensor is represented by a node, each index by a line or \emph{leg}, and contraction by joining the corresponding legs.
For a matrix $A$, the left leg carries its row (output) index and the right leg its column (input) index.
Ket and bra vectors have a leg on the left and right, respectively, while a scalar has no free legs.
The same convention applies to composite diagrams: their uncontracted legs specify their remaining indices.
Figure~\ref{fig:tensor_basic_objects} summarizes these basic elements.

\begin{figure}[htbp]
    \centering

    \begin{tikzpicture}[baseline=(current bounding box.center)]
        \node[tensorbox] (A) at (0,0) {$A$};
        \draw[leg] (-1.0,0) -- (A.west);
        \draw[leg] (A.east) -- (1.0,0);
        \node at (0,-0.95) {matrix};
    \end{tikzpicture}
    \hspace{1.0cm}
    \begin{tikzpicture}[baseline=(current bounding box.center)]
        \node[ketnode] (vket) at (0,0) {$v$};
        \draw[leg] (-1.0,0) -- (vket.west);
        \node at (0,-0.95) {ket vector};
    \end{tikzpicture}
    \hspace{1.0cm}
    \begin{tikzpicture}[baseline=(current bounding box.center)]
        \node[branode] (vbra) at (0,0) {$v$};
        \draw[leg] (vbra.east) -- (1.0,0);
        \node at (0,-0.95) {bra vector};
    \end{tikzpicture}
    \hspace{1.0cm}
    \begin{tikzpicture}[baseline=(current bounding box.center)]
        \node[scalarnode] (c) at (0,0) {$c$};
        \node at (0,-0.95) {scalar};
    \end{tikzpicture}

    \caption{Basic graphical elements in the tensor-graph notation: a matrix, a ket vector, a bra vector, and a scalar.}
    \label{fig:tensor_basic_objects}
\end{figure}

Operators on several subsystems have one input and one output leg for each subsystem.
Tensor products are represented by stacking diagrams without connecting their legs.
Bending the two matrix legs to exchange their left and right positions represents the transpose in the chosen basis; it does not complex-conjugate the entries.
Matrix multiplication joins the input leg of the left factor to the output leg of the right factor, so the diagram with $A$ to the left of $B$ represents $AB$.
Joining the two remaining legs of a matrix takes its trace.
These elementary operations are illustrated in Fig.~\ref{fig:tensor_basic_operations}.

\begin{figure}[htbp]
    \centering

    \begin{tikzpicture}[baseline=(current bounding box.center)]
        \node[tensorbox, minimum width=1.15cm, minimum height=1.0cm] (T) at (0,0) {$T$};
        \draw[leg] (-1.0,0.30) -- (T.west |- -1.0,0.30);
        \draw[leg] (-1.0,-0.30) -- (T.west |- -1.0,-0.30);
        \draw[leg] (T.east |- 1.0,0.30) -- (1.0,0.30);
        \draw[leg] (T.east |- 1.0,-0.30) -- (1.0,-0.30);
        \node at (0,-1.15) {many-body tensor};
    \end{tikzpicture}
    \hspace{0.55cm}
    \begin{tikzpicture}[baseline=(current bounding box.center)]
        \node[tensorbox] (A) at (0,0.55) {$A$};
        \node[tensorbox] (B) at (0,-0.55) {$B$};
        \draw[leg] (-0.95,0.55) -- (A.west);
        \draw[leg] (A.east) -- (0.95,0.55);
        \draw[leg] (-0.95,-0.55) -- (B.west);
        \draw[leg] (B.east) -- (0.95,-0.55);
        \node at (0,-1.45) {$A \otimes B$};
    \end{tikzpicture}
    \hspace{0.55cm}
\begin{tikzpicture}[baseline=(current bounding box.center)]
    \node[tensorbox] (A) at (0,0) {$A$};

    \coordinate (L) at ($(A.west)+(-0.25,0)$);
    \coordinate (R) at ($(A.east)+(0.25,0)$);
    \draw[leg] (A.west) -- (L);
    \draw[leg] (A.east) -- (R);

    \draw[leg] (R) .. controls ($(R)+(0.40,0)$) and ($(R)+(0.40,0.60)$) .. ($(R)+(-0.30,0.60)$);
    \draw[leg] ($(R)+(-0.30,0.60)$) -- ($(R)+(-1.20,0.60)$);

    \draw[leg] (L) .. controls ($(L)+(-0.40,0)$) and ($(L)+(-0.40,-0.60)$) .. ($(L)+(0.30,-0.60)$);
    \draw[leg] ($(L)+(0.30,-0.60)$) -- ($(L)+(1.20,-0.60)$);

    \node at (0,-1.20) {$A^{\mathrm{T}}$};
\end{tikzpicture}
\hspace{0.55cm}
    \begin{tikzpicture}[baseline=(current bounding box.center)]
        \node[tensorbox] (A) at (-0.75,0) {$A$};
        \node[tensorbox] (B) at (0.75,0) {$B$};
        \draw[leg] (-1.8,0) -- (A.west);
        \draw[leg] (A.east) -- ($(A.east)!0.5!(B.west)$);
        \draw[leg] ($(A.east)!0.5!(B.west)$) -- (B.west);
        \draw[leg] (B.east) -- (1.8,0);
        \node at (0,-0.95) {$AB$};
    \end{tikzpicture}

    \vspace{0.6cm}

    \begin{tikzpicture}[baseline=(current bounding box.center)]
        \node[tensorbox] (A) at (0,0) {$A$};

        \coordinate (L) at ($(A.west)+(-0.28,0)$);
        \coordinate (R) at ($(A.east)+(0.28,0)$);
        \draw[leg] (A.west) -- (L);
        \draw[leg] (A.east) -- (R);

        \draw[leg] (L) arc[start angle=-90,end angle=-270,radius=0.38];

        \draw[leg] (R) arc[start angle=-90,end angle=90,radius=0.38];

        \draw[leg] ($(L)+(0,0.76)$) -- ($(R)+(0,0.76)$);
    \end{tikzpicture}
    \hspace{0.4cm}
    {\Large $=$}
    \hspace{0.4cm}
    \begin{tikzpicture}[baseline=(current bounding box.center)]
        \node[tensorbox] (A) at (0,0) {$A$};

        \coordinate (L) at ($(A.west)+(-0.42,0)$);
        \coordinate (R) at ($(A.east)+(0.42,0)$);
        \draw[leg] (A.west) -- (L);
        \draw[leg] (A.east) -- (R);

        \draw[gray, dashed, thick] (L) -- ($(L)+(0,0.5)$);
        \draw[gray, dashed, thick] (L) -- ($(L)+(0,-0.5)$);
        \draw[gray, dashed, thick] (R) -- ($(R)+(0,0.5)$);
        \draw[gray, dashed, thick] (R) -- ($(R)+(0,-0.5)$);
    \end{tikzpicture}

    \vspace{0.2cm}

    \centerline{$\mathrm{Tr}(A)$}

    \caption{Elementary tensor-graph operations: a many-body tensor, tensor product, transpose, matrix multiplication, and two equivalent graphical representations of $\mathrm{Tr}(A)$, where the grey dashed lines represent the periodic boundary condition.}
    \label{fig:tensor_basic_operations}
\end{figure}

The identity is a straight line, and the SWAP operator consists of two crossing lines.
The unnormalized maximally entangled vector
\begin{equation}
    \ket{\Omega_d}=\sum_{i=0}^{d-1}\ket{ii}=\sqrt d\ket{\Phi_d}
\end{equation}
is a bent line with both free legs on the left; its dual has both legs on the right.
These bends carry no normalization factor, so a closed identity loop evaluates to $d$.
Figure~\ref{fig:tensor_special_objects} shows these special tensors.

\begin{figure}[htbp]
    \centering

    \begin{tikzpicture}[baseline=(current bounding box.center)]
        \draw[leg] (-1.0,0) -- (1.0,0);
        \node at (0,-0.75) {identity};
    \end{tikzpicture}
    \hspace{1.1cm}
    \begin{tikzpicture}[baseline=(current bounding box.center)]
        \draw[leg] (-1.0,0.35) -- (1.0,-0.35);
        \draw[leg] (-1.0,-0.35) -- (1.0,0.35);
        \node at (0,-0.95) {SWAP};
    \end{tikzpicture}
    \hspace{1.1cm}
    \begin{tikzpicture}[baseline=(current bounding box.center)]
        \draw[leg] (-0.75,0.28) .. controls (0.25,0.28) and (0.25,-0.28) .. (-0.75,-0.28);
        \node at (-0.15,-0.95) {maximally entangled ket};
    \end{tikzpicture}
    \hspace{1.1cm}
    \begin{tikzpicture}[baseline=(current bounding box.center)]
        \draw[leg] (0.75,0.28) .. controls (-0.25,0.28) and (-0.25,-0.28) .. (0.75,-0.28);
        \node at (0.15,-0.95) {maximally entangled bra};
    \end{tikzpicture}

    \caption{Special tensors in the tensor-graph notation: the identity, the SWAP operator, and the unnormalized maximally entangled state vector and its dual.}
    \label{fig:tensor_special_objects}
\end{figure}

\subsection{Unitary ensembles}

We collect the ensemble definitions and integration formulas used in the moment calculations below.
Throughout, $d$ denotes the total query dimension, and $d$ is even whenever the symplectic form $J$ is used.
We first describe Haar integration on the unitary, orthogonal, and compact symplectic groups, then turn to the circular ensembles that model dynamical TRS.
The orthogonal and symplectic groups satisfy antiunitary operator invariance, whereas the corresponding circular ensembles satisfy the forward--backward symmetry in \eqref{eq:main-dynamical-trs}.
The integration formulas follow from Weingarten calculus~\cite{CollinsMatsumoto2009OrthogonalWeingarten,Matsumoto2013SymmetricSpaces}.
When the invariant tensors are linearly dependent, the corresponding Gram matrices are inverted in the Moore--Penrose sense.

\subsubsection{Haar-random unitary ensemble}

Let $\mathcal{G}$ be a compact group.
The normalized Haar measure $\mu$ is the unique probability measure invariant under both left and right multiplication:
\begin{equation}
    \int_{\mathcal{G}} f(g)d\mu(g)
    =
    \int_{\mathcal{G}} f(g_0g)d\mu(g)
    =
    \int_{\mathcal{G}} f(gg_0)d\mu(g)
\end{equation}
for any fixed $g_0\in \mathcal{G}$ and any integrable function $f$.

For the unitary group,
\begin{equation}
    \mathcal{G}=\mathrm{U}(d)=\{U\in \mathrm{GL}(d,\mathbb C):UU^\dagger=\mathbb{I}_d\},
\end{equation}
sampling from this measure gives the Haar unitary ensemble, also called the circular unitary ensemble (CUE).

Let $U=(u_{ij})_{0\leq i,j\leq d-1}$ be Haar-random in $\mathrm{U}(d)$.
The unitary Weingarten formula states that, for
$\mathbf{i}=(i_1,\ldots,i_T)$,
$\mathbf{j}=(j_1,\ldots,j_T)$,
$\mathbf{i}'=(i'_1,\ldots,i'_T)$, and
$\mathbf{j}'=(j'_1,\ldots,j'_T)$,
\begin{equation}
    \mathbb E_U
    \left[
        u_{i_1j_1}\cdots u_{i_Tj_T}
        \overline{u_{i'_1j'_1}}\cdots \overline{u_{i'_Tj'_T}}
    \right]
    =
    \sum_{\sigma,\tau\in \mathcal{S}_T}
    \delta_\sigma(\mathbf{i},\mathbf{i}')
    \delta_\tau(\mathbf{j},\mathbf{j}')
    \mathrm{Wg}^{\mathrm U}(\sigma^{-1}\tau;d),
\end{equation}
where $\mathcal{S}_T$ denotes the permutation group on $T$ elements,
$\mathrm{Wg}^{\mathrm U}$ denotes the Weingarten coefficients for the Haar-random unitary ensemble, and
\begin{equation}
    \delta_\sigma(\mathbf{i},\mathbf{i}')
    =
    \prod_{s=1}^{T}\delta_{i_{\sigma(s)},i'_s}.
\end{equation}
If the numbers of $u$ factors and $\overline{u}$ factors are different, the average vanishes.

The associated $T$-copy twirling map is
\begin{equation}
    \mathcal T_{\mathrm U}^{(T)}(X)
    :=
    \mathbb E_{U\sim \mathrm{Haar}}
    \left[
        U^{\otimes T} X (U^\dagger)^{\otimes T}
    \right].
\end{equation}
Figure~\ref{fig:haar_unitary_pairing_rule} illustrates the index contractions for $T=3$.
Each filled marker is paired with a hollow marker of the same color, corresponding to the two permutations in the Weingarten formula.
Contracting the indices attached to $X$ gives scalar coefficients, while the remaining contractions give permutation operators.
Thus the twirl projects onto the span of permutation operators on the $T$ copies.

\begin{figure}[htbp]
    \centering
    \begin{tikzpicture}[baseline=(current bounding box.center), scale=1.0]

        \node[tensorbox, minimum width=1.6cm, minimum height=2.8cm] (X) at (0,0) {$X$};

        \node[tensorbox, minimum width=0.95cm, minimum height=0.80cm] (U1) at (-2.6, 1.0) {$U$};
        \node[tensorbox, minimum width=0.95cm, minimum height=0.80cm] (U2) at (-2.6, 0.0) {$U$};
        \node[tensorbox, minimum width=0.95cm, minimum height=0.80cm] (U3) at (-2.6,-1.0) {$U$};

        \draw[leg] (-3.6, 1.0) -- (U1.west);
        \draw[leg] (-3.6, 0.0) -- (U2.west);
        \draw[leg] (-3.6,-1.0) -- (U3.west);

        \draw[leg] (U1.east) -- ($(X.west)+(0,1.0)$);
        \draw[leg] (U2.east) -- ($(X.west)+(0,0.0)$);
        \draw[leg] (U3.east) -- ($(X.west)+(0,-1.0)$);

        \fill[forestgreen] ($(U1.west)+(-0.18,0)$) circle (2.2pt);
        \fill[forestgreen] ($(U2.west)+(-0.18,0)$) circle (2.2pt);
        \fill[forestgreen] ($(U3.west)+(-0.18,0)$) circle (2.2pt);

        \fill[red] ($(U1.east)+(0.18,0)$) circle (2.2pt);
        \fill[red] ($(U2.east)+(0.18,0)$) circle (2.2pt);
        \fill[red] ($(U3.east)+(0.18,0)$) circle (2.2pt);

        \node[tensorbox, minimum width=0.95cm, minimum height=0.80cm] (Ud1) at (2.6, 1.0) {$U^\dagger$};
        \node[tensorbox, minimum width=0.95cm, minimum height=0.80cm] (Ud2) at (2.6, 0.0) {$U^\dagger$};
        \node[tensorbox, minimum width=0.95cm, minimum height=0.80cm] (Ud3) at (2.6,-1.0) {$U^\dagger$};

        \draw[leg] ($(X.east)+(0,1.0)$) -- (Ud1.west);
        \draw[leg] ($(X.east)+(0,0.0)$) -- (Ud2.west);
        \draw[leg] ($(X.east)+(0,-1.0)$) -- (Ud3.west);

        \draw[leg] (Ud1.east) -- (3.6, 1.0);
        \draw[leg] (Ud2.east) -- (3.6, 0.0);
        \draw[leg] (Ud3.east) -- (3.6,-1.0);

        \draw[red, thick, fill=white] ($(Ud1.west)+(-0.18,0)$) circle (2.7pt);
        \draw[red, thick, fill=white] ($(Ud2.west)+(-0.18,0)$) circle (2.7pt);
        \draw[red, thick, fill=white] ($(Ud3.west)+(-0.18,0)$) circle (2.7pt);

        \draw[forestgreen, thick, fill=white] ($(Ud1.east)+(0.18,0)$) circle (2.7pt);
        \draw[forestgreen, thick, fill=white] ($(Ud2.east)+(0.18,0)$) circle (2.7pt);
        \draw[forestgreen, thick, fill=white] ($(Ud3.east)+(0.18,0)$) circle (2.7pt);

    \end{tikzpicture}
    \caption{Graphical illustration of the Haar-random unitary integration rule for the three-copy twirling expression $\mathbb E_U[U^{\otimes 3} X (U^\dagger)^{\otimes 3}]$.}
    \label{fig:haar_unitary_pairing_rule}
\end{figure}

For one copy, the twirl is the completely depolarizing map:
\begin{equation}
    \mathcal T_{\mathrm U}^{(1)}(X)
    =
    \frac{\operatorname{Tr}(X)}{d}\mathbb{I}_d .
\end{equation}

Diagrammatically, this contracts $X$ to its trace and leaves an identity operator as
\begin{equation}
    \mathcal T_{\mathrm U}^{(1)}(X)
    =
    \frac{1}{d}
    \left(
    \vcenter{\hbox{
    \begin{tikzpicture}[baseline=(current bounding box.center)]
        \node[tensorbox, minimum height=0.95cm] (X) at (0,0) {$X$};
        \coordinate (L) at ($(X.west)+(-0.42,0)$);
        \coordinate (R) at ($(X.east)+(0.42,0)$);

        \draw[leg] (X.west) -- (L);
        \draw[leg] (X.east) -- (R);

        \draw[gray, dashed, thick] ($(L |- X.north)$) -- ($(L |- X.south)$);
        \draw[gray, dashed, thick] ($(R |- X.north)$) -- ($(R |- X.south)$);
    \end{tikzpicture}
    }}
    \;\times\;
    \vcenter{\hbox{
    \begin{tikzpicture}[baseline=(current bounding box.center)]
        \draw[leg] (-0.36,0) -- (0.36,0);
    \end{tikzpicture}
    }}
    \right).
\end{equation}

For $T=2$ and $d\geq2$, the commutant of $U^{\otimes2}$ is spanned by the identity $\mathbb I$ and the swap operator $\mathbb S$.
Using
\begin{equation}
    \mathrm{Wg}^{\mathrm U}(\mathrm{id}_2;d)=\frac{1}{d^2-1},
\qquad
    \mathrm{Wg}^{\mathrm U}((12);d)=-\frac{1}{d(d^2-1)},
\end{equation}
the Weingarten formula gives
\begin{equation}
\begin{aligned}
    \mathcal T_{\mathrm U}^{(2)}(X)
    &=
    \mathrm{Wg}^{\mathrm U}(\mathrm{id}_2;d)
    \left[
        \operatorname{Tr}(X)\mathbb{I}
        +
        \operatorname{Tr}(\mathbb{S}X)\mathbb{S}
    \right] \\
    &\quad+
    \mathrm{Wg}^{\mathrm U}((12);d)
    \left[
        \operatorname{Tr}(\mathbb{S}X)\mathbb{I}
        +
        \operatorname{Tr}(X)\mathbb{S}
    \right].
\end{aligned}
\end{equation}
Equivalently,
\begin{equation}
    \mathcal T_{\mathrm U}^{(2)}(X)
    =
    \frac{\operatorname{Tr}(X)-\operatorname{Tr}(\mathbb{S}X)/d}{d^2-1}\mathbb{I}
    +
    \frac{\operatorname{Tr}(\mathbb{S}X)-\operatorname{Tr}(X)/d}{d^2-1}\mathbb{S}.
\end{equation}
The four terms in the two-copy Haar-unitary twirl can be represented diagrammatically as
\begin{equation}
\begin{aligned}
    \mathcal T_{\mathrm U}^{(2)}(X)
    =&
    \mathrm{Wg}^{\mathrm U}(\mathrm{id}_2;d)
    \Bigl(
    \vcenter{\hbox{
    \begin{tikzpicture}[baseline=(current bounding box.center)]
        \node[tensorbox, minimum height=0.95cm] (X1) at (0,0) {$X$};

        \coordinate (L11) at ($(X1.west)+(-0.42,0.25)$);
        \coordinate (L12) at ($(X1.west)+(-0.42,-0.25)$);
        \coordinate (R11) at ($(X1.east)+(0.42,0.25)$);
        \coordinate (R12) at ($(X1.east)+(0.42,-0.25)$);
        \draw[leg] (X1.west |- L11) -- (L11);
        \draw[leg] (X1.west |- L12) -- (L12);
        \draw[leg] (X1.east |- R11) -- (R11);
        \draw[leg] (X1.east |- R12) -- (R12);

        \draw[gray, dashed, thick] ($(L11 |- X1.north)$) -- ($(L11 |- X1.south)$);
        \draw[gray, dashed, thick] ($(R11 |- X1.north)$) -- ($(R11 |- X1.south)$);
    \end{tikzpicture}
    }}
    \;\times\;
    \vcenter{\hbox{
    \begin{tikzpicture}[baseline=(current bounding box.center)]
        \draw[leg] (-0.38,0.18) -- (0.38,0.18);
        \draw[leg] (-0.38,-0.18) -- (0.38,-0.18);
    \end{tikzpicture}
    }}
    \;+\;
    \vcenter{\hbox{
    \begin{tikzpicture}[baseline=(current bounding box.center)]
        \node[tensorbox, minimum height=0.95cm] (X2) at (0,0) {$X$};

        \coordinate (LN21) at ($(X2.west)+(-0.18,0.25)$);
        \coordinate (LN22) at ($(X2.west)+(-0.18,-0.25)$);
        \coordinate (LF21) at ($(X2.west)+(-0.80,0.25)$);
        \coordinate (LF22) at ($(X2.west)+(-0.80,-0.25)$);

        \coordinate (R21) at ($(X2.east)+(0.42,0.25)$);
        \coordinate (R22) at ($(X2.east)+(0.42,-0.25)$);

        \draw[leg] (X2.west |- LN21) -- (LN21);
        \draw[leg] (X2.west |- LN22) -- (LN22);
        \draw[leg] (LN21) -- (LF22);
        \draw[leg] (LN22) -- (LF21);

        \draw[leg] (X2.east |- R21) -- (R21);
        \draw[leg] (X2.east |- R22) -- (R22);

        \draw[gray, dashed, thick] ($(LF21 |- X2.north)$) -- ($(LF21 |- X2.south)$);
        \draw[gray, dashed, thick] ($(R21 |- X2.north)$) -- ($(R21 |- X2.south)$);
    \end{tikzpicture}
    }}
    \;\times\;
    \vcenter{\hbox{
    \begin{tikzpicture}[baseline=(current bounding box.center)]
        \draw[leg] (-0.38,0.18) -- (0.38,-0.18);
        \draw[leg] (-0.38,-0.18) -- (0.38,0.18);
    \end{tikzpicture}
    }}
    \Bigr)
    \\
    &+
    \mathrm{Wg}^{\mathrm U}((12);d)
    \Bigl(
    \vcenter{\hbox{
    \begin{tikzpicture}[baseline=(current bounding box.center)]
        \node[tensorbox, minimum height=0.95cm] (X3) at (0,0) {$X$};

        \coordinate (LN31) at ($(X3.west)+(-0.18,0.25)$);
        \coordinate (LN32) at ($(X3.west)+(-0.18,-0.25)$);
        \coordinate (LF31) at ($(X3.west)+(-0.80,0.25)$);
        \coordinate (LF32) at ($(X3.west)+(-0.80,-0.25)$);

        \coordinate (R31) at ($(X3.east)+(0.42,0.25)$);
        \coordinate (R32) at ($(X3.east)+(0.42,-0.25)$);

        \draw[leg] (X3.west |- LN31) -- (LN31);
        \draw[leg] (X3.west |- LN32) -- (LN32);
        \draw[leg] (LN31) -- (LF32);
        \draw[leg] (LN32) -- (LF31);

        \draw[leg] (X3.east |- R31) -- (R31);
        \draw[leg] (X3.east |- R32) -- (R32);

        \draw[gray, dashed, thick] ($(LF31 |- X3.north)$) -- ($(LF31 |- X3.south)$);
        \draw[gray, dashed, thick] ($(R31 |- X3.north)$) -- ($(R31 |- X3.south)$);
    \end{tikzpicture}
    }}
    \;\times\;
    \vcenter{\hbox{
    \begin{tikzpicture}[baseline=(current bounding box.center)]
        \draw[leg] (-0.38,0.18) -- (0.38,0.18);
        \draw[leg] (-0.38,-0.18) -- (0.38,-0.18);
    \end{tikzpicture}
    }}
    \;+\;
    \vcenter{\hbox{
    \begin{tikzpicture}[baseline=(current bounding box.center)]
        \node[tensorbox, minimum height=0.95cm] (X4) at (0,0) {$X$};

        \coordinate (L41) at ($(X4.west)+(-0.42,0.25)$);
        \coordinate (L42) at ($(X4.west)+(-0.42,-0.25)$);
        \coordinate (R41) at ($(X4.east)+(0.42,0.25)$);
        \coordinate (R42) at ($(X4.east)+(0.42,-0.25)$);
        \draw[leg] (X4.west |- L41) -- (L41);
        \draw[leg] (X4.west |- L42) -- (L42);
        \draw[leg] (X4.east |- R41) -- (R41);
        \draw[leg] (X4.east |- R42) -- (R42);

        \draw[gray, dashed, thick] ($(L41 |- X4.north)$) -- ($(L41 |- X4.south)$);
        \draw[gray, dashed, thick] ($(R41 |- X4.north)$) -- ($(R41 |- X4.south)$);
    \end{tikzpicture}
    }}
    \;\times\;
    \vcenter{\hbox{
    \begin{tikzpicture}[baseline=(current bounding box.center)]
        \draw[leg] (-0.38,0.18) -- (0.38,-0.18);
        \draw[leg] (-0.38,-0.18) -- (0.38,0.18);
    \end{tikzpicture}
    }}
    \Bigr).
\end{aligned}
\end{equation}
The dashed boundaries close the scalar trace contractions.
Crossed legs on the left of $X$ insert $\mathbb S$ in $\operatorname{Tr}(\mathbb S X)$.

\subsubsection{Orthogonal group}

The real orthogonal group is
\begin{equation}
    \mathrm O(d)=\{O\in \mathrm{GL}(d,\mathbb R):OO^{\mathrm{T}}=\mathbb I_d\}.
\end{equation}
We equip this group with its normalized Haar measure.
Its elements satisfy $KOK^{-1}=O$, which is operator invariance under complex conjugation.

Let $O=(o_{ij})_{0\leq i,j\leq d-1}$ be Haar-random in $\mathrm O(d)$.
The orthogonal Weingarten formula states that, for sequences
$\mathbf{i}=(i_1,\ldots,i_{2T})$ and
$\mathbf{j}=(j_1,\ldots,j_{2T})$,
\begin{equation}
    \mathbb E_O
    \left[
        o_{i_1j_1}\cdots o_{i_{2T}j_{2T}}
    \right]
    =
    \sum_{\sigma,\tau\in P_{2T}}
    \Delta_\sigma(\mathbf{i})\Delta_\tau(\mathbf{j})
    \mathrm{Wg}^{\mathrm O}(\sigma^{-1}\tau;d),
\end{equation}
where $P_{2T}$ denotes the set of pairings of $\{1,\ldots,2T\}$, represented by permutations $\sigma$ satisfying $\sigma(2r-1)<\sigma(2r)$ and $\sigma(1)<\sigma(3)<\cdots<\sigma(2T-1)$, and
\begin{equation}
    \Delta_\sigma(\mathbf{i})
    =
    \prod_{s=1}^{T}
    \delta_{i_{\sigma(2s-1)},i_{\sigma(2s)}}.
\end{equation}
Equivalently, $P_{2T}$ can be identified with a set of representatives of
$\mathcal S_{2T}/H_T$, where $H_T$ is the hyperoctahedral subgroup that preserves
the standard pairing
\begin{equation}
\{\{1,2\},\{3,4\},\ldots,\{2T-1,2T\}\}.
\end{equation}
Thus $H_T$ permutes the $T$ pairs and swaps the elements within each pair.
All odd moments vanish because Haar measure is invariant under $O\mapsto-O$:
\begin{equation}
    \mathbb E_O
    \left[
        o_{i_1j_1}\cdots o_{i_{2T+1}j_{2T+1}}
    \right]
    =
    0.
\end{equation}

The corresponding $T$-copy twirling map is
\begin{equation}
    \mathcal T_{\mathrm O}^{(T)}(X)
    =
    \mathbb E_O
    \left[
        O^{\otimes T}X(O^{\mathrm{T}})^{\otimes T}
    \right].
\end{equation}
For $T=1$, one has
\begin{equation}
    \mathcal T_{\mathrm O}^{(1)}(X)
    =
    \frac{\operatorname{Tr}(X)}{d}\mathbb I_d .
\end{equation}

Figure~\ref{fig:orthogonal_pairing_rule} shows the corresponding contractions for $T=3$.
Unlike unitary integration, orthogonal integration allows arbitrary pairings within each set of $2T$ indices.
The pairings attached to $X$ produce scalar contractions, while the external pairings produce Brauer diagrams, including permutations and cup--cap contractions.

\begin{figure}[htbp]
    \centering
    \begin{tikzpicture}[baseline=(current bounding box.center), scale=1.0]

        \node[tensorbox, minimum width=1.6cm, minimum height=2.8cm] (X) at (0,0) {$X$};

        \node[tensorbox, minimum width=0.95cm, minimum height=0.80cm] (O1) at (-2.6, 1.0) {$O$};
        \node[tensorbox, minimum width=0.95cm, minimum height=0.80cm] (O2) at (-2.6, 0.0) {$O$};
        \node[tensorbox, minimum width=0.95cm, minimum height=0.80cm] (O3) at (-2.6,-1.0) {$O$};

        \draw[leg] (-3.6, 1.0) -- (O1.west);
        \draw[leg] (-3.6, 0.0) -- (O2.west);
        \draw[leg] (-3.6,-1.0) -- (O3.west);

        \draw[leg] (O1.east) -- ($(X.west)+(0,1.0)$);
        \draw[leg] (O2.east) -- ($(X.west)+(0,0.0)$);
        \draw[leg] (O3.east) -- ($(X.west)+(0,-1.0)$);

        \node[tensorbox, minimum width=0.95cm, minimum height=0.80cm] (OT1) at (2.6, 1.0) {$O^{\mathrm T}$};
        \node[tensorbox, minimum width=0.95cm, minimum height=0.80cm] (OT2) at (2.6, 0.0) {$O^{\mathrm T}$};
        \node[tensorbox, minimum width=0.95cm, minimum height=0.80cm] (OT3) at (2.6,-1.0) {$O^{\mathrm T}$};

        \draw[leg] ($(X.east)+(0,1.0)$) -- (OT1.west);
        \draw[leg] ($(X.east)+(0,0.0)$) -- (OT2.west);
        \draw[leg] ($(X.east)+(0,-1.0)$) -- (OT3.west);

        \draw[leg] (OT1.east) -- (3.6, 1.0);
        \draw[leg] (OT2.east) -- (3.6, 0.0);
        \draw[leg] (OT3.east) -- (3.6,-1.0);

        \draw[forestgreen, thick, fill=white] ($(O1.west)+(-0.16,0)$) circle (3.0pt);
        \fill[forestgreen] ($(O1.west)+(-0.16,0)$) circle (1.5pt);
        \draw[forestgreen, thick, fill=white] ($(O2.west)+(-0.16,0)$) circle (3.0pt);
        \fill[forestgreen] ($(O2.west)+(-0.16,0)$) circle (1.5pt);
        \draw[forestgreen, thick, fill=white] ($(O3.west)+(-0.16,0)$) circle (3.0pt);
        \fill[forestgreen] ($(O3.west)+(-0.16,0)$) circle (1.5pt);

        \draw[forestgreen, thick, fill=white] ($(OT1.east)+(0.16,0)$) circle (3.0pt);
        \fill[forestgreen] ($(OT1.east)+(0.16,0)$) circle (1.5pt);
        \draw[forestgreen, thick, fill=white] ($(OT2.east)+(0.16,0)$) circle (3.0pt);
        \fill[forestgreen] ($(OT2.east)+(0.16,0)$) circle (1.5pt);
        \draw[forestgreen, thick, fill=white] ($(OT3.east)+(0.16,0)$) circle (3.0pt);
        \fill[forestgreen] ($(OT3.east)+(0.16,0)$) circle (1.5pt);

        \draw[red, thick, fill=white] ($(O1.east)+(0.16,0)$) circle (3.0pt);
        \fill[red] ($(O1.east)+(0.16,0)$) circle (1.5pt);
        \draw[red, thick, fill=white] ($(O2.east)+(0.16,0)$) circle (3.0pt);
        \fill[red] ($(O2.east)+(0.16,0)$) circle (1.5pt);
        \draw[red, thick, fill=white] ($(O3.east)+(0.16,0)$) circle (3.0pt);
        \fill[red] ($(O3.east)+(0.16,0)$) circle (1.5pt);

        \draw[red, thick, fill=white] ($(OT1.west)+(-0.16,0)$) circle (3.0pt);
        \fill[red] ($(OT1.west)+(-0.16,0)$) circle (1.5pt);
        \draw[red, thick, fill=white] ($(OT2.west)+(-0.16,0)$) circle (3.0pt);
        \fill[red] ($(OT2.west)+(-0.16,0)$) circle (1.5pt);
        \draw[red, thick, fill=white] ($(OT3.west)+(-0.16,0)$) circle (3.0pt);
        \fill[red] ($(OT3.west)+(-0.16,0)$) circle (1.5pt);

    \end{tikzpicture}
    \caption{Orthogonal Haar integration for $\mathbb E_O[O^{\otimes3}X(O^{\mathrm T})^{\otimes3}]$.
    The six markers of each color are partitioned into three pairs, independently for the two colors.
    Internal pairings contract $X$ to scalar coefficients; external pairings give Brauer diagrams.}
    \label{fig:orthogonal_pairing_rule}
\end{figure}

For $T=2$ and $d\geq2$, the commutant is spanned by three Brauer diagrams,
\begin{equation}
    B_1=\mathbb I,
\qquad    
    B_2=\mathbb S,
\qquad
    B_3=\Phi ,
\end{equation}
where $\ket{\Omega}=\sum_{j=0}^{d-1}\ket{j,j}$ and $\Phi=\ketbra{\Omega}{\Omega}$.
In particular, $\Phi$ is unnormalized, with $\operatorname{Tr}\Phi=d$.
The Hilbert--Schmidt Gram matrix is
\begin{equation}
    G_{\mathrm O}
    =
    \left(
        \operatorname{Tr}(B_a^\dagger B_b)
    \right)_{a,b=1}^{3}
    =
    \begin{bmatrix}
        d^2 & d & d \\
        d & d^2 & d \\
        d & d & d^2
    \end{bmatrix}.
\end{equation}
The twirl is the Hilbert--Schmidt orthogonal projection onto this commutant,
\begin{equation}
    \mathcal T_{\mathrm O}^{(2)}(X)
    =
    \sum_{a,b=1}^{3}
    (G_{\mathrm O}^{-1})_{ab}
    \operatorname{Tr}(B_a^\dagger X)B_b.
\end{equation}
Expanding the scalar contractions and output operators gives
\begin{equation}
\begin{aligned}
    &\mathcal T_{\mathrm O}^{(2)}(X)
    =
    (G_{\mathrm O}^{-1})_{11}
    \Bigl(
        \OTrXDiag
        \times
        \OIdDiag
    \Bigr)
    +
    (G_{\mathrm O}^{-1})_{12}
    \Bigl(
        \OTrXDiag
        \times
        \OSwapDiag
    \Bigr)
    +
    (G_{\mathrm O}^{-1})_{13}
    \Bigl(
        \OTrXDiag
        \times
        \OPhiDiag
    \Bigr)
    \\
    &+
    (G_{\mathrm O}^{-1})_{21}
    \Bigl(
        \OTrSXDiag
        \times
        \OIdDiag
    \Bigr)
    +
    (G_{\mathrm O}^{-1})_{22}
    \Bigl(
        \OTrSXDiag
        \times
        \OSwapDiag
    \Bigr)
    +
    (G_{\mathrm O}^{-1})_{23}
    \Bigl(
        \OTrSXDiag
        \times
        \OPhiDiag
    \Bigr)
    \\
    &+
    (G_{\mathrm O}^{-1})_{31}
    \Bigl(
        \OTrPhiXDiag
        \times
        \OIdDiag
    \Bigr)
    +
    (G_{\mathrm O}^{-1})_{32}
    \Bigl(
        \OTrPhiXDiag
        \times
        \OSwapDiag
    \Bigr)
    +
    (G_{\mathrm O}^{-1})_{33}
    \Bigl(
        \OTrPhiXDiag
        \times
        \OPhiDiag
    \Bigr).
\end{aligned}
\end{equation}

The scalar diagrams represent $\operatorname{Tr}(X)$, $\operatorname{Tr}(\mathbb S X)$, and $\operatorname{Tr}(\Phi X)$.
The output diagrams represent $\mathbb I$, $\mathbb S$, and $\Phi$, respectively.

\subsubsection{Symplectic group}

Assume that $d$ is even and use the canonical form from \eqref{eq:main-trs-classes}:
\begin{equation}
    J=J_d=i\sigma_y\otimes\mathbb I_{d/2}=
    \begin{bmatrix}
        0 & \mathbb I_{d/2} \\
        -\mathbb I_{d/2} & 0
    \end{bmatrix}.
\end{equation}
This real matrix satisfies $J^{\mathrm T}=J^\dagger=-J$ and $J^2=-\mathbb I_d$.
For a $d\times d$ matrix $X$, the symplectic dual is
\begin{equation}
    X^{\mathrm D}=JX^{\mathrm T}J^\dagger.
\end{equation}
The compact symplectic group is realized as
\begin{equation}
    \mathrm{Sp}(d)
    =
    \{S\in \mathrm U(d):SS^{\mathrm D}=\mathbb I_{d}\}.
\end{equation}
Equivalently, $SJS^{\mathrm T}=J$, or $(JK)S(JK)^{-1}=S$.
Our notation $\mathrm{Sp}(d)$ refers to $d\times d$ matrices; the group has rank $d/2$ and is also commonly denoted $\mathrm{USp}(d)$.

Let $S=(s_{ij})_{0\leq i,j\leq d-1}$ be Haar-random in $\mathrm{Sp}(d)$.
The symplectic Weingarten formula states that, for sequences
$\mathbf{i}=(i_1,\ldots,i_{2T})$ and
$\mathbf{j}=(j_1,\ldots,j_{2T})$,
\begin{equation}
    \mathbb E_S
    \left[
        s_{i_1j_1}\cdots s_{i_{2T}j_{2T}}
    \right]
    =
    \sum_{\sigma,\tau\in P_{2T}}
    \Delta'_\sigma(\mathbf{i})\Delta'_\tau(\mathbf{j})
    \mathrm{Wg}^{\mathrm{Sp}}(\sigma^{-1}\tau;d/2),
\end{equation}
where the pairing representatives are ordered as above and
\begin{equation}
    \Delta'_\sigma(\mathbf{i})
    =
    \prod_{r=1}^{T}
    \langle i_{\sigma(2r-1)},i_{\sigma(2r)}\rangle,
\qquad    
    \langle a,b\rangle=J_{ab}.
\end{equation}
Again, all odd moments vanish by invariance under $S\mapsto-S$, since
\begin{equation}
    \mathbb E_S
    \left[
        s_{i_1j_1}\cdots s_{i_{2T+1}j_{2T+1}}
    \right]
    =
    0.
\end{equation}
The symplectic Weingarten function uses the conventional rank parameter $z=d/2$, rather than the total dimension $d$.
Writing $\epsilon(\sigma)$ for the sign of a permutation, its relation to the orthogonal Weingarten function is
\begin{equation}
    \mathrm{Wg}^{\mathrm{Sp}}(\sigma;z)
    =
    (-1)^T\epsilon(\sigma)
    \mathrm{Wg}^{\mathrm O}(\sigma;-2z).
\end{equation}

The graphical rule has the same pairing structure as orthogonal integration, but each pairing contracts indices through $J$ instead of the Kronecker delta.
In the diagrams below, a black dot marks an insertion of $J$; the orientation of each contraction follows the displayed algebraic formulas.

The corresponding $T$-copy twirling map is
\begin{equation}
    \mathcal T_{\mathrm{Sp}}^{(T)}(X)
    =
    \mathbb E_S
    \left[
        S^{\otimes T}X(S^\dagger)^{\otimes T}
    \right].
\end{equation}
For one copy, $S^\dagger=S^{\mathrm D}=JS^{\mathrm T}J^{\mathrm T}$ and the invariant contraction is
\begin{equation}
    \mathbb E_S[SAS^{\mathrm T}]
    =
    \frac{\operatorname{Tr}(J^{\mathrm T}A)}{d}J.
\end{equation}
Taking $A=XJ$ gives
\begin{equation}
    \mathcal T_{\mathrm{Sp}}^{(1)}(X)
    =
    \mathbb E_S[S(XJ)S^{\mathrm T}]J^{\mathrm T}
    =
    \frac{\operatorname{Tr}(X)}{d}\mathbb I_d.
\end{equation}
Thus all three group ensembles have the same one-copy conjugation average.

For $T=2$, define
\begin{equation}
    \ket{\Omega_J}
    =
    (\mathbb I_{d}\otimes J)\ket{\Omega},
\qquad    
    \Phi_J
    =
    \ketbra{\Omega_J}{\Omega_J}.
\end{equation}
The commutant is spanned by
\begin{equation}
    C_1=\mathbb I,
    \qquad
    C_2=\mathbb S,
    \qquad
    C_3=\Phi_J .
\end{equation}
Their Hilbert--Schmidt Gram matrix is
\begin{equation}
    G_{\mathrm{Sp}}
    =
    \left(
        \operatorname{Tr}(C_a^\dagger C_b)
    \right)_{a,b=1}^{3}
    =
    \begin{bmatrix}
        d^2 & d & d \\
        d & d^2 & -d \\
        d & -d & d^2
    \end{bmatrix}.
\end{equation}
The negative off-diagonal entries follow from $\mathbb S\ket{\Omega_J}=-\ket{\Omega_J}$.
For $d\geq4$, this Gram matrix is invertible.
For $d=2$, the spanning operators are linearly dependent, and $G_{\mathrm{Sp}}^{-1}$ below denotes the Moore--Penrose pseudoinverse.
As in the orthogonal case, the twirl is the Hilbert--Schmidt orthogonal projection onto the commutant:
\begin{equation}
    \mathcal T_{\mathrm{Sp}}^{(2)}(X)
    =
    \sum_{a,b=1}^{3}
    (G_{\mathrm{Sp}}^{-1})_{ab}
    \operatorname{Tr}(C_a^\dagger X)
    C_b .
\end{equation}

Its diagrammatic expansion is
\begin{equation}
\begin{aligned}
    &\mathcal T_{\mathrm{Sp}}^{(2)}(X)
    =
    (G_{\mathrm{Sp}}^{-1})_{11}
    \Bigl(
        \OTrXDiag
        \times
        \OIdDiag
    \Bigr)
    +
    (G_{\mathrm{Sp}}^{-1})_{12}
    \Bigl(
        \OTrXDiag
        \times
        \OSwapDiag
    \Bigr)
    +
    (G_{\mathrm{Sp}}^{-1})_{13}
    \Bigl(
        \OTrXDiag
        \times
        \SpPhiDiag
    \Bigr)
    \\
    &+
    (G_{\mathrm{Sp}}^{-1})_{21}
    \Bigl(
        \OTrSXDiag
        \times
        \OIdDiag
    \Bigr)
    +
    (G_{\mathrm{Sp}}^{-1})_{22}
    \Bigl(
        \OTrSXDiag
        \times
        \OSwapDiag
    \Bigr)
    +
    (G_{\mathrm{Sp}}^{-1})_{23}
    \Bigl(
        \OTrSXDiag
        \times
        \SpPhiDiag
    \Bigr)
    \\
    &+
    (G_{\mathrm{Sp}}^{-1})_{31}
    \Bigl(
        \SpTrPhiXDiag
        \times
        \OIdDiag
    \Bigr)
    +
    (G_{\mathrm{Sp}}^{-1})_{32}
    \Bigl(
        \SpTrPhiXDiag
        \times
        \OSwapDiag
    \Bigr)
    +
    (G_{\mathrm{Sp}}^{-1})_{33}
    \Bigl(
        \SpTrPhiXDiag
        \times
        \SpPhiDiag
    \Bigr).
\end{aligned}
\end{equation}
The scalar diagrams represent $\operatorname{Tr}(X)$, $\operatorname{Tr}(\mathbb S X)$, and $\operatorname{Tr}(\Phi_J X)$; the output diagrams represent $\mathbb I$, $\mathbb S$, and $\Phi_J$.
The black dots distinguish the $J$ contractions from the orthogonal case.

\subsubsection{Circular orthogonal ensemble}

The circular orthogonal ensemble (COE) is the invariant probability distribution on the symmetric unitary matrices, identified with the compact symmetric space $\mathrm U(d)/\mathrm O(d)$.
If $W$ is Haar distributed on $\mathrm U(d)$, the construction
\begin{equation}
    V=W^{\mathrm T}W,
     V^{\mathrm T}=V,
     VV^\dagger=I_d
\end{equation}
realizes this distribution, consistently with \eqref{eq:coe-cse-definitions}.
We first give its entry-moment formula and then derive the operator decomposition used in the query lower bound.

Let $V=(v_{ij})_{1\leq i,j\leq d}$ be a COE matrix.
For $1\leq T\leq d$ and multi-indices $\mathbf{i},\mathbf{j}\in\{1,\ldots,d\}^{2T}$, the COE Weingarten formula is~\cite{Matsumoto2012COEMoments,Matsumoto2013SymmetricSpaces}
\begin{equation}
    \mathbb E_V
    \left[
        v_{i_1i_2}v_{i_3i_4}\cdots v_{i_{2T-1}i_{2T}}
        \overline{
        v_{j_1j_2}v_{j_3j_4}\cdots v_{j_{2T-1}j_{2T}}
        }
    \right]
    =
    \sum_{\sigma\in S_{2T}}
    \delta_\sigma(\mathbf{i},\mathbf{j})
    \mathrm{Wg}^{\mathrm{COE}}(\sigma;d),
\end{equation}
where
\begin{equation}
    \delta_\sigma(\mathbf{i},\mathbf{j})
    =
    \prod_{r=1}^{2T}
    \delta_{i_{\sigma(r)},j_r}.
\end{equation}
The COE Weingarten function is directly related to the orthogonal one by
\begin{equation}
    \mathrm{Wg}^{\mathrm{COE}}(\sigma;d)
    =
    \mathrm{Wg}^{\mathrm O}(\sigma;d+1).
\end{equation}
The shifted parameter $d+1$ will determine the eigenvalues of the COE moment operator.
The permutation sum can also be organized by pairings.
Let $H_T\subseteq S_{2T}$ be the subgroup preserving the standard pairing $\{\{1,2\},\ldots,\{2T-1,2T\}\}$, and choose pairing representatives $P_{2T}$ for $S_{2T}/H_T$.
Each $\sigma\in S_{2T}$ has a unique decomposition $\sigma=mh$ with $m\in P_{2T}$ and $h\in H_T$, where
\begin{equation}
    |H_T|=2^T T!,
     |P_{2T}|=(2T-1)!!,
     |P_{2T}||H_T|=(2T)!.
\end{equation}
The subgroup permutes complete pairs and swaps the two elements of each pair.
Since the orthogonal Weingarten function is $H_T$-bi-invariant, the coefficient depends only on the pairing representative:
\begin{equation}
    \mathrm{Wg}^{\mathrm{COE}}(mh;d)
    =\mathrm{Wg}^{\mathrm O}(m;d+1).
\end{equation}

Figure~\ref{fig:coe_pairing_rule} illustrates the entry contractions for the three-copy twirl.
Each of the $2T$ indices of the $V$ factors is matched to an index of the $V^\dagger$ factors, including indices on either side of each matrix box.
No contractions join two indices belonging only to the $V$ factors or only to the $V^\dagger$ factors.
These contractions produce traces and transposed tensor structures involving $X$, as illustrated by the one-copy formula below.

\begin{figure}[htbp]
    \centering
    \begin{tikzpicture}[baseline=(current bounding box.center), scale=1.0]

        \node[tensorbox, minimum width=1.6cm, minimum height=2.8cm] (X) at (0,0) {$X$};

        \node[tensorbox, minimum width=1.15cm, minimum height=0.80cm] (V1) at (-2.6, 1.0) {$V$};
        \node[tensorbox, minimum width=1.15cm, minimum height=0.80cm] (V2) at (-2.6, 0.0) {$V$};
        \node[tensorbox, minimum width=1.15cm, minimum height=0.80cm] (V3) at (-2.6,-1.0) {$V$};

        \draw[leg] (-3.6, 1.0) -- (V1.west);
        \draw[leg] (-3.6, 0.0) -- (V2.west);
        \draw[leg] (-3.6,-1.0) -- (V3.west);

        \draw[leg] (V1.east) -- ($(X.west)+(0,1.0)$);
        \draw[leg] (V2.east) -- ($(X.west)+(0,0.0)$);
        \draw[leg] (V3.east) -- ($(X.west)+(0,-1.0)$);

        \node[tensorbox, minimum width=1.15cm, minimum height=0.80cm] (VD1) at (2.7, 1.0) {$V^\dagger$};
        \node[tensorbox, minimum width=1.15cm, minimum height=0.80cm] (VD2) at (2.7, 0.0) {$V^\dagger$};
        \node[tensorbox, minimum width=1.15cm, minimum height=0.80cm] (VD3) at (2.7,-1.0) {$V^\dagger$};

        \draw[leg] ($(X.east)+(0,1.0)$) -- (VD1.west);
        \draw[leg] ($(X.east)+(0,0.0)$) -- (VD2.west);
        \draw[leg] ($(X.east)+(0,-1.0)$) -- (VD3.west);

        \draw[leg] (VD1.east) -- (3.9, 1.0);
        \draw[leg] (VD2.east) -- (3.9, 0.0);
        \draw[leg] (VD3.east) -- (3.9,-1.0);

        \fill[forestgreen] ($(V1.west)+(-0.16,0)$) circle (2.2pt);
        \fill[forestgreen] ($(V2.west)+(-0.16,0)$) circle (2.2pt);
        \fill[forestgreen] ($(V3.west)+(-0.16,0)$) circle (2.2pt);

        \fill[forestgreen] ($(V1.east)+(0.16,0)$) circle (2.2pt);
        \fill[forestgreen] ($(V2.east)+(0.16,0)$) circle (2.2pt);
        \fill[forestgreen] ($(V3.east)+(0.16,0)$) circle (2.2pt);

        \draw[forestgreen, thick, fill=white] ($(VD1.west)+(-0.16,0)$) circle (2.8pt);
        \draw[forestgreen, thick, fill=white] ($(VD2.west)+(-0.16,0)$) circle (2.8pt);
        \draw[forestgreen, thick, fill=white] ($(VD3.west)+(-0.16,0)$) circle (2.8pt);

        \draw[forestgreen, thick, fill=white] ($(VD1.east)+(0.16,0)$) circle (2.8pt);
        \draw[forestgreen, thick, fill=white] ($(VD2.east)+(0.16,0)$) circle (2.8pt);
        \draw[forestgreen, thick, fill=white] ($(VD3.east)+(0.16,0)$) circle (2.8pt);

    \end{tikzpicture}
    \caption{Graphical illustration of the COE integral for the three-copy twirling expression $\mathbb E_V[V^{\otimes 3}X(V^\dagger)^{\otimes 3}]$. The solid green markers and hollow green markers may be paired with one another.}
    \label{fig:coe_pairing_rule}
\end{figure}

At one copy, the two permutations give
\begin{equation}
    \mathcal T_{\mathrm{COE}}^{(1)}(X)
    =
    \mathbb E_V[VXV^\dagger]
    =
    \frac{\operatorname{Tr}(X)I_d+X^{\mathrm T}}{d+1}.
\end{equation}
The transpose term is the one-query distinction between the COE mean channel and the completely depolarizing channel.
The same identity has the graphical form
\begin{equation}
    \mathcal T_{\mathrm{COE}}^{(1)}(X)
    =
    \frac{1}{d+1}
    \left(
        \COETraceDiag
        \times
        \COEIdDiag
        +
        \COETransposeDiag
    \right).
\end{equation}
Here the first term represents $\operatorname{Tr}(X)I_d$, while the second term represents the transpose contribution $X^{\mathrm T}$.

\subsubsection{Spectral decomposition of the COE moment operator}
\label{sec:coe-moment-operator}

The lower-bound proof uses moments of the vectorized unitary rather than the twirling map acting on an arbitrary operator.
The following decomposition identifies the support and eigenvalues of these moments.
For a COE matrix $V=(v_{ab})_{1\leq a,b\leq d}$, define the unnormalized vectorization and its $T$-copy moment operator by
\begin{equation}
    \ket{v}:=\operatorname{vec}(V)
    =\sum_{a,b=1}^{d}v_{ab}\ket{a}_{\mathrm{out}}\otimes\ket{b}_{\mathrm{in}},
\qquad    
    M_T^{\mathrm{COE}}:=\mathbb{E}_{V} \left[
        \bigl(\ketbra{v}{v}\bigr)^{\otimes T}
    \right].
    \label{eq:coe-moment-definition}
\end{equation}
Its tensor-network representation is
\begin{equation}
M_T^{\mathrm{COE}}
=
\mathbb E_{V\sim\mathrm{COE}} \left[
\;\vcenter{\hbox{%
\begin{tikzpicture}[
  x=1cm,y=1cm,
  line width=0.6pt,
  font=\small,
  op/.style={
    draw,
    fill=white,
    minimum width=0.64cm,
    minimum height=0.46cm,
    inner sep=2pt
  }
]

\foreach \y in {0,-2.40}{
  \begin{scope}[shift={(0,\y)}]

    \draw (0,0.35)
      arc[start angle=90,end angle=-90,
          x radius=0.45cm,y radius=0.35cm];

    \draw (-1.20,0.35) -- (0,0.35);
    \draw (-1.20,-0.35) -- (0,-0.35);
    \node[op] at (-0.60,-0.35) {$V$};

    \draw (1.40,0.35)
      arc[start angle=90,end angle=270,
          x radius=0.45cm,y radius=0.35cm];

    \draw (1.40,0.35) -- (2.60,0.35);
    \draw (1.40,-0.35) -- (2.60,-0.35);
    \node[op] at (2.00,-0.35) {$V^\dagger$};

  \end{scope}
}

\node at (-0.35,-1.20) {$\vdots$};
\node at (1.75,-1.20) {$\vdots$};
\node[font=\scriptsize] at (0.70,-1.20) {$T$ copies};

\end{tikzpicture}%
}}\;
\right].
\label{eq:coe-moment-diagram}
\end{equation}
The tensor factors are ordered as $(\mathrm{out}_1,\mathrm{in}_1,\ldots,\mathrm{out}_T,\mathrm{in}_T)$, so $M_T^{\mathrm{COE}}$ acts on $(\mathbb{C}^{d})^{\otimes 2T}$ and satisfies $\operatorname{Tr}(M_T^{\mathrm{COE}})=d^T$.
The corresponding moment operator for normalized Choi states is $d^{-T}M_T^{\mathrm{COE}}$.

For the pairing subgroup $H_T$ defined above, introduce the tensor-permutation operators and their group average:
\begin{equation}
    P_\sigma\ket{i_1,\ldots,i_{2T}}
    =\ket{i_{\sigma^{-1}(1)},\ldots,i_{\sigma^{-1}(2T)}},
\qquad    
    \Pi_H:=\frac{1}{|H_T|}\sum_{\eta\in H_T}P_\eta.
    \label{eq:coe-pairing-projector}
\end{equation}
Thus $\Pi_H$ projects onto tensors that are symmetric within each pair and invariant under permutations of complete pairs.
Because $V$ is symmetric and its $T$ copies are identical, $\ket{v}^{\otimes T}$ belongs to this subspace. 
For later use, we recast standard results on spherical-function expansions of Weingarten functions as the following operator spectral decomposition.

\begin{lemma}[Spectral decomposition of the COE moment operator]
\label{lem:coe-moment-spectrum}
Let $1\leq T\leq d$.
There exist mutually orthogonal projectors $Q_\lambda$, indexed by the partitions $\lambda\vdash T$, such that
\begin{equation}
    M_T^{\mathrm{COE}}
    =2^T T!\sum_{\lambda\vdash T}
        \frac{Q_\lambda}{C_\lambda(d+1)},
\qquad    
    C_\lambda(d+1)
    :=\prod_{(i,j)\in\lambda}(d+2j-i),
    \label{eq:coe-moment-spectrum}
\end{equation}
where $(i,j)$ ranges over the boxes in the Young diagram of $\lambda$.
These projectors satisfy
\begin{equation}
    Q_\lambda Q_\mu=\delta_{\lambda\mu}Q_\lambda,
\qquad    
    \sum_{\lambda\vdash T}Q_\lambda=\Pi_H.
    \label{eq:coe-moment-projectors}
\end{equation}
\end{lemma}

\begin{proof}
We first rewrite the COE moment formula in operator form.
For multi-indices $\boldsymbol{i},\boldsymbol{j}\in\{1,\ldots,d\}^{2T}$, our permutation convention gives
\begin{equation}
    \bra{\boldsymbol{i}}P_\sigma\ket{\boldsymbol{j}}
    =\prod_{r=1}^{2T}\delta_{i_{\sigma(r)},j_r}
    =\delta_\sigma(\boldsymbol{i},\boldsymbol{j}).
    \label{eq:coe-permutation-matrix-elements}
\end{equation}
Consequently, the COE Weingarten formula and the identity
$\operatorname{Wg}^{\mathrm{COE}}(\sigma;d)
=\operatorname{Wg}^{\mathrm{O}}(\sigma;d+1)$ yield~\cite{Matsumoto2012COEMoments,Matsumoto2013SymmetricSpaces}
\begin{equation}
    M_T^{\mathrm{COE}}=\sum_{\sigma\in S_{2T}}
        \operatorname{Wg}^{\mathrm{O}}(\sigma;d+1)P_\sigma.
    \label{eq:coe-moment-permutation-expansion}
\end{equation}
Since $V^\mathrm{T}=V$ and all $T$ copies of $\ket{v}$ are identical, $P_\eta\ket{v}^{\otimes T}=\ket{v}^{\otimes T}$ for every $\eta\in H_T$.
Hence $M_T^{\mathrm{COE}}=\Pi_H M_T^{\mathrm{COE}}\Pi_H$.

We next intersect the $H_T$-invariant subspace with the isotypic components of the ambient tensor representation of $S_{2T}$.
Let $\chi^\nu$ be the irreducible character indexed by $\nu\vdash 2T$, and let $f^\nu=\chi^\nu(e)$ be its dimension, where $e$ is the identity permutation.
The central isotypic projectors in the tensor representation are
\begin{equation}
    E_\nu:=\frac{f^\nu}{(2T)!}
        \sum_{\sigma\in S_{2T}}
        \chi^\nu(\sigma^{-1})P_\sigma.
    \label{eq:coe-central-projectors}
\end{equation}
Character orthogonality gives $E_\nu E_\kappa=\delta_{\nu\kappa}E_\nu$ and $\sum_{\nu\vdash 2T}E_\nu=I$.
Moreover, each $E_\nu$ commutes with $\Pi_H$.
The Gelfand-pair property of $(S_{2T},H_T)$ implies that, within each irreducible $S_{2T}$-representation, the $H_T$-invariant subspace is nonzero  precisely when the Young diagram has even row lengths, an is then one-dimensional~\cite{CollinsMatsumoto2009OrthogonalWeingarten}.
Writing these diagrams as $2\lambda=(2\lambda_1,2\lambda_2,\ldots)$, we therefore have
\begin{equation}
    Q_\lambda:=E_{2\lambda}\Pi_H,
\qquad    
    Q_\lambda Q_\mu=\delta_{\lambda\mu}Q_\lambda,
    \qquad
    \sum_{\lambda\vdash T}Q_\lambda=\Pi_H.
    \label{eq:coe-projector-construction}
\end{equation}
In particular, $Q_\lambda$ is an orthogonal projector because it is the product of two commuting orthogonal projectors.

It remains to determine the coefficient on each projector.
Define the zonal spherical functions and the associated denominator polynomials by
\begin{equation}
    \omega^\lambda(\sigma)
    :=\frac{1}{|H_T|}\sum_{\eta\in H_T}
        \chi^{2\lambda}(\sigma\eta),
\qquad    
    C_\lambda(z):=
        \prod_{(i,j)\in\lambda}(z+2j-i-1).
    \label{eq:coe-spherical-functions}
\end{equation}
For $z=d+1$, the spherical-function expansion of the orthogonal Weingarten function is~\cite{CollinsMatsumoto2009OrthogonalWeingarten,Matsumoto2012COEMoments}
\begin{equation}
    \operatorname{Wg}^{\mathrm{O}}(\sigma;z)
    =\frac{|H_T|}{(2T)!}
        \sum_{\lambda\vdash T}
        \frac{f^{2\lambda}}{C_\lambda(z)}
        \omega^\lambda(\sigma).
    \label{eq:coe-weingarten-spherical-expansion}
\end{equation}
The range $T\leq d$ makes every factor $d+2j-i\geq d+2-T$ strictly positive, so the expansion requires no singular-parameter prescription.
To convert the spherical functions into the projectors above, substitute $\tau=\sigma\eta$ and use $\chi^{2\lambda}(\tau)=\chi^{2\lambda}(\tau^{-1})$ to obtain
\begin{equation}
    \begin{aligned}
        \sum_{\sigma\in S_{2T}}\omega^\lambda(\sigma)P_\sigma
        &={}
        \left(\sum_{\tau\in S_{2T}}
            \chi^{2\lambda}(\tau)P_\tau\right)
        \left(\frac{1}{|H_T|}\sum_{\eta\in H_T}
            P_{\eta^{-1}}\right)\\
        &=\frac{(2T)!}{f^{2\lambda}}E_{2\lambda}\Pi_H
        =\frac{(2T)!}{f^{2\lambda}}Q_\lambda.
    \end{aligned}
    \label{eq:coe-spherical-projector-identity}
\end{equation}
Substituting \eqref{eq:coe-weingarten-spherical-expansion} and~\eqref{eq:coe-spherical-projector-identity} into \eqref{eq:coe-moment-permutation-expansion} gives
\begin{equation}
    \begin{aligned}
        M_T^{\mathrm{COE}}
        &=\frac{|H_T|}{(2T)!}
          \sum_{\lambda\vdash T}
          \frac{f^{2\lambda}}{C_\lambda(d+1)}
          \frac{(2T)!}{f^{2\lambda}}Q_\lambda\\
        &=2^T T!\sum_{\lambda\vdash T}
          \frac{Q_\lambda}{C_\lambda(d+1)}.
    \end{aligned}
    \label{eq:coe-spectrum-proof-conclusion}
\end{equation}
All coefficients are positive, so the support of $M_T^{\mathrm{COE}}$ is exactly the range of $\Pi_H$.
\end{proof}

To illustrate this decomposition, consider $T=1$, for which
$Q_{(1)}=(I+\mathbb S)/2$ and $C_{(1)}(d+1)=d+1$.
The first moment therefore has the following graphical representation:
\begin{equation}
M_1^{\mathrm{COE}}
=
\mathbb E_V \left[\;
\vcenter{\hbox{%
\begin{tikzpicture}[
  line width=0.6pt,
  font=\small,
  op/.style={
    draw,fill=white,
    minimum width=0.55cm,
    minimum height=0.40cm,
    inner sep=1pt
  }
]
\draw (0,0.26)
  arc[start angle=90,end angle=-90,
      x radius=0.35cm,y radius=0.26cm];
\draw (-1,0.26) -- (0,0.26);
\draw (-1,-0.26) -- (0,-0.26);
\node[op] at (-0.5,-0.26) {$V$};

\draw (1.1,0.26)
  arc[start angle=90,end angle=270,
      x radius=0.35cm,y radius=0.26cm];
\draw (1.1,0.26) -- (2.1,0.26);
\draw (1.1,-0.26) -- (2.1,-0.26);
\node[op] at (1.6,-0.26) {$V^\dagger$};
\end{tikzpicture}%
}}\;\right]
=
\frac{1}{d+1}\left[\;
\vcenter{\hbox{%
\begin{tikzpicture}[line width=0.6pt]
\draw (0,0.26) -- (1,0.26);
\draw (0,-0.26) -- (1,-0.26);
\end{tikzpicture}%
}}
\;+\;
\vcenter{\hbox{%
\begin{tikzpicture}[line width=0.6pt]
\draw (0,0.26) -- (1,-0.26);
\draw[preaction={draw=white,line width=2.6pt}]
  (0,-0.26) -- (1,0.26);
\end{tikzpicture}%
}}\;\right].
\label{eq:coe-first-moment-diagram}
\end{equation}
The two diagrams on the right represent the identity $I$ and the swap operator $\mathbb S$, respectively.

\subsubsection{Circular symplectic ensemble}

For even total dimension $d$, the circular symplectic ensemble (CSE) is the invariant probability distribution on self-dual unitaries, identified with $\mathrm U(d)/\mathrm{Sp}(d)$.
Here $\mathrm{Sp}(d)$ denotes the compact symplectic group acting on $\mathbb C^d$.
We use the same canonical form and symplectic dual as in the main text:
\begin{equation}
    J=i\sigma_y\otimes I_{d/2}
    =\begin{bmatrix}0&I_{d/2}\\-I_{d/2}&0\end{bmatrix},
     J^{\mathrm T}=J^\dagger=-J,
     X^{\mathrm D}:=JX^{\mathrm T}J^\dagger.
\end{equation}
For Haar-distributed $W\in\mathrm U(d)$,
\begin{equation}
    V=W^{\mathrm D}W,
     V^{\mathrm D}=V,
     VV^\dagger=I_d.
\end{equation}
It is convenient to express the moments through the antisymmetric matrix $JV$.
Writing $e_i$ for the standard basis vectors, define
\begin{equation}
    \widetilde v_{ij}:=e_i^{\mathrm T}JV e_j=(JV)_{ij}.
\end{equation}
Thus the bilinear form used below is explicitly $x^{\mathrm T}Jy$, with no complex conjugation.
For $1\leq T\leq d/2$ and multi-indices $\mathbf{i},\mathbf{j}\in\{1,\ldots,d\}^{2T}$, the CSE Weingarten formula is~\cite{Matsumoto2013SymmetricSpaces}
\begin{equation}
    \mathbb E_V
    \left[
        \widetilde v_{i_1i_2}
        \widetilde v_{i_3i_4}
        \cdots
        \widetilde v_{i_{2T-1}i_{2T}}
        \overline{
        \widetilde v_{j_1j_2}
        \widetilde v_{j_3j_4}
        \cdots
        \widetilde v_{j_{2T-1}j_{2T}}
        }
    \right]
    =
    \sum_{\sigma\in S_{2T}}
    \delta_\sigma(\mathbf{i},\mathbf{j})
    \mathrm{Wg}^{\mathrm{CSE}}(\sigma;d/2),
\end{equation}
where $\delta_\sigma$ is the same index-matching factor as for the COE.
The conventional dimension argument of the symplectic Weingarten function is the half-dimension $d/2$, although $d$ always denotes the full matrix dimension here.
The CSE Weingarten function is directly related to the symplectic one by
\begin{equation}
    \mathrm{Wg}^{\mathrm{CSE}}(\sigma;d/2)
    =
    \mathrm{Wg}^{\mathrm{Sp}}
    \left(
        \sigma;
        \frac{d-1}{2}
    \right).
\end{equation}
Unlike the orthogonal function, the symplectic Weingarten function acquires the sign character under multiplication by an element of $H_T$.
For the decomposition $\sigma=mh$ used above, this gives
\begin{equation}
    \mathrm{Wg}^{\mathrm{CSE}}(mh;d/2)
    =\operatorname{sgn}(h)
    \mathrm{Wg}^{\mathrm{Sp}} \left(m;\frac{d-1}{2}\right).
\end{equation}
The entry contractions for $JV$ therefore have the same matching pattern as in Fig.~\ref{fig:coe_pairing_rule}, with signed coefficients.
Returning to the ordinary entries of $V$ introduces the fixed $J$ rotations, yielding the symplectic-dual term at one copy.

For the usual matrix representation, the one-copy CSE twirling formula is
\begin{equation}
    \mathcal T_{\mathrm{CSE}}^{(1)}(X)
    =
    \mathbb E_V[VXV^\dagger]
    =
    \frac{
        \operatorname{Tr}(X)I_d
        -
        X^{\mathrm D}
    }{d-1}.
\end{equation}
The symplectic-dual term has the opposite sign to the transpose term in the COE mean channel.
Its graphical form is
\begin{equation}
    \mathcal T_{\mathrm{CSE}}^{(1)}(X)
    =
    \frac{1}{d-1}
    \left(
        \COETraceDiag
        \times
        \COEIdDiag
        -
        \CSEDualDiag
    \right).
\end{equation}
Here the first term represents $\operatorname{Tr}(X)I_d$.
The second term represents $X^{\mathrm D}=JX^{\mathrm T}J^\dagger$.
The black dots mark the $J$ and $J^\dagger$ insertions relative to the transpose diagram.

\subsubsection{Spectral decomposition of the CSE moment operator}
\label{sec:cse-moment-operator}

We derive the CSE analogue of Lemma~\ref{lem:coe-moment-spectrum} using the same permutation convention and tensor ordering.
For a $d\times d$ CSE matrix $V=W^{\mathrm D}W$, define
\begin{equation}
    \ket{v}:=\operatorname{vec}(V),
\qquad    
    M_T^{\mathrm{CSE}}:=\mathbb{E}_V \left[
        \bigl(\ketbra{v}{v}\bigr)^{\otimes T}
    \right].
    \label{eq:cse-moment-definition}
\end{equation}
Vectorization is again unnormalized, so $\operatorname{Tr}(M_T^{\mathrm{CSE}})=d^T$.
The relevant symmetry is most transparent after the fixed transformation
\begin{equation}
    A:=JV=W^\mathrm{T}JW,
     A^\mathrm{T}=-A,
     \mathcal J_T:=(J\otimes I_{d})^{\otimes T}.
    \label{eq:cse-antisymmetric-representation}
\end{equation}
Using the same pairing subgroup and tensor-permutation convention as in \eqref{eq:coe-pairing-projector}, define the signed average
\begin{equation}
    \Pi_H^-:=\frac{1}{|H_T|}
        \sum_{\eta\in H_T}\operatorname{sgn}(\eta)P_\eta.
    \label{eq:cse-signed-pairing-projector}
\end{equation}
This orthogonal projector selects tensors that are antisymmetric within each pair and symmetric under permutations of complete pairs.
In particular, $\operatorname{vec}(A)^{\otimes T}$ belongs to its range. 
We again collect standard results on the Weingarten decompositions as a convenient operator spectral decomposition for later use.

\begin{lemma}[Spectral decomposition of the CSE moment operator]
\label{lem:cse-moment-spectrum}
Let $d$ be even and $1\leq T\leq d/2$.
There exist mutually orthogonal projectors $Q_\lambda^{\mathrm{CSE}}$, indexed by $\lambda\vdash T$, such that
\begin{equation}
    M_T^{\mathrm{CSE}}
    =2^T T!\sum_{\lambda\vdash T}
        \frac{Q_\lambda^{\mathrm{CSE}}}{C_\lambda^{\mathrm{CSE}}(d)},
\qquad    
    C_\lambda^{\mathrm{CSE}}(d)
    :=\prod_{(i,j)\in\lambda}(d+j-2i).
    \label{eq:cse-moment-spectrum}
\end{equation}
They satisfy
\begin{equation}
    Q_\lambda^{\mathrm{CSE}}Q_\mu^{\mathrm{CSE}}
    =\delta_{\lambda\mu}Q_\lambda^{\mathrm{CSE}},
\qquad    
    \sum_{\lambda\vdash T}Q_\lambda^{\mathrm{CSE}}
    =\mathcal J_T^\dagger\Pi_H^-\mathcal J_T.
    \label{eq:cse-moment-projectors}
\end{equation}
\end{lemma}

\begin{proof}
Since $\operatorname{vec}(A)=(J\otimes I_{d})\operatorname{vec}(V)$, the CSE Weingarten formula gives~\cite{Matsumoto2013SymmetricSpaces}
\begin{equation}
    \widetilde M_T:=\mathcal J_T M_T^{\mathrm{CSE}}\mathcal J_T^\dagger
    =\sum_{\sigma\in S_{2T}}
        \operatorname{Wg}^{\mathrm{Sp}} \left(\sigma;\tfrac{d-1}{2}\right)P_\sigma.
    \label{eq:cse-moment-permutation-expansion}
\end{equation}
We use the central projectors $E_\nu$ from \eqref{eq:coe-central-projectors}, now in the tensor representation on $(\mathbb C^{d})^{\otimes 2T}$.
For the twisted Gelfand pair $(S_{2T},H_T,\operatorname{sgn}|_{H_T})$, the sign character of $H_T$ occurs with multiplicity one precisely in the irreducible representations with diagrams $\lambda\cup\lambda=(\lambda_1,\lambda_1,\lambda_2,\lambda_2,\ldots)$~\cite{Matsumoto2013SymmetricSpaces}.
Thus the commuting-projector argument for the COE gives mutually orthogonal projectors $\widetilde Q_\lambda:=E_{\lambda\cup\lambda}\Pi_H^-$ satisfying $\sum_\lambda\widetilde Q_\lambda=\Pi_H^-$.

Define the twisted spherical functions by $\pi^\lambda(\sigma):=|H_T|^{-1}\sum_{\eta\in H_T}\operatorname{sgn}(\eta)\chi^{\lambda\cup\lambda}(\sigma\eta)$.
Their Weingarten expansion is~\cite{Matsumoto2013SymmetricSpaces}
\begin{equation}
    \operatorname{Wg}^{\mathrm{Sp}} \left(\sigma;\tfrac{d-1}{2}\right)
    =\frac{|H_T|}{(2T)!}
        \sum_{\lambda\vdash T}
        \frac{f^{\lambda\cup\lambda}}{C_\lambda^{\mathrm{CSE}}(d)}
        \pi^\lambda(\sigma).
    \label{eq:cse-weingarten-spherical-expansion}
\end{equation}
The range $T\leq d/2$ ensures $d+j-2i\geq d+1-2T\geq1$, so every denominator is positive and no singular-parameter prescription is needed.
The same change of variables as in \eqref{eq:coe-spherical-projector-identity}, with the additional factor $\operatorname{sgn}(\eta)$, yields
\begin{equation}
    \sum_{\sigma\in S_{2T}}\pi^\lambda(\sigma)P_\sigma
    =\frac{(2T)!}{f^{\lambda\cup\lambda}}\widetilde Q_\lambda.
    \label{eq:cse-spherical-projector-identity}
\end{equation}
Substitution into \eqref{eq:cse-moment-permutation-expansion} gives $\widetilde M_T=2^TT!\sum_{\lambda\vdash T}\widetilde Q_\lambda/C_\lambda^{\mathrm{CSE}}(d)$.
Conjugating by $\mathcal J_T^\dagger$ proves the lemma with $Q_\lambda^{\mathrm{CSE}}:=\mathcal J_T^\dagger\widetilde Q_\lambda\mathcal J_T$.
The positivity of the coefficients also identifies the support as the range of $\mathcal J_T^\dagger\Pi_H^-\mathcal J_T$.
\end{proof}

At $T=1$, the decomposition reduces to
\begin{equation}
    M_1^{\mathrm{CSE}}
    =\frac{I+\mathbb S_J}{d-1},
\qquad    
    \mathbb S_J:=\mathbb S(J\otimes J).
\end{equation}
Indeed, $\Pi_H^-=(I-\mathbb S)/2$ and $(J^\dagger\otimes I)\mathbb S(J\otimes I)=-\mathbb S_J$.
This agrees with the self-dual exchange symmetry used in the main text.

\subsection{Classical-memory protocols and channel-property lower bounds}
\label{sec:channel-learning-preliminaries}

\subsubsection{Query model}
We use the classical-memory query model of Sec.~\ref{sec:limited-entanglement}, allowing a general unknown channel $\mathcal E:\mathsf L(\mathbb C^d)\to\mathsf L(\mathbb C^d)$ with $d=2^n$.
In round $t$, conditioned on the preceding classical record $h_{t-1}$, the learner prepares a possibly mixed state $\rho_{Q,A}^{(t)}(h_{t-1})$, applies $\mathcal E\otimes\operatorname{id}_A$, and performs a joint POVM $\mathsf M_{Q,A}^{(t)}(h_{t-1})$.
We write $\rho_{Q,A}^{(t)}$ and $\mathsf M_{Q,A}^{(t)}$ when the conditioning is implicit; the superscript labels the query round.
The query register $Q$ has dimension $d$, and $A$ includes every system that bypasses the channel during that round.
The register $A$ may have arbitrary finite dimension, which may depend on the round and the classical record.
The resource constraints concern entanglement across $Q{:}A$: on every adaptive branch, the actual probe and joint POVM satisfy
\[
    E_N^{\mathrm s}\bigl(\rho_{Q,A}^{(t)}\bigr)\leq e_{\mathrm s},
\qquad    
    E_N^{\mathrm m}\bigl(\mathsf M_{Q,A}^{(t)}\bigr)\leq e_{\mathrm m},
\]
with the negativities defined in \eqref{eq:main-negativity} and \eqref{eq:main-measurement-negativity}.
No upper bound is imposed on the size of $A$.
The same unknown channel is reused, but all quantum registers are discarded after each round.
Only the classical record, including any retained classical randomness, determines later preparations and measurements, as illustrated in Fig.~\ref{fig:classical-memory-channel-model}.

Without entanglement constraints, mixed probes and general POVMs can be refined into pure probes and rank-one measurement outcomes without reducing their distinguishing power.
Indeed, given a decomposition $\rho_{Q,A}^{(t)}=\sum_a p_a\ketbra{\phi_a}{\phi_a}$, one can prepare $\ket{\phi_a}$ with probability $p_a$ and retain the label $a$ in the classical record.
Similarly, decomposing each POVM effect as $M_y^{(t)}=\sum_b\lambda_{yb}\ketbra{\psi_{yb}}{\psi_{yb}}$ gives a finer measurement whose outcome is $(y,b)$.
Ignoring the additional labels $a,b$ and choosing subsequent operations using only the original outcomes reproduces the original protocol exactly.
Retaining these labels therefore cannot decrease the optimal discrimination success probability.

This refinement, however, need not preserve the entanglement constraints: the pure components and refined POVM may have larger negativity than the original mixed probe and full POVM.
We therefore use this reduction only in the resource-independent parts of the analysis.
When applying the negativity bounds, we retain the actual conditional states $\rho_{Q,A}^{(t)}$ and complete POVMs $\mathsf M_{Q,A}^{(t)}$, without assuming that they are rank one.

\begin{figure}[htbp]
\centering
\resizebox{\linewidth}{!}{%
\begin{tikzpicture}[
    x=1cm,
    y=1cm,
    font=\small,
    line width=0.7pt,
    >=stealth
]

\foreach \offset/\roundindex in {0/1,6.1/2,13.3/T}{
    \begin{scope}[shift={(\offset,0)}]

        \draw[fill=white]
            (0,0)
            -- (0.95,0.85)
            -- (0.95,-0.85)
            -- cycle;
        \node at (0.54,0)
            {$\rho_{Q,A}^{(\roundindex)}$};

        \draw (0.95,0.4) -- (1.65,0.4);
        \draw (2.45,0.4) -- (3.15,0.4);
        \draw (1.22,0.32) -- (1.36,0.48);
        \node[above] at (1.29,0.48) {$Q$};

        \draw (0.95,-0.4) -- (3.15,-0.4);
        \draw (1.22,-0.48) -- (1.36,-0.32);
        \node[above] at (1.29,-0.32) {$A$};

        \draw[fill=black!5]
            (1.65,0.1) rectangle (2.45,0.7);
        \node at (2.05,0.4) {$\mathcal E$};

        \draw[fill=white]
            (3.15,0.85)
            -- (4.10,0)
            -- (3.15,-0.85)
            -- cycle;
        \node at (3.56,0)
            {$\mathsf M_{Q,A}^{(\roundindex)}$};

    \end{scope}
}

\draw[double,double distance=1pt,->]
    (4.50,0) -- (5.70,0);
\node[above,font=\scriptsize] at (5.10,0.18)
    {classical memory};

\draw[double,double distance=1pt]
    (10.60,0) -- (11.40,0);
\node at (11.75,0) {$\cdots$};
\draw[double,double distance=1pt,->]
    (12.10,0) -- (12.90,0);
\node[above,font=\scriptsize] at (11.75,0.18)
    {classical memory};

\end{tikzpicture}%
}

\caption{Adaptive channel queries with classical memory.
Round $t$ prepares a possibly mixed probe $\rho_{Q,A}^{(t)}$, applies $\mathcal E$ to the $n$-qubit register $Q$ while $A$ bypasses the channel, and performs a joint POVM $\mathsf M_{Q,A}^{(t)}$.
The register $A$ has unrestricted finite dimension; the probe and measurement negativities across $Q{:}A$ are bounded by $e_{\mathrm s}$ and $e_{\mathrm m}$, respectively.
The history dependence of both $\rho_{Q,A}^{(t)}$ and $\mathsf M_{Q,A}^{(t)}$ is implicit in the figure.
Only classical information is retained between rounds.}
\label{fig:classical-memory-channel-model}
\end{figure}

\subsubsection{Learning trees and channel moments}
Following Ref.~\cite{Chen2021LearningWithoutQuantumMemory}, represent a $T$-query protocol as a depth-$T$ learning tree.
A node $u$ at depth $t-1$ represents the preceding record $h_{t-1}$; write $\rho_{Q,A}^{(t)}(u)$ and $\mathsf M_{Q,A}^{(t)}(u)$ for the corresponding conditional probe and POVM.
Write $p_{\mathcal E}(u)$ for the probability of reaching $u$, and let $E_{v\mid u}$ denote the effect of $\mathsf M_{Q,A}^{(t)}(u)$ associated with outcome $v$.
The transition to a child $v$ obeys
\[
    p_{\mathcal E}(v)=p_{\mathcal E}(u)
    \operatorname{Tr} \left[
        E_{v\mid u}(\mathcal E\otimes\operatorname{id}_A)
        (\rho_{Q,A}^{(t)}(u))
    \right],
\qquad    
    \sum_{v\in\operatorname{child}(u)}E_{v\mid u}=I_{Q,A}.
\]
To obtain the rank-one building blocks, first consider a normalized state vector $\ket{\phi_u}\in Q\otimes A$ and a POVM with normalized vectors $\ket{\psi_v}$, indexed by the children $v$ of $u$:
\begin{equation}
    E_{v\mid u}=\lambda_{v\mid u}\ketbra{\psi_v}{\psi_v},
    \qquad
    \sum_{v\in\operatorname{child}(u)}E_{v\mid u}=I_{Q,A}.
    \label{eq:learning-tree-povm}
\end{equation}
Here $\lambda_{v\mid u}=\operatorname{Tr}E_{v\mid u}\geq0$; these coefficients are fixed by the POVM effects and need not sum to one.
The corresponding transition rule becomes
\begin{equation}
    p_{\mathcal E}(v)
    =p_{\mathcal E}(u)\lambda_{v\mid u}
    \bra{\psi_v}(\mathcal E\otimes\operatorname{id}_A)
    (\ketbra{\phi_u}{\phi_u})\ket{\psi_v}.
    \label{eq:learning-tree-transition}
\end{equation}
Protocols that stop earlier can be padded with ignored queries.
POVM completeness normalizes the leaf distribution and permits recursive summation over adaptive records.
We write sums for discrete outcomes; integrals give the same statements for general outcome spaces.

For later moment estimates, we rewrite each learning-tree transition as a contraction with the unnormalized Choi operator
\begin{equation}
    \Gamma(\mathcal E):=(\mathcal E\otimes\operatorname{id})
    (\ketbra{\Omega_d}{\Omega_d}),
\qquad    
    \ket{\Omega_d}:=\sum_{a=1}^{d}\ket{a}\ket{a}.
    \label{eq:learning-choi-definition}
\end{equation}
With the output--input ordering of the Choi factors, its matrix elements satisfy
$\Gamma(\mathcal E)_{ab,a'b'}
=\bra{a}\mathcal E(\ketbra{b}{b'})\ket{a'}$.
For an edge $u\to v$, contracting the input and measurement tensors over the auxiliary-system index defines
$\ket{z_{u,v}}:=\operatorname{vec}(Z_{u,v})$, where
\begin{equation}
    (Z_{u,v})_{ab}
    :=\sum_{\alpha}(\psi_v)_{a\alpha}
         \overline{(\phi_u)_{b\alpha}}.
    \label{eq:learning-edge-vector}
\end{equation}
Here $\alpha$ ranges over an orthonormal basis of the bypassing register $A$.
The matrix element in \eqref{eq:learning-tree-transition} can then be rearranged as follows, with
$\phi=\phi_u$, $\psi=\psi_v$, and $z=z_{u,v}$ in the diagrams:
\begin{equation}
\begin{aligned}
&\bra{\psi}(\mathcal E\otimes\operatorname{id}_A)
  (\ketbra{\phi}{\phi})\ket{\psi}
\\
&\quad=
\vcenter{\hbox{%
\begin{tikzpicture}[
  line width=0.6pt,
  font=\small,
  tensor/.style={
    draw,fill=white,
    minimum width=0.70cm,
    minimum height=0.42cm,
    inner sep=1pt
  },
  choi/.style={
    draw,fill=white,
    minimum width=1.10cm,
    minimum height=1.30cm,
    inner sep=2pt
  },
  idx/.style={font=\scriptsize,inner sep=1pt}
]

\draw[dashed,draw=black!45,rounded corners=2pt]
  (-2.65,-0.82) rectangle (-1.02,0.82);
\draw[dashed,draw=black!45,rounded corners=2pt]
  (1.02,-0.82) rectangle (2.65,0.82);

\draw (-1.5,0.45) -- (1.5,0.45);
\draw (-1.5,-0.45) -- (1.5,-0.45);
\node[choi] at (0,0) {$\Gamma(\mathcal E)$};

\node[tensor] (lp) at (-1.5,0.45) {$\overline\psi$};
\node[tensor] (lf) at (-1.5,-0.45) {$\phi$};
\node[tensor] (rp) at (1.5,0.45) {$\psi$};
\node[tensor] (rf) at (1.5,-0.45) {$\overline\phi$};

\draw (lp.west)
  .. controls (-2.35,0.45) and (-2.35,-0.45)
  .. (lf.west);
\draw (rp.east)
  .. controls (2.35,0.45) and (2.35,-0.45)
  .. (rf.east);

\node[idx] at (-2.43,0) {$\alpha$};
\node[idx] at (2.43,0) {$\beta$};
\node[idx] at (-0.80,0.62) {$a$};
\node[idx] at (-0.80,-0.62) {$b$};
\node[idx] at (0.80,0.62) {$a'$};
\node[idx] at (0.80,-0.62) {$b'$};

\end{tikzpicture}%
}}
\;=\;
\vcenter{\hbox{%
\begin{tikzpicture}[
  line width=0.6pt,
  font=\small,
  choi/.style={
    draw,fill=white,
    minimum width=1.10cm,
    minimum height=0.78cm,
    inner sep=2pt
  }
]

\draw (-1.4,0.24) -- (1.4,0.24);
\draw (-1.4,-0.24) -- (1.4,-0.24);
\node[choi] at (0,0) {$\Gamma(\mathcal E)$};

\draw[fill=white]
  (-2.3,0) -- (-1.4,0.52)
  -- (-1.4,-0.52) -- cycle;
\node[inner sep=0pt] at (-1.73,0) {$\bra{z}$};

\draw[fill=white]
  (2.3,0) -- (1.4,0.52)
  -- (1.4,-0.52) -- cycle;
\node[inner sep=0pt] at (1.73,0) {$\ket{z}$};

\end{tikzpicture}%
}}
\\
&\quad=\bra{z}\Gamma(\mathcal E)\ket{z}.
\end{aligned}
\label{eq:learning-edge-choi-diagram}
\end{equation}
Here bars denote componentwise complex conjugation, and connected indices are summed.
The left dashed group contracts $\overline\psi$ with $\phi$ to produce $\bra{z}$, while the right group contracts $\psi$ with $\overline\phi$ to produce $\ket{z}$.
No rank restriction on $Z_{u,v}$ is needed for the moment representation.

For a leaf $\ell$ with path $u_0\to\cdots\to u_T=\ell$, write
$z_t:=z_{u_{t-1},u_t}$ and set
$\ket{Z_\ell}:=\bigotimes_{t=1}^{T}\ket{z_t}$ and
$\kappa_\ell:=\prod_{t=1}^{T}\lambda_{u_t\mid u_{t-1}}$.
Multiplying the transition probabilities along this path gives
\begin{equation}
    \begin{aligned}
        P_{\mathcal E}(\ell)
        &=\kappa_\ell\prod_{t=1}^{T}
          \bra{z_t}\Gamma(\mathcal E)\ket{z_t}\\
        &=\kappa_\ell\bra{Z_\ell}
          \Gamma(\mathcal E)^{\otimes T}\ket{Z_\ell}.
    \end{aligned}
    \label{eq:learning-path-choi-product}
\end{equation}
If a channel is sampled once from an ensemble $\mu$ and reused in all rounds, averaging over that channel yields
\begin{equation}
    \begin{aligned}
        P_\mu(\ell)
        &:=\mathbb E_{\mathcal E\sim\mu}[p_{\mathcal E}(\ell)]
        =\kappa_\ell\bra{Z_\ell}M_T^\mu\ket{Z_\ell},\\
        M_T^\mu
        &:=\mathbb E_{\mathcal E\sim\mu}
          [\Gamma(\mathcal E)^{\otimes T}].
    \end{aligned}
    \label{eq:learning-leaf-moment}
\end{equation}
The expectation is taken after multiplying the $T$ factors because the sampled channel is reused.
Along a fixed path, all adaptive choices are fixed, and the absence of quantum memory between rounds gives the product vector $\ket{Z_\ell}$.
For mixed probes and general POVMs, each leaf probability is a nonnegative sum of these rank-one contributions, with decomposition weights independent of the channel.
For a unitary channel $\mathcal U(X)=UXU^\dagger$, one has
$\Gamma(\mathcal U)=\ketbra{\operatorname{vec}U}{\operatorname{vec}U}$, recovering the moment operators used for COE and CSE.

\subsubsection{From property estimation to binary testing}
Choose two priors $\mu_0,\mu_1$ over channels.
A hidden label $b\in\{0,1\}$ is chosen with equal probability, a channel is sampled once from $\mu_b$, and the learner uses $T$ queries to infer $b$.
For a fixed learning tree, the two leaf distributions are $P_b:=P_{\mu_b}$.

\begin{lemma}[Learning-to-testing reduction]
\label{lem:channel-learning-to-testing}
Suppose $\mu_b$ is supported on $\mathfrak C_b$ and a real-valued channel property $f$ satisfies
\begin{equation}
    \inf_{\substack{\mathcal E_0\in\mathfrak C_0\\
                     \mathcal E_1\in\mathfrak C_1}}
    |f(\mathcal E_0)-f(\mathcal E_1)|>2\varepsilon.
    \label{eq:channel-property-gap}
\end{equation}
Any $T$-query estimator with additive error at most $\varepsilon$ and success probability at least $1-\eta$ on every channel in $\mathfrak C_0\cup\mathfrak C_1$ yields a $T$-query distinguisher with success probability at least $1-\eta$.
\end{lemma}
The reduction assigns the estimate to the closer of the two separated sets of property values.
If the separation holds only on subsets of prior probabilities at least $1-\beta_0$ and $1-\beta_1$, the same rule succeeds with probability at least $1-\eta-(\beta_0+\beta_1)/2$.

\begin{lemma}[Le Cam's binary-testing bound~\cite{LeCam1986AsymptoticMethods}]
\label{lem:channel-le-cam}
For a fixed learning tree, the optimal decision rule has success probability
\begin{equation}
    p_{\mathrm{succ}}^{\mathrm{opt}}
    =\frac{1+d_{\mathrm{TV}}(P_0,P_1)}{2},
\qquad    
    d_{\mathrm{TV}}(P,Q):=\frac12\sum_\ell|P(\ell)-Q(\ell)|.
    \label{eq:channel-le-cam}
\end{equation}
In particular, success probability at least $2/3$ requires $d_{\mathrm{TV}}(P_0,P_1)\geq1/3$.
\end{lemma}

\section{Query lower bounds with limited probe and measurement entanglement}
\label{app:proof}
\label{app:entanglement-limited-lower-bound}

This appendix proves Theorem~\ref{thm:main-negativity-lower-bound} for the classical-memory model of Sec.~\ref{sec:channel-learning-preliminaries}.
The query dimension is $d$, and no bound is imposed on the finite auxiliary dimension.
We constrain the actual conditional probe $\rho_{Q,A}^{(u)}$ and the full joint POVM $\mathsf M^{(u)}=\{E_{v\mid u}\}_v$ at each node $u$.
As in \eqref{eq:main-negativity}, the probe's logarithmic negativity is~\cite{EisertPhD2001,VidalWerner2002Negativity}
\begin{equation}
    E_N^{\mathrm s}(\rho_{Q,A})
    :=\log_2\|\rho_{Q,A}^{\mathrm T_Q}\|_1,
    \label{eq:probe-logarithmic-negativity}
\end{equation}
where $\mathrm T_Q$ denotes partial transpose over the queried system.
Using the measurement negativity $E_N^{\mathrm m}$ defined in \eqref{eq:measurement-logarithmic-negativity} below, we assume
\begin{equation}
    E_N^{\mathrm s}\bigl(\rho_{Q,A}^{(u)}\bigr)\leq e_{\mathrm s},
\qquad    
    E_N^{\mathrm m}\bigl(\mathsf M_{Q,A}^{(u)}\bigr)\leq e_{\mathrm m},
    \qquad
    0\leq e_{\mathrm s},e_{\mathrm m}\leq\log_2d.
    \label{eq:probe-entanglement-constraint}
\end{equation}
These constraints apply to the actual conditional state and the full POVM, at every node of the protocol tree, including histories with zero probability under a particular hypothesis.
They therefore also hold under the reference channels used below.
For brevity, write $e_*:=\min\{e_{\mathrm s},e_{\mathrm m}\}$.

The proof separates two contributions to the transcript distance.
Moment bounds control the correlations created by reusing one random unitary, while the two negativity bounds control the distinguishability of its averaged channel from $\mathcal D(X)=\operatorname{Tr}(X)I_d/d$ in each query.
The completely depolarizing channel then provides a common reference for COE, CSE, and Haar.
We first establish the measurement-negativity properties needed for this argument, together with their interpretation in terms of exact LOCC implementation.
All transcript comparisons below use the same learning protocol.

\subsection{Measurement negativity and entanglement cost}
\label{app:measurement-negativity}

For a POVM $\mathsf M=\{M_y\}_y$ on $Q\otimes A$, recall \eqref{eq:main-measurement-negativity}:
\begin{equation}
    E_N^{\mathrm m}(\mathsf M)
    :=
    \log_2
    \sup_{\substack{\sigma_{Q,A}\geq0\\
                    \operatorname{Tr}\sigma_{Q,A}=1}}
    \sum_y
    \left|
        \operatorname{Tr} \left(
            M_y^{\mathrm T_Q}\sigma_{Q,A}
        \right)
    \right|.
    \label{eq:measurement-logarithmic-negativity}
\end{equation}
The partially transposed effects sum to the identity but need not be positive, so the sum of the absolute values can exceed one.
The supremum is over states on $Q\otimes A$ itself, without an additional reference system.

\begin{lemma}[Probe and measurement bounds]
\label{lem:measurement-negativity-properties}
For $\dim Q=d$, every POVM satisfies $0\leq E_N^{\mathrm m}(\mathsf M)\leq\log_2d$, independently of $\dim A$.
Moreover, $E_N^{\mathrm m}(\mathsf M)=0$ if and only if $M_y^{\mathrm T_Q}\geq0$ for every outcome $y$.
The quantity is invariant under local unitary conjugation of all effects.
For every bipartite state $\rho_{Q,A}$,
\begin{equation}
    \sum_y\left|
        \operatorname{Tr}(M_y\rho_{Q,A}^{\mathrm T_Q})
    \right|
    \leq
    \min\left\{
        2^{E_N^{\mathrm s}(\rho_{Q,A})},
        2^{E_N^{\mathrm m}(\mathsf M)}
    \right\}.
    \label{eq:joint-negativity-bound}
\end{equation}
\end{lemma}

\begin{proof}
For any Hermitian operator $X$, positivity and completeness of the POVM give
\begin{equation}
    \sum_y|\operatorname{Tr}(M_yX)|
    \leq
    \sum_y\operatorname{Tr}(M_y|X|)
    =\|X\|_1.
    \label{eq:povm-trace-norm-contraction}
\end{equation}
Applying this to $X=\rho_{Q,A}^{\mathrm T_Q}$ gives the first bound in \eqref{eq:joint-negativity-bound}.
The second follows from
$\operatorname{Tr}(M_y\rho_{Q,A}^{\mathrm T_Q})
=\operatorname{Tr}(M_y^{\mathrm T_Q}\rho_{Q,A})$
and \eqref{eq:measurement-logarithmic-negativity}.
Since $\sum_y\operatorname{Tr}(M_y^{\mathrm T_Q}\sigma)=1$, the absolute sum in the definition is at least one.
It is at most $\|\sigma^{\mathrm T_Q}\|_1\leq d$ by \eqref{eq:povm-trace-norm-contraction}.
The latter bound follows for pure states from their Schmidt coefficients,
$\|\ketbra{\psi}{\psi}^{\mathrm T_Q}\|_1=(\sum_j\sqrt{\lambda_j})^2\leq d$,
and extends to mixed states by convexity.

If all effects are PPT, every term in the absolute sum is nonnegative, so the sum equals one.
Conversely, if some $M_y^{\mathrm T_Q}$ has a negative eigenvalue, choosing its eigenstate as $\sigma$ makes one term negative; because the terms still sum to one, their absolute sum exceeds one.
Finally, for $M'_y=(V_Q\otimes V_A)M_y(V_Q\otimes V_A)^\dagger$,
\[
    (M'_y)^{\mathrm T_Q}
    =(V_Q^*\otimes V_A)M_y^{\mathrm T_Q}
      (V_Q^*\otimes V_A)^\dagger.
\]
Unitary conjugation preserves the set of density operators in the supremum, proving local unitary invariance.
\end{proof}

Outcome relabeling and proportional splitting of effects leave $E_N^{\mathrm m}$ unchanged.
More generally, classical postprocessing cannot increase it, by the triangle inequality.

To interpret \eqref{eq:measurement-logarithmic-negativity} operationally, consider the entanglement cost of implementing a nonlocal measurement~\cite{Bandyopadhyay2009MeasurementCost,GourScandolo2021BipartiteChannel}.
The following LOCC model concerns implementation cost; it is not an additional restriction on the TRS learning model.
Alice holds $Q$ and Bob holds $A$.
Before receiving the input, they share a fixed resource state $\omega_{Q', A'}$, with $Q'$ held by Alice and $A'$ by Bob, independent of the input.
They may use arbitrary local operations, local auxiliary systems initially in product states, and unrestricted classical communication, but no quantum communication or additional shared entanglement.
An exact deterministic implementation must produce the target probabilities $\operatorname{Tr}(M_y\rho_{Q,A})$ for every input state $\rho_{Q,A}$, including entangled inputs, without postselection.
Only the outcome probabilities are specified; no particular postmeasurement quantum state is required.

\begin{lemma}[Entanglement cost of exact measurement implementation]
\label{lem:measurement-epr-cost}
If $\mathsf M$ admits an exact implementation in this model using $\omega_{Q', A'}$, then
\begin{equation}
    E_N^{\mathrm m}(\mathsf M)
    \leq
    E_N^{\mathrm s}(\omega_{Q', A'})
    =\log_2\|\omega_{Q', A'}^{\mathrm T_{Q'}}\|_1.
    \label{eq:measurement-resource-negativity-cost}
\end{equation}
In particular, an implementation using $m$ preshared EPR pairs requires
\begin{equation}
    m\geq\left\lceil E_N^{\mathrm m}(\mathsf M)\right\rceil.
    \label{eq:measurement-epr-pair-cost}
\end{equation}
\end{lemma}

\begin{proof}
After absorbing the local auxiliary preparations, the LOCC protocol defines a POVM $\{F_y\}_y$ on $QQ':AA'$ such that
\begin{equation}
    \operatorname{Tr}(M_y\rho_{Q,A})
    =\operatorname{Tr} \left[
        F_y(\rho_{Q,A}\otimes\omega_{Q', A'})
    \right]
    \label{eq:measurement-locc-simulation}
\end{equation}
for every input state.
Each $F_y$ is a separable effect across $QQ':AA'$ and hence
$F_y^{\mathrm T_{QQ'}}\geq0$.
Since $\sum_yF_y^{\mathrm T_{QQ'}}=I$, the partially transposed effects form a POVM.
By linearity, \eqref{eq:measurement-locc-simulation} also holds for Hermitian inputs, so for every density operator $\sigma_{Q,A}$,
\begin{equation}
    \operatorname{Tr}(M_y^{\mathrm T_Q}\sigma_{Q,A})
    =\operatorname{Tr} \left[
        F_y^{\mathrm T_{QQ'}}
        (\sigma_{Q,A}\otimes\omega_{Q', A'}^{\mathrm T_{Q'}})
    \right].
    \label{eq:measurement-locc-partial-transpose}
\end{equation}
Applying \eqref{eq:povm-trace-norm-contraction} to this POVM yields
\begin{equation}
    \sum_y\left|\operatorname{Tr}(M_y^{\mathrm T_Q}\sigma_{Q,A})\right|
    \leq
    \|\sigma_{Q,A}\otimes\omega_{Q', A'}^{\mathrm T_{Q'}}\|_1
    =\|\omega_{Q', A'}^{\mathrm T_{Q'}}\|_1.
\end{equation}
Taking the supremum over $\sigma_{Q,A}$ proves \eqref{eq:measurement-resource-negativity-cost}.
For $m$ EPR pairs, the resource has logarithmic negativity $m$, proving \eqref{eq:measurement-epr-pair-cost}.
\end{proof}

The bound counts the entanglement initially supplied to the implementation, rather than a net consumption after recovering a resource or an asymptotic approximate cost.
It also applies to measurement localization, which further restricts when classical communication can be used~\cite{Pauwels2025MeasurementLocalization}.
The bound need not be tight for general POVMs: PPT effects do not by themselves guarantee LOCC implementability.

The bound is tight for the two measurements used in Sec.~\ref{sec:quantifying-trs}.
For $\mathsf M_{\mathbb S}=\{P_+,P_-\}$ with $P_\pm=(I\pm\mathbb S)/2$ on two $d$-dimensional systems, write $\Phi_d=\ketbra{\Phi_d}{\Phi_d}$.
Then $P_\pm^{\mathrm T_Q}=(I\pm d\Phi_d)/2$, and the choice $\sigma=\Phi_d$ in \eqref{eq:measurement-logarithmic-negativity} gives the absolute sum $(d+1)/2+(d-1)/2=d$.
Together with Lemma~\ref{lem:measurement-negativity-properties}, this proves $E_N^{\mathrm m}(\mathsf M_{\mathbb S})=\log_2d$.
For even $d$, let $\mathbb S_J=\mathbb S(J\otimes J)$ with the canonical $J=i\sigma_y\otimes I_{d/2}$, and set $P_\pm^J=(I\pm\mathbb S_J)/2$.
Since $J^\dagger=-J$,
\begin{equation}
    P_\pm^J
    =(J\otimes I_d)P_\mp(J^\dagger\otimes I_d),
\qquad    
    E_N^{\mathrm m}(\mathsf M_{\mathbb S_J})
    =E_N^{\mathrm m}(\mathsf M_{\mathbb S})
    =\log_2d,
    \label{eq:swap-measurement-negativity}
\end{equation}
by local unitary invariance and outcome relabeling.
For $d=2^n$, exact LOCC implementation of either POVM therefore requires at least $n$ EPR pairs, and $n$ suffice by teleporting $Q$ to Bob and measuring locally.

\subsection{From moment bounds to transcript bounds}
\label{sec:moment-to-transcript-bounds}

Fix a depth-$T$ learning tree as in Sec.~\ref{sec:channel-learning-preliminaries}.
Write $P_\mathcal E(\ell)=p_\mathcal E(\ell)$ for its leaf distribution under a fixed channel $\mathcal E$, and $P_\mu(\ell)=\mathbb E_{\mathcal E\sim\mu}P_{\mathcal E}(\ell)$ when one channel is drawn from $\mu$ and reused throughout.

An intermediate comparison will be with the averaged channel
\begin{equation}
    \overline{\mathcal E}_\mu
    :=\mathbb E_{\mathcal E\sim\mu}\mathcal E,
\qquad    
    \Gamma(\overline{\mathcal E}_\mu)
    =\mathbb E_{\mathcal E\sim\mu}\Gamma(\mathcal E)
    =M_1^\mu.
    \label{eq:learning-averaged-channel}
\end{equation}
Using $\overline{\mathcal E}_\mu$ in every round gives the same transcript distribution as independently resampling the channel at each query, because each conditional outcome probability is linear in the queried channel.
The difference between $P_\mu$ and $P_{\overline{\mathcal E}_\mu}$ therefore isolates correlations from reusing a single draw.

The key probabilistic observation is that a uniform lower bound on individual leaf probabilities yields an upper bound on total variation distance.
For normalized distributions $P,Q$ and $0\leq\alpha\leq1$,
\begin{equation}
    P(\ell)\geq\alpha Q(\ell)
    \quad\text{for every leaf }\ell
    \quad\Longrightarrow\quad
    d_{\mathrm{TV}}(P,Q)
    =1-\sum_\ell\min\{P(\ell),Q(\ell)\}
    \leq1-\alpha.
    \label{eq:learning-pointwise-domination}
\end{equation}
This follows by summing $\min\{P(\ell),Q(\ell)\}\geq\alpha Q(\ell)$ and also applies to leaves of zero probability.

\begin{lemma}[Moment comparison for learning trees]
\label{lem:learning-moment-domination}
Suppose $M_1^\mu=cK$, where $c>0$ and $K$ is an orthogonal projector on the one-query Choi space
$\mathbb C^d\otimes\mathbb C^d$.
Define
\begin{equation}
    \mathsf S_T:=\frac{1}{T!}\sum_{\pi\in S_T}W_\pi,
    \label{eq:learning-copy-symmetrizer}
\end{equation}
where $W_\pi$ permutes the $T$ complete output-input pairs.
Thus, $\mathsf S_T$ projects onto the symmetric subspace of the $T$ Choi copies and commutes with $K^{\otimes T}$.
If, for some $\alpha_T\in[0,1]$,
\begin{equation}
    M_T^\mu\succeq
    \alpha_T c^T T!\mathsf S_T K^{\otimes T},
    \label{eq:learning-moment-comparison}
\end{equation}
then every $T$-query learning tree with only classical memory between rounds satisfies
\begin{equation}
    P_\mu(\ell)
    \geq\alpha_T P_{\overline{\mathcal E}_\mu}(\ell)
    \quad\text{for every leaf }\ell,
    \qquad
    d_{\mathrm{TV}}
    \bigl(P_\mu,P_{\overline{\mathcal E}_\mu}\bigr)
    \leq1-\alpha_T.
    \label{eq:learning-ensemble-to-average}
\end{equation}
\end{lemma}

\begin{proof}
First suppose the probes are pure and the effects have rank one.
Fix a leaf $\ell$, with vectors $z_t$ and coefficient $\kappa_\ell$ as in \eqref{eq:learning-leaf-moment}.
These data are fixed along its path and are the same under both hypotheses, even for an adaptive protocol.
Set $\ket{Z_\ell}=\bigotimes_{t=1}^T\ket{z_t}$ and $x_t=Kz_t$.
Since $\Gamma(\overline{\mathcal E}_\mu)=cK$, its probability under the averaged channel is
\begin{equation}
    \begin{aligned}
        P_{\overline{\mathcal E}_\mu}(\ell)
        &=
        \kappa_\ell
        \prod_{t=1}^T
        \bra{z_t}
        \Gamma(\overline{\mathcal E}_\mu)
        \ket{z_t}\\
        &=
        \kappa_\ell c^T
        \prod_{t=1}^T\|x_t\|^2.
    \end{aligned}
    \label{eq:learning-average-leaf-product}
\end{equation}

To compare this product with the symmetrized expression in the moment bound, introduce the Gram matrix
$G_{st}:=\braket{x_s}{x_t}$.
Using $G\succeq0$ and the definition of $\mathsf S_T$, we obtain
\begin{equation}
    \begin{aligned}
        T!\bra{Z_\ell}
        \mathsf S_T K^{\otimes T}
        \ket{Z_\ell}
        &=
        \sum_{\pi\in S_T}
        \left(\bigotimes_{t=1}^T\bra{x_t}\right)
        W_\pi
        \left(\bigotimes_{t=1}^T\ket{x_t}\right)\\
        &=
        \sum_{\pi\in S_T}
        \prod_{t=1}^T G_{t,\pi(t)}\\
        &=
        \operatorname{per}(G)
        \geq
        \prod_{t=1}^T G_{tt}
        =
        \prod_{t=1}^T\|x_t\|^2.
    \end{aligned}
    \label{eq:learning-gram-permanent}
\end{equation}
The last inequality is the permanent analogue of Hadamard's inequality, valid for all positive semidefinite Gram matrices, including singular ones~\cite{Marcus1964PermanentHadamard}.

Combining the leaf-moment representation with the assumed operator inequality now gives
\begin{equation}
    \begin{aligned}
        P_\mu(\ell)
        &=
        \kappa_\ell
        \bra{Z_\ell}M_T^\mu\ket{Z_\ell}\\
        &\geq
        \alpha_T\kappa_\ell c^T T!
        \bra{Z_\ell}
        \mathsf S_TK^{\otimes T}
        \ket{Z_\ell}\\
        &\geq
        \alpha_T\kappa_\ell c^T
        \prod_{t=1}^T\|x_t\|^2\\
        &=
        \alpha_T P_{\overline{\mathcal E}_\mu}(\ell).
    \end{aligned}
    \label{eq:learning-leaf-domination-proof}
\end{equation}
For mixed probes and general POVMs, expand each leaf probability into its nonnegative rank-one contributions as in \eqref{eq:learning-leaf-moment}.
The decomposition weights are identical under both hypotheses, so summing the rank-one inequality gives pointwise domination for the original protocol.
No component of this algebraic decomposition is required to satisfy \eqref{eq:probe-entanglement-constraint}: the moment comparison is unrestricted, and the resource bounds will be applied to the actual conditional state and full POVM below.
Finally, \eqref{eq:learning-pointwise-domination} gives the total-variation bound.
\end{proof}

It remains to compare the transcript distributions generated by the averaged channels.
For two such channels $\overline{\mathcal E}_0$ and $\overline{\mathcal E}_1$, define
\begin{equation}
    \delta_{e_{\mathrm s},e_{\mathrm m}}
    (\overline{\mathcal E}_0,\overline{\mathcal E}_1)
    :=
    \frac12
    \sup_{\substack{
        A,\rho_{Q,A},\mathsf M=\{M_y\}_y\\
        E_N^{\mathrm s}(\rho_{Q,A})\leq e_{\mathrm s}\\
        E_N^{\mathrm m}(\mathsf M)\leq e_{\mathrm m}
    }}
    \sum_y
    \left|
        \operatorname{Tr} \left[
            M_y
            [(\overline{\mathcal E}_0-\overline{\mathcal E}_1)
            \otimes\operatorname{id}_A](\rho_{Q,A})
        \right]
    \right|.
    \label{eq:learning-single-query-distance}
\end{equation}
This is the largest total variation distance between the two single-query outcome distributions allowed by the probe and measurement entanglement constraints.
The auxiliary system $A$ may have arbitrary finite dimension.

For the same allowed adaptive protocol run with either channel, the complete transcript distributions satisfy
\begin{equation}
    d_{\mathrm{TV}}
    \bigl(P_{\overline{\mathcal E}_0},
          P_{\overline{\mathcal E}_1}\bigr)
    \leq
    T\delta_{e_{\mathrm s},e_{\mathrm m}}
    (\overline{\mathcal E}_0,\overline{\mathcal E}_1).
    \label{eq:learning-fixed-channel-hybrid}
\end{equation}
To see this, replace the channel in one query at a time.
At any fixed preceding history, each replacement changes the next-outcome distribution by at most $\delta_{e_{\mathrm s},e_{\mathrm m}}$.
Subsequent classical processing cannot increase this distance, so the triangle inequality gives the factor $T$.

Now suppose that two ensembles $\mu_0,\mu_1$ satisfy
Lemma~\ref{lem:learning-moment-domination}
with parameters $\alpha_{0,T},\alpha_{1,T}$.
Comparing each ensemble with its averaged channel and then applying
\eqref{eq:learning-fixed-channel-hybrid} gives
\begin{equation}
        d_{\mathrm{TV}}(P_{\mu_0},P_{\mu_1})
        \leq{}
        (1-\alpha_{0,T})+(1-\alpha_{1,T})+
        T\delta_{e_{\mathrm s},e_{\mathrm m}}
        \bigl(\overline{\mathcal E}_{\mu_0},
              \overline{\mathcal E}_{\mu_1}\bigr).
    \label{eq:learning-master-tv-bound}
\end{equation}
The first two terms account for correlations from reusing a sampled channel, while the last term bounds the distinguishability of the averaged channels.

For a fixed reference channel, there is no ensemble-to-average error on that side.
In particular, taking the completely depolarizing channel $\mathcal D$ as the reference yields
\begin{equation}
    d_{\mathrm{TV}}(P_\mu,P_{\mathcal D})
    \leq
    1-\alpha_T
    +T\delta_{e_{\mathrm s},e_{\mathrm m}}
    (\overline{\mathcal E}_\mu,\mathcal D).
    \label{eq:learning-ensemble-depolarizing}
\end{equation}
We now apply this bound to the COE and CSE.

\subsection{COE channels versus the completely depolarizing channel}
\label{sec:coe-depolarizing-bound}

\begin{lemma}[COE versus depolarizing]
\label{lem:coe-depolarizing-bound}
Consider a $T$-query protocol in the above classical-memory model whose conditional probes and POVMs satisfy \eqref{eq:probe-entanglement-constraint}.
Let $P_{\mathrm{COE}}$ denote its transcript distribution when a $d\times d$ COE matrix $V$ is sampled once and the channel $X\mapsto VXV^\dagger$ is queried in every round.
Let $P_{\mathcal D}$ denote the corresponding distribution under the completely depolarizing channel.
For $1\leq T\leq d$,
\begin{equation}
\begin{aligned}
d_{\mathrm{TV}}(P_{\mathrm{COE}},P_{\mathcal D})
&\leq
1-\prod_{j=0}^{T-1}\frac{d+1}{d+2j+1}
+\frac{T(d-1)(2^{e_*}+1)}{2d(d+1)}
\\
&\leq
\frac{T(T-1)}{d+1}
+\frac{T(d-1)(2^{e_*}+1)}{2d(d+1)}.
\end{aligned}
\label{eq:coe-depolarizing-tv}
\end{equation}
\end{lemma}

\begin{proof}
\emph{Fixed draw versus averaged channel.}
The averaged COE channel is
\begin{equation}
\overline{\mathcal E}(X)
:=
\mathbb E_V[VXV^\dagger]
=
\frac{\operatorname{Tr}(X)I_d+X^\mathrm{T}}{d+1}.
\label{eq:coe-average-channel-bound}
\end{equation}
Its unnormalized Choi operator is $M_1=2K_+/(d+1)$, where $K_+=(I+\mathbb S)/2$ and $\mathbb S$ swaps the output and input factors.
Moreover,
$\Pi_H=\mathsf S_TK_+^{\otimes T}$,
where $\mathsf S_T$ symmetrizes the $T$ complete output-input pairs, as defined in Lemma~\ref{lem:learning-moment-domination}.

For every partition $\lambda\vdash T$, the denominator in Lemma~\ref{lem:coe-moment-spectrum}, applied in dimension $d$, satisfies
\begin{equation}
C_\lambda(d+1)
=
\prod_{(i,j)\in\lambda}(d+2j-i)
\leq
\prod_{j=0}^{T-1}(d+2j+1).
\label{eq:coe-denominator-upper-bound}
\end{equation}
Indeed, when the boxes are enumerated row by row, the $m$-th box has column index at most $m$ and row index at least one, so its factor is at most $d+2m-1$.
All factors are positive for $T\leq d$.
The spectral decomposition and $\sum_\lambda Q_\lambda=\Pi_H$ therefore give
\begin{equation}
M_T
\succeq
\alpha_T
\left(\frac{2}{d+1}\right)^T
T!\Pi_H,
\qquad
\alpha_T
:=
\prod_{j=0}^{T-1}\frac{d+1}{d+2j+1}.
\label{eq:coe-moment-lower-bound}
\end{equation}
Applying Lemma~\ref{lem:learning-moment-domination}, we obtain
\begin{equation}
P_{\mathrm{COE}}(\ell)
\geq
\alpha_TP_{\overline{\mathcal E}}(\ell)
\label{eq:coe-leaf-domination}
\end{equation}
for every leaf $\ell$, and hence
\begin{equation}
d_{\mathrm{TV}}
(P_{\mathrm{COE}},P_{\overline{\mathcal E}})
\leq
1-\alpha_T.
\label{eq:coe-fixed-to-average}
\end{equation}

\emph{Averaged channel versus depolarization.}
For any input state $\rho_{Q,A}$, let
$\rho_A=\operatorname{Tr}_Q\rho_{Q,A}$.
Using \eqref{eq:coe-average-channel-bound}, we have
\begin{equation}
    \bigl[
        (\overline{\mathcal E}-\mathcal D)
        \otimes\operatorname{id}_A
    \bigr](\rho_{Q,A})
    =
    \frac{
        \rho_{Q,A}^{\mathrm T_Q}-I_d\otimes\rho_A/d
    }{d+1}.
    \label{eq:coe-one-query-difference}
\end{equation}
Let $p_y$ and $q_y$ denote the outcome probabilities obtained by applying
$\overline{\mathcal E}$ and $\mathcal D$, respectively, with the same
probe $\rho_{Q,A}$ and POVM $\mathsf M=\{M_y\}_y$.
Set $a_y=\operatorname{Tr}[M_y(I_d\otimes\rho_A)]$, $b_y=\operatorname{Tr}(M_y\rho_{Q,A}^{\mathrm T_Q})$, and $\kappa=2^{e_*}$.
The completely positive maps $X\mapsto\operatorname{Tr}(X)I_d\pm X^{\mathrm T}$ imply $|b_y|\leq a_y$.
Moreover, $\sum_y a_y=d$, $\sum_y b_y=1$, and $\sum_y|b_y|\leq\kappa$.
For $|x|\leq1$ and $c=1/d$, the piecewise-linear inequality
\begin{equation}
 |x-c|\leq c+(1-c)|x|-cx
 \label{eq:je-centered-absolute-bound}
\end{equation}
therefore gives
\begin{equation}
 \sum_y\left|b_y-\frac{a_y}{d}\right|
 \leq\left(1-\frac1d\right)(\kappa+1).
 \label{eq:je-centered-negativity-bound}
\end{equation}
Terms with $a_y=0$ vanish.
Since $p_y-q_y=(b_y-a_y/d)/(d+1)$, we obtain the sharpened bound
\begin{equation}
 \frac12\sum_y|p_y-q_y|
 \leq\frac{(d-1)(2^{e_*}+1)}{2d(d+1)}.
 \label{eq:coe-one-query-tv-bound}
 \csname ltx@label\endcsname{eq:coe-negativity-one-query-bound}
\end{equation}
This holds for every allowed mixed probe and full POVM, so it also bounds $\delta_{e_{\mathrm s},e_{\mathrm m}}(\overline{\mathcal E},\mathcal D)$.

Applying \eqref{eq:learning-fixed-channel-hybrid} therefore gives
\begin{equation}
    d_{\mathrm{TV}}
    (P_{\overline{\mathcal E}},P_{\mathcal D})
    \leq
    T\delta_{e_{\mathrm s},e_{\mathrm m}}
    (\overline{\mathcal E},\mathcal D)
    \leq
    \frac{T(d-1)(2^{e_*}+1)}{2d(d+1)}.
    \label{eq:coe-average-to-depolarizing}
\end{equation}

Combining \eqref{eq:coe-fixed-to-average}
and~\eqref{eq:coe-average-to-depolarizing}
through the triangle inequality proves the first bound.
Finally, using
$1-\prod_j(1-x_j)\leq\sum_jx_j$
for $x_j\in[0,1]$, we have
\begin{equation}
1-\alpha_T
\leq
\sum_{j=0}^{T-1}\frac{2j}{d+2j+1}
\leq
\frac{T(T-1)}{d+1},
\label{eq:coe-correlation-error}
\end{equation}
which proves the second bound.
\end{proof}

In particular, if every conditional probe is PPT, or every conditional POVM has only PPT effects, then $e_*=0$ and
\begin{equation}
d_{\mathrm{TV}}(P_{\mathrm{COE}},P_{\mathcal D})
\leq
\frac{T(dT-1)}{d(d+1)}.
\label{eq:coe-depolarizing-no-ancilla}
\csname ltx@label\endcsname{eq:coe-depolarizing-ppt}
\end{equation}
By Lemma~\ref{lem:channel-le-cam}, distinguishing these hypotheses with equal prior probabilities and success probability at least $2/3$ requires
$T=\Omega(\sqrt d)$.
For an $n$-qubit channel, $d=2^n$, giving the exponential lower bound $\Omega(2^{n/2})$ even with an arbitrarily large auxiliary system.

\subsection{CSE channels versus the completely depolarizing channel}
\label{sec:cse-depolarizing-bound}

The CSE case follows the same argument as the COE case, with the symmetric projector replaced by its symplectic counterpart.
Throughout this subsection, $d$ is even and $J$ is the fixed symplectic matrix defining the physical time-reversal transformation.

\begin{lemma}[CSE versus depolarizing]
\label{lem:cse-depolarizing-bound}
Consider a $T$-query protocol in the above classical-memory model whose conditional probes and POVMs satisfy \eqref{eq:probe-entanglement-constraint}.
Let $P_{\mathrm{CSE}}$ denote its transcript distribution when a $d\times d$ CSE matrix is sampled once and its unitary channel is queried in every round.
For $1\leq T\leq d/2$,
\begin{equation}
\begin{aligned}
d_{\mathrm{TV}}(P_{\mathrm{CSE}},P_{\mathcal D})
&\leq
1-\prod_{j=0}^{T-1}\frac{d-1}{d+j-1}
+\frac{T(2^{e_*}+1)}{2d}
\\
&\leq
\frac{T(T-1)}{2(d-1)}+\frac{T(2^{e_*}+1)}{2d}.
\end{aligned}
\label{eq:cse-depolarizing-tv}
\end{equation}
\end{lemma}

\begin{proof}
\emph{Fixed draw versus averaged channel.}
The averaged CSE channel is
\begin{equation}
\overline{\mathcal S}(X)
:=
\mathbb E_V[VXV^\dagger]
=
\frac{\operatorname{Tr}(X)I_d-J^\dagger X^\mathrm{T} J}{d-1}.
\label{eq:cse-average-channel-bound}
\end{equation}
Its Choi operator is $M_1^{\mathrm{CSE}}=2K_J/(d-1)$, where
\[
K_J
:=
(J^\dagger\otimes I_d)
\frac{I-\mathbb S}{2}
(J\otimes I_d).
\]
The projectors in Lemma~\ref{lem:cse-moment-spectrum} sum to
$\mathsf S_TK_J^{\otimes T}$.

As in \eqref{eq:coe-denominator-upper-bound}, enumerating the Young diagram boxes row by row gives
\begin{equation}
C_\lambda^{\mathrm{CSE}}(d)
=
\prod_{(i,j)\in\lambda}(d+j-2i)
\leq
\prod_{j=0}^{T-1}(d+j-1).
\label{eq:cse-denominator-upper-bound}
\end{equation}
All factors are positive for $T\leq d/2$.
Writing
$\beta_T:=\prod_{j=0}^{T-1}(d-1)/(d+j-1)$,
the CSE spectral decomposition therefore implies
\begin{equation}
M_T^{\mathrm{CSE}}
\succeq
\beta_T
\left(\frac{2}{d-1}\right)^T
T!\mathsf S_TK_J^{\otimes T}.
\label{eq:cse-moment-lower-bound}
\end{equation}
Lemma~\ref{lem:learning-moment-domination} consequently yields
\begin{equation}
d_{\mathrm{TV}}
(P_{\mathrm{CSE}},P_{\overline{\mathcal S}})
\leq
1-\beta_T.
\label{eq:cse-fixed-to-average}
\end{equation}

\emph{Averaged channel versus depolarization.}
For an actual conditional probe $\rho_{Q,A}$, \eqref{eq:cse-average-channel-bound} gives
\begin{equation}
\begin{aligned}
&\bigl[(\overline{\mathcal S}-\mathcal D)
\otimes\operatorname{id}_A\bigr](\rho_{Q,A})\\
&=
\frac{I_d\otimes\rho_A/d
-(J^\dagger\otimes I_A)\rho_{Q,A}^{\mathrm T_Q}(J\otimes I_A)}{d-1}.
\end{aligned}
\label{eq:cse-negativity-difference}
\end{equation}
For the actual POVM $\mathsf M=\{M_y\}_y$, define
$\widetilde M_y=(J\otimes I_A)M_y(J^\dagger\otimes I_A)$.
Local unitary invariance gives $E_N^{\mathrm m}(\widetilde{\mathsf M})=E_N^{\mathrm m}(\mathsf M)$.
Writing $p_y^- =\operatorname{Tr}[M_y(\overline{\mathcal S}\otimes\operatorname{id}_A)(\rho_{Q,A})]$,
$q_y=\operatorname{Tr}[M_y(I_d\otimes\rho_A/d)]$, and
$a_y^J=\operatorname{Tr}(\widetilde M_y\rho_{Q,A}^{\mathrm T_Q})$,
we have $p_y^--q_y=(q_y-a_y^J)/(d-1)$.
Applying \eqref{eq:je-centered-negativity-bound} to the locally rotated POVM, with $a_y=dq_y$ and $b_y=a_y^J$, yields
\begin{equation}
 \frac12\sum_y|p_y^--q_y|
 \leq\frac{(1-1/d)(2^{e_*}+1)}{2(d-1)}
 =\frac{2^{e_*}+1}{2d}.
 \label{eq:cse-one-query-tv-bound}
 \csname ltx@label\endcsname{eq:cse-negativity-one-query-bound}
\end{equation}
Applying \eqref{eq:learning-fixed-channel-hybrid} gives
\begin{equation}
d_{\mathrm{TV}}
(P_{\overline{\mathcal S}},P_{\mathcal D})
\leq
\frac{T(2^{e_*}+1)}{2d}.
\label{eq:cse-average-to-depolarizing}
\end{equation}
The triangle inequality proves the first bound.
Finally,
\begin{equation}
1-\beta_T
\leq
\sum_{j=0}^{T-1}\frac{j}{d+j-1}
\leq
\frac{T(T-1)}{2(d-1)},
\label{eq:cse-correlation-error}
\end{equation}
which gives the second bound.
\end{proof}

\subsection{Distinguishing circular ensembles from Haar-random channels}
\label{sec:circular-ensembles-versus-haar}

Let $P_{\mathrm H}$ denote the transcript distribution when a Haar-random unitary on $\mathbb C^d$ is sampled once and reused throughout the protocol.
Kean Chen et al.~\cite{chen2023unitarity} established an $\Omega(\sqrt d)$ query lower bound for distinguishing this hypothesis from the completely depolarizing channel under incoherent access.
Their model allows arbitrary finite-dimensional auxiliary systems, unrestricted probe entanglement and joint measurements, and classical adaptation, while retaining no quantum memory between queries.

We use the following finite-query bound, whose validity over the full range $1\leq T\leq d$ is verified below:
\begin{equation}
    d_{\mathrm{TV}}(P_{\mathrm H},P_{\mathcal D})
    \leq
    1-\prod_{j=0}^{T-1}\frac{d}{d+j}
    \leq
    \frac{T(T-1)}{2d}.
    \label{eq:haar-depolarizing-tv}
\end{equation}
With our unnormalized Choi convention, the Haar moment operator satisfies
\begin{equation}
    M_T^{\mathrm H}
    \succeq
    \frac{T!}{\prod_{j=0}^{T-1}(d+j)}
    \mathsf S_T.
    \label{eq:haar-moment-lower-bound}
\end{equation}
Indeed, its eigenvalues on the orthogonal blocks summing to $\mathsf S_T$ are
$T!/\prod_{(i,j)\in\lambda}(d+j-i)$,
and each denominator is at most $\prod_{j=0}^{T-1}(d+j)$.
Since $M_1^{\mathrm H}=I_{d^2}/d$, Lemma~\ref{lem:learning-moment-domination} gives the first inequality in \eqref{eq:haar-depolarizing-tv}.
The second follows from
$1-\prod_j(1-x_j)\leq\sum_jx_j$
with $x_j=j/(d+j)$.
The averaged Haar channel is exactly $\mathcal D$, so no additional one-query error term is needed.
In particular, this bound is independent of the auxiliary dimension and requires no negativity constraint on either the probes or the measurements.

\begin{lemma}[Circular ensembles versus Haar]
\label{lem:circular-ensembles-versus-haar}
For any $T$-query protocol in the above classical-memory model satisfying \eqref{eq:probe-entanglement-constraint}, the COE transcript distribution obeys, for $1\leq T\leq d$,
\begin{equation}
d_{\mathrm{TV}}(P_{\mathrm{COE}},P_{\mathrm H})
\leq{}
\frac{T(T-1)}{d+1}
+\frac{T(d-1)(2^{e_*}+1)}{2d(d+1)}
+
\frac{T(T-1)}{2d}.
\label{eq:coe-haar-tv}
\csname ltx@label\endcsname{eq:coe-haar-entanglement-bound}
\end{equation}
For even $d$ and $1\leq T\leq d/2$, the CSE transcript distribution satisfies
\begin{equation}
d_{\mathrm{TV}}(P_{\mathrm{CSE}},P_{\mathrm H})
\leq
\frac{T(T-1)}{2(d-1)}+\frac{T(2^{e_*}+1)}{2d}
+
\frac{T(T-1)}{2d}.
\label{eq:cse-haar-tv}
\csname ltx@label\endcsname{eq:cse-haar-entanglement-bound}
\end{equation}
\end{lemma}

\begin{proof}
For either ensemble $\mathsf E\in\{\mathrm{COE},\mathrm{CSE}\}$, the triangle inequality gives
\begin{equation}
d_{\mathrm{TV}}(P_{\mathsf E},P_{\mathrm H})
\leq
d_{\mathrm{TV}}(P_{\mathsf E},P_{\mathcal D})
+
d_{\mathrm{TV}}(P_{\mathrm H},P_{\mathcal D}).
\label{eq:circular-haar-triangle}
\end{equation}
Substituting Lemma~\ref{lem:coe-depolarizing-bound} or Lemma~\ref{lem:cse-depolarizing-bound}, together with \eqref{eq:haar-depolarizing-tv}, proves the claims.
\end{proof}

The following theorem gives both binary discrimination bounds and the three-class statement of Theorem~\ref{thm:main-negativity-lower-bound}.

\begin{theorem}[Bounded probe and measurement entanglement]
\label{thm:entanglement-limited-lower-bound}
Consider protocols with only classical memory between rounds and arbitrary auxiliary dimension, with every conditional probe and POVM satisfying \eqref{eq:probe-entanglement-constraint}.
Let $e_* = \min\{e_{\mathrm s},e_{\mathrm m}\}$.
The unknown unitary is sampled once from the specified ensemble and its channel is reused in all rounds.
Distinguishing the COE from Haar, or the CSE from Haar when $d$ is even, with equal priors and success probability at least $2/3$ requires
\begin{equation}
    T=\Omega \left(
        \min\left\{\sqrt d,\frac{d}{2^{e_*}}\right\}
    \right).
    \label{eq:circular-haar-query-lower-bound}
    \csname ltx@label\endcsname{eq:entanglement-limited-query-lower-bound}
\end{equation}
For even $d$, the same lower bound holds for identifying a COE, CSE, or Haar draw with equal prior probability and average success probability at least $2/3$.
In particular, for an $n$-qubit channel with $d=2^n$,
\begin{equation}
    T=\Omega \left(
        \min\left\{2^{n/2},2^{n-\min\{e_{\mathrm s},e_{\mathrm m}\}}\right\}
    \right).
    \label{eq:entanglement-limited-qubit-lower-bound}
\end{equation}
\end{theorem}

\begin{proof}
Both transcript bounds in Lemma~\ref{lem:circular-ensembles-versus-haar} are of order $(T^2+T2^{e_*})/d$.
For binary discrimination with equal priors, Lemma~\ref{lem:channel-le-cam} implies that success probability at least $2/3$ requires total variation distance at least $1/3$.
The stated lower bound follows because both terms are small whenever $T$ is a sufficiently small constant times $\min\{\sqrt d,d/2^{e_*}\}$.
The required ranges $T\leq d$ and $T\leq d/2$ are satisfied at this scale for sufficiently large $d$.

For the three-class problem, let $a_{\mathrm{COE}}(\ell)$, $a_{\mathrm{CSE}}(\ell)$, and $a_{\mathrm H}(\ell)$ be the conditional probabilities of the final decisions given transcript $\ell$.
These numbers lie in $[0,1]$ and sum to one for every leaf.
Using $P_{\mathcal D}$ as a common reference gives the sharper direct bound
\begin{equation}
\begin{aligned}
 p_{\mathrm{succ}}
 &=\frac13\sum_{\mathsf E\in\{\mathrm{COE},\mathrm{CSE},\mathrm H\}}
   \sum_\ell P_{\mathsf E}(\ell)a_{\mathsf E}(\ell)\\
 &\leq\frac13+\frac13
   \sum_{\mathsf E\in\{\mathrm{COE},\mathrm{CSE},\mathrm H\}}
   d_{\mathrm{TV}}(P_{\mathsf E},P_{\mathcal D}).
\end{aligned}
\label{eq:three-class-transcript-success-bound}
\end{equation}
Indeed, the reference contribution is $\sum_\ell P_{\mathcal D}(\ell)\sum_{\mathsf E}a_{\mathsf E}(\ell)=1$, and each remaining summand is bounded by its total variation distance.
Success probability at least $2/3$ requires the sum of the three total variation distances to be at least one, which, together with \eqref{eq:coe-depolarizing-tv}, \eqref{eq:cse-depolarizing-tv}, and \eqref{eq:haar-depolarizing-tv}, yields the same asymptotic query lower bound.
The qubit expression follows by setting $d=2^n$.
\end{proof}

\section{TRS signatures and certification protocols}
\label{app:protocol-extensions}

This appendix establishes the properties of the TRS signatures and proves the certification guarantees in the main text.
We first derive the exchange identities and show that the signatures separate the COE, CSE, and Haar ensembles with high probability.
We then prove their dynamical interpretation, including Observation~\ref{obs:main-trs-exchange} and Theorem~\ref{thm:main-trs-curvature}.
The remaining subsections analyze the estimator of Theorem~\ref{thm:gfp-testing} with a fixed probe and fixed measurements, and the protocol without auxiliary systems of Theorem~\ref{thm:main-limited-ancilla-protocol}.
Throughout, the same unknown unitary channel is reused in every query, and only classical information is retained between runs.

\subsection{Exchange signatures and ensemble separation}
\label{app:protocol-signatures}

Throughout this appendix, $d$ denotes the total query dimension; it is even whenever the symplectic signature or the CSE is used.
We fix
\begin{equation}
    J=i\sigma_y\otimes I_{d/2},
     J^*=J,\quad J^{\mathrm T}=J^\dagger=-J,\quad J^2=-I_d,
     U^{\mathrm D}:=JU^{\mathrm T}J^\dagger.
    \label{eq:protocol-canonical-J}
\end{equation}
The normalized Choi state vector and state are
\begin{equation}
    \dket{U}=(U\otimes I)\ket{\Phi_d},
\qquad    
    \ket{\Phi_d}=\frac1{\sqrt d}\sum_{j=0}^{d-1}\ket{j,j},
    \qquad
    \rho_{\mathcal U}=\dket{U}\dbra{U}.
    \label{eq:unitary-choi-vector}
\end{equation}
Let $\mathbb S$ be the swap and $\mathbb S_J=\mathbb S(J\otimes J)$, as in \eqref{eq:main-trs-signatures}.

\paragraph{Exchange identities.}
Both $\mathbb S$ and $\mathbb S_J$ are Hermitian involutions, so their expectations are real and lie in $[-1,1]$.
Using $(A\otimes B)\operatorname{vec}(X)=\operatorname{vec}(AXB^{\mathrm T})$ and the convention in \eqref{eq:unitary-choi-vector}, we obtain
\begin{equation}
    \mathbb S\dket{U}=\dket{U^{\mathrm T}},
\qquad    
    \mathbb S_J\dket{U}=\dket{U^{\mathrm D}}.
    \label{eq:protocol-exchange-action}
\end{equation}
In the second identity, $(JUJ^{\mathrm T})^{\mathrm T}=JU^{\mathrm T}J^\dagger$.
Consequently,
\begin{equation}
    f_+(\mathcal U)=\frac1d\operatorname{Tr}(U^\dagger U^{\mathrm T})
        =\frac1d\operatorname{Tr}(UU^*),
\qquad    
    f_-(\mathcal U)=\frac1d\operatorname{Tr}(U^\dagger U^{\mathrm D}).
    \label{eq:protocol-signature-overlaps}
\end{equation}
An expectation of a Hermitian involution equals $1$ exactly when the state vector belongs to its $+1$ eigenspace.
Thus \eqref{eq:protocol-exchange-action} gives
\begin{equation}
    f_+(\mathcal U)=1\ \Longleftrightarrow\ U=U^{\mathrm T},
\qquad    
    f_-(\mathcal U)=1\ \Longleftrightarrow\ U=U^{\mathrm D}.
\end{equation}
These are precisely the two dynamical TRS conditions in the canonical bases.
The quantitative Frobenius-distance identity of Observation~\ref{obs:main-trs-exchange} is proved in Appendix~\ref{app:trs-dynamical-interpretation}.

\paragraph{Moments of the signatures.}
For Haar-random $U\in \mathrm{U}(d)$, the first two moments are
\begin{equation}
    \mathbb E_{\mathrm H}f_+=\frac1d,
\qquad    
    \mathbb E_{\mathrm H}f_-=-\frac1d,
    \qquad
    \mathbb E_{\mathrm H}f_+^2
      =\mathbb E_{\mathrm H}f_-^2=\frac2{d^2}.
    \label{eq:protocol-Haar-signature-moments}
\end{equation}
For the circular ensembles, the exact moments of the other signature are
\begin{align}
    &f_+=1\quad\text{on COE},
    &\mathbb E_{\mathrm{COE}}f_-&=-\frac1{d+1},
    &\mathbb E_{\mathrm{COE}}f_-^2
        &=\frac{3d+8}{d(d+1)(d+3)},
        \label{eq:protocol-COE-cross-moments}\\
    &f_-=1\quad\text{on CSE},
    &\mathbb E_{\mathrm{CSE}}f_+&=\frac1{d-1},
    &\mathbb E_{\mathrm{CSE}}f_+^2
        &=\frac{3d-8}{d(d-1)(d-3)}.
        \label{eq:protocol-CSE-cross-moments}
\end{align}
The Haar formula for $f_+$ holds for $d\ge2$; all formulas involving $f_-$ or a circular cross signature are stated here for even $d\ge4$.
The exceptional case $d=2$ is not needed for the asymptotic classification guarantee.

To verify \eqref{eq:protocol-Haar-signature-moments}, write $\chi(U)=\operatorname{Tr}(UU^*)=\sum_{i,j}U_{ij}\overline{U_{ji}}$.
The one- and two-copy Haar formulas give
\begin{equation}
    \mathbb E_{\mathrm H}\chi=1,
\qquad    
    \mathbb E_{\mathrm H}\chi^2
       =\frac{2d^2}{d^2-1}-\frac{2}{d^2-1}=2.
\end{equation}
Moreover, \eqref{eq:protocol-signature-overlaps} gives $f_-(\mathcal U)=-f_+(\mathcal U\circ\mathcal J)$, where $\mathcal J(X)=JXJ^\dagger$.
Right multiplication by $J$ preserves Haar measure, proving the remaining Haar moments.

We derive the circular-ensemble moments by contracting the one- and two-copy moment operators from Lemmas~\ref{lem:coe-moment-spectrum} and \ref{lem:cse-moment-spectrum}.
Let $\varepsilon\in\{+1,-1\}$ and set $X=U$ for COE and $X=JU$ for CSE, respectively.
Then $X^{\mathrm T}=\varepsilon X$; its distribution is that of $WW^{\mathrm T}$ or $WJW^{\mathrm T}$, respectively, for Haar-random $W\in \mathrm{U}(d)$.
Writing $\ket{x}=\operatorname{vec}(X)$ without normalization, the signature not fixed by the ensemble symmetry is
\begin{equation}
    h_\varepsilon(X)
    :=\frac1d\operatorname{Tr}(X^\dagger JXJ^\dagger)
    =\frac1d\bra{ x}J\otimes J\ket{x}
    =\begin{cases}
       f_-(\mathcal U),&\varepsilon=+1,\\
       f_+(\mathcal U),&\varepsilon=-1.
     \end{cases}
    \label{eq:protocol-cross-signature-contraction}
\end{equation}
The one-copy moment is $\mathbb E\ketbra{x}{ x}=(I+\varepsilon\mathbb S)/(d+\varepsilon)$.
Using $\operatorname{Tr}J=0$ and $\operatorname{Tr}J^2=-d$ gives $\mathbb E h_\varepsilon=-\varepsilon/(d+\varepsilon)$.

For the second moment, let $H_2\subset S_4$ preserve the pairing $\{\{1,2\},\{3,4\}\}$.
Set $\xi_+(\sigma)=1$, $\xi_-(\sigma)=\operatorname{sgn}(\sigma)$, and define
\begin{equation}
    \Pi_H^\varepsilon:=\frac18\sum_{\eta\in H_2}
        \xi_\varepsilon(\eta)P_\eta,
\qquad    
    P_4^\varepsilon:=\frac1{24}\sum_{\sigma\in S_4}
        \xi_\varepsilon(\sigma)P_\sigma.
\end{equation}
Here $P_4^+$ and $P_4^-$ are the fully symmetric and fully antisymmetric projectors.
Specializing the aforementioned moment spectra to two copies, and writing their dimensions in terms of the total query dimension $d$, gives
\begin{equation}
    \mathbb E(\ketbra{x}{ x})^{\otimes2}
    =\frac8{d(d+\varepsilon)}\Pi_H^\varepsilon
     -\frac{24\varepsilon}{d(d+\varepsilon)(d+3\varepsilon)}
        P_4^\varepsilon.
    \label{eq:protocol-two-copy-circular-moment}
\end{equation}
For a permutation $\sigma$, $\operatorname{Tr}(J^{\otimes4}P_\sigma) =\prod_{c}\operatorname{Tr}(J^{|c|})$, where $c$ runs over its cycles.
This trace vanishes unless the cycle type is $(2,2)$ or $(4)$, in which cases it equals $d^2$ or $d$, respectively.
All three double transpositions and two of the six four-cycles belong to $H_2$.
Therefore
\begin{equation}
    \operatorname{Tr}(J^{\otimes4}\Pi_H^\varepsilon)
       =\frac{3d^2+2\varepsilon d}{8},
\qquad    
    \operatorname{Tr}(J^{\otimes4}P_4^\varepsilon)
       =\frac{d(d+2\varepsilon)}8.
\end{equation}
Contracting \eqref{eq:protocol-two-copy-circular-moment} with $d^{-2}J^{\otimes4}$ yields
\begin{equation}
    \mathbb E h_\varepsilon^2
       =\frac{3d+8\varepsilon}
        {d(d+\varepsilon)(d+3\varepsilon)},
\end{equation}
which proves \eqref{eq:protocol-COE-cross-moments} and \eqref{eq:protocol-CSE-cross-moments}.

\paragraph{A constant-query three-class protocol.}
Markov's inequality applied to the squared signatures, together with a union bound for Haar, gives, for even $d\ge4$ and every $\eta>0$,
\begin{equation}\label{eq:protocol-signature-tails}
\begin{aligned}
    \Pr_{\mathrm H}\{\max(|f_+|,|f_-|)>\eta\}
       &\le\frac4{d^2\eta^2},\\
    \Pr_{\mathrm{COE}}\{|f_-|>\eta\}
       &\le\frac3{d^2\eta^2},\\
    \Pr_{\mathrm{CSE}}\{|f_+|>\eta\}
       &\le\frac6{d^2\eta^2}.
\end{aligned}
\end{equation}
Thus the typical signature pairs approach $(1,0)$, $(0,1)$, and $(0,0)$ for COE, CSE, and Haar, respectively.
To distinguish them, prepare $m$ fresh Choi states for each signature and perform the corresponding two-outcome measurement
\begin{equation}
    \mathsf M_O=\left\{\frac{I+O}{2},\frac{I-O}{2}\right\},
     O\in\{\mathbb S,\mathbb S_J\},
\end{equation}
assigning outcomes $+1$ and $-1$.
Each Choi state uses one invocation of the same unknown channel.
Conditional on this fixed $U$, the outcomes are independent, and Hoeffding's inequality bounds the sample means $\widehat f_\pm$ by
\begin{equation}
    \Pr\left\{\max_{s\in\{+,-\}}
       |\widehat f_s-f_s|>\frac14\middle|U\right\}
       \le4e^{-m/32}.
\end{equation}
Output COE if only $\widehat f_+$ exceeds $1/2$, CSE if only $\widehat f_-$ exceeds $1/2$, and Haar if neither does.
If both exceed $1/2$, output any fixed label.
On the typical event in \eqref{eq:protocol-signature-tails} with $\eta=1/4$, estimation errors at most $1/4$ guarantee the correct label.
Under each ensemble, and hence also under equal priors, the error probability satisfies
\begin{equation}
    p_{\mathrm{err}}\le\frac{96}{d^2}+4e^{-m/32}.
    \label{eq:protocol-three-class-error}
\end{equation}
For $0<\delta<1$, taking $m\ge32\log(8/\delta)$ and $d^2\ge192/\delta$ gives success probability at least $1-\delta$ using $2m=\mathcal{O}(\log(1/\delta))$ queries.
The query complexity is therefore constant at fixed confidence for sufficiently large $d$.
This guarantee concerns classification under the ensemble promises: the two exact TRS conditions need not be mutually exclusive, as $U=I_d$ illustrates.

The resource requirements agree with Theorem~\ref{thm:main-negativity-lower-bound}.
For $d=2^n$, the maximally entangled probe has $E_N^{\mathrm s}=n$, and Appendix~\ref{app:measurement-negativity} gives $E_N^{\mathrm m}(\mathsf M_{\mathbb S})=n$.
The identity $\mathbb S_J=-(J\otimes I)\mathbb S(J^\dagger\otimes I)$ shows that $\mathsf M_{\mathbb S_J}$ differs from the swap POVM only by a local unitary and an outcome relabeling, so its measurement negativity is also $n$.
Hence this protocol attains the constant-query scale at $e_{\mathrm s}=e_{\mathrm m}=n$.

\subsection{Dynamical interpretation of the TRS signatures}
\label{app:trs-dynamical-interpretation}

We write $\Xi$ for the physical time-reversal operator, represented in a canonical basis by $\Xi_+=K$ or $\Xi_-=JK$.

\paragraph{Proof of Observation~\ref{obs:main-trs-exchange}.}
Let $\mathcal R_\pm(U):=\Xi_\pm U^\dagger\Xi_\pm^{-1}$.
The overlap identities established above give $f_\pm(\mathcal U)=d^{-1}\operatorname{Tr}[U^\dagger\mathcal R_\pm(U)]\in\mathbb R$.
Since both $U$ and $\mathcal R_\pm(U)$ are unitary,
\begin{equation}
    \begin{aligned}
    \|\Xi_\pm U\Xi_\pm^{-1}-U^\dagger\|_{\mathrm F}^2&=
        \|U-\mathcal R_\pm(U)\|_{\mathrm F}^2
        \\&=2d-2\operatorname{Re}\operatorname{Tr}
          [U^\dagger\mathcal R_\pm(U)]\\
        &=2d\bigl[1-f_\pm(\mathcal U)\bigr].
    \end{aligned}
    \label{eq:app-trs-defect-proof}
\end{equation}
Thus $f_\pm(\mathcal U)=1$ if and only if $U=\Xi_\pm U^\dagger\Xi_\pm^{-1}$, equivalently $\Xi_\pm U\Xi_\pm^{-1}=U^\dagger$.

\paragraph{Proof of Theorem~\ref{thm:main-trs-curvature}.}
For a time-independent Hamiltonian $H$, let $U(t)=e^{-iHt}$ and $\mathcal U_t(X)=U(t)XU(t)^\dagger$.
Define $H_\pm:=(H-\Xi_\pm H\Xi_\pm^{-1})/2$.
Antiunitarity implies $\Xi_\pm U(t)^\dagger\Xi_\pm^{-1}=e^{-i\Xi_\pm H\Xi_\pm^{-1}t}$, so
\begin{equation}
    f_\pm(\mathcal U_t)
    =\frac1d\operatorname{Tr}
      \left[e^{itH}e^{-i\Xi_\pm H\Xi_\pm^{-1}t}\right].
    \label{eq:app-trs-fidelity-amplitude}
\end{equation}
Differentiating twice at $t=0$ and using cyclicity of the trace yields
\begin{equation}
    \begin{aligned}
        \left.\frac{d^2}{dt^2}f_\pm(\mathcal U_t)\right|_{t=0}
        &=\frac1d\operatorname{Tr}
          \left[2H\Xi_\pm H\Xi_\pm^{-1}-H^2-(\Xi_\pm H\Xi_\pm^{-1})^2\right]\\
        &=-\frac1d\operatorname{Tr}(H-\Xi_\pm H\Xi_\pm^{-1})^2
         =-\frac4d\operatorname{Tr}(H_\pm^2).
    \end{aligned}
    \label{eq:app-trs-curvature-proof}
\end{equation}
Since $H_\pm$ is Hermitian, the curvature vanishes if and only if $H_\pm=0$, equivalently $\Xi_\pm H\Xi_\pm^{-1}=H$.
Moreover, cyclicity of the trace gives $f_\pm(\mathcal U_{-t})=\overline{f_\pm(\mathcal U_t)}=f_\pm(\mathcal U_t)$, so the signature is even in $t$.
Its short-time expansion is therefore $1-f_\pm(\mathcal U_t)=2t^2\operatorname{Tr}(H_\pm^2)/d+\mathcal{O}(t^4)$.

\subsection{One-sided testing with fixed probes and measurements}
\label{app:gfp-testing}

We prove Theorem~\ref{thm:gfp-testing} for a fixed, known probe
$\rho_{Q,A}$ and a fixed joint POVM
$\mathsf M=\{M_y\}_y$.
The query register has dimension $d=2^n\geq2$, while the auxiliary
register may have arbitrary finite dimension.
The probe may be pure or mixed.
Every run uses the same preparation and POVM, with independently
sampled local control unitaries on $Q$ before and after the unknown
evolution.
For brevity, write $\rho\equiv\rho_{Q,A}$.

\begin{proof}[Proof of Theorem~\ref{thm:gfp-testing}]
\ \\
\noindent\textit{Support and the averaged experiment.}\quad
Denote
\begin{equation}
\begin{gathered}
A_\rho=I_Q\otimes\rho_A,
\qquad
B_\rho=\rho_{Q,A}^{\mathrm T_Q},\\
a_y=\operatorname{Tr}
\left[M_y(I_Q\otimes\rho_A)\right],\qquad
b_y=\operatorname{Tr}
\left(M_y\rho_{Q,A}^{\mathrm{T}_Q}\right),
\end{gathered}
\end{equation}
and
\begin{equation}
\begin{aligned}
    \mathcal F_\tau(\rho,\mathsf M)
    &:=
    \{y\mid a_y>0,\ b_y=-\tau a_y\},\\
    \lambda_\tau(\rho,\mathsf M)
    &:=
    \sum_{y\in\mathcal F_\tau(\rho,\mathsf M)}a_y .
\end{aligned}
\label{eq:gfp-lambda}
\end{equation}
The completely positive maps
$X\mapsto\operatorname{Tr}(X)I_Q\pm X^{\mathrm T}$
have unnormalized Choi operators $I\pm \mathbb{S}\geq0$.
Hence
\begin{equation}
    -A_\rho\leq B_\rho\leq A_\rho,
    \qquad
    a_y\geq0,
    \qquad
    |b_y|\leq a_y.
    \label{eq:gfp-domination}
\end{equation}
If $a_y=0$, positivity implies that $M_y$ annihilates
$\operatorname{supp}(A_\rho)$.
Every state obtained from $\rho_{Q,A}$ by applying a unitary on $Q$
is supported on this subspace.
Such an outcome therefore has zero probability for every control
unitary and may be omitted.

Choose a unitary representative $U$ of $\mathcal U$ and set
\begin{equation}
    U_\tau=
    \begin{cases}
        U,&\tau=+1,\\
        UJ,&\tau=-1,
    \end{cases}
    \qquad
    \mathcal U_\tau(X)=U_\tau XU_\tau^\dagger .
\end{equation}
Since
\begin{equation}
    f_+(\mathcal U\circ\mathcal J)
    =
    -f_-(\mathcal U),
    \qquad
    \mathcal J(X)=JXJ^\dagger,
\end{equation}
we have
\begin{equation}
    g:=f_+(\mathcal U_\tau)
    =
    \tau f_\tau(\mathcal U).
\end{equation}
In each round, independently sample $W$ from an exact unitary
$2$-design, apply $W$ or $JW$ before the query according to $\tau$,
and apply $W^{\mathrm T}$ afterwards.
The effective unitary is $W^{\mathrm T}U_\tau W$ and uses one
invocation of $\mathcal U$.
Let $\mathcal V_W$ denote the unitary channel represented by
$V_W=W^{\mathrm T}U_\tau W$, and let
$\rho_{\mathcal V_W}
=|V_W\rangle\rangle\langle\langle V_W|$
denote its normalized Choi state. Then the averaged output state
and outcome probabilities are
\begin{equation}
\label{eq:gfp-averaged}
\begin{aligned}
\sigma_g
&:=
\mathbb E_W\left[
(V_W\otimes I_A)\rho_{Q,A}
(V_W^\dagger\otimes I_A)
\right]
=
\frac{(d-g)A_\rho+(dg-1)B_\rho}{d^2-1},\\
p_y(g)
&:=
\operatorname{Tr}(M_y\sigma_g)
=
\frac{(d-g)a_y+(dg-1)b_y}{d^2-1}.
\end{aligned}
\end{equation}
For completeness, this follows by twirling the normalized Choi state
of $U_\tau$:
\begin{equation}
    \mathbb E_W\rho_{\mathcal V_W}
    =
    \frac{1+g}{d(d+1)}\Pi_+
    +
    \frac{1-g}{d(d-1)}\Pi_-,
    \qquad
    \Pi_\pm=\frac{I\pm \mathbb{S}}{2}.
\end{equation}
The Choi-state matrix elements have degree two in $W$ and degree two
in its complex conjugate, so an exact unitary $2$-design suffices.

\noindent\textit{Forbidden outcomes and the upper bound.}\quad
For $y\in\mathcal F_\tau(\rho,\mathsf M)$,
$b_y=-\tau a_y$.
Substituting this relation and
$g=\tau f_\tau(\mathcal U)$ into
~\eqref{eq:gfp-averaged} gives
\begin{equation}
    p_y(g)
    =
    \frac{a_y[1-f_\tau(\mathcal U)]}{d-\tau}.
    \label{eq:gfp-individual-forbidden}
\end{equation}
Therefore, if the protocol rejects whenever an outcome in
$\mathcal F_\tau(\rho,\mathsf M)$ occurs, its one-round rejection
probability is
\begin{equation}
    q_\tau(\mathcal U)
    =
    \frac{\lambda_\tau(\rho,\mathsf M)}{d-\tau}
    [1-f_\tau(\mathcal U)].
    \label{eq:gfp-rejection}
\end{equation}

Under $H_0^{(\tau)}$, $q_\tau(\mathcal U)=0$, so the protocol accepts
with probability one.
Under $H_1^{(\tau)}$,
\begin{equation}
    q_\tau(\mathcal U)
    \geq
    \frac{\lambda_\tau(\rho,\mathsf M)\epsilon}{d-\tau}.
\end{equation}
Since the unknown channel is fixed while the controls and measurements
are sampled independently in different rounds,
\begin{equation}
\begin{aligned}
    \Pr(\mathrm{accept}\mid H_1^{(\tau)})
    &=
    (1-q_\tau)^{T_\tau}\\
    &\leq
    \exp\left[
        -\frac{T_\tau
        \lambda_\tau(\rho,\mathsf M)\epsilon}{d-\tau}
    \right]
    \leq\delta,
\end{aligned}
\end{equation}
for the choice in~\eqref{eq:gfp-query}.
Every round is counted, including complementary outcomes; there is no
postselection or uncounted restart.
All controls act locally on $Q$, so the effective probe and measurement
are local-unitary conjugates of the specified pair, and no quantum
memory is retained between channel uses.

\noindent\textit{Matching accuracy lower bound.}\quad
We now prove the converse statement in the theorem.
The lower bound allows arbitrary finite auxiliary dimension,
unrestricted probe and measurement entanglement, and arbitrary
classical adaptation between queries.

Set
\begin{equation}
    U_0=I_d,
    \qquad
    U_1=e^{-i\phi H},
    \qquad
    \phi=\frac12\arccos(1-\epsilon),
\end{equation}
where
\begin{equation}
    H=
    \begin{cases}
        \sigma_y\otimes I_{d/2},&\tau=+1,\\
        \sigma_z\otimes I_{d/2},&\tau=-1.
    \end{cases}
    \label{eq:gfp-hard-pair}
\end{equation}
Let $\mathcal U_b(X)=U_bXU_b^\dagger$ for $b\in\{0,1\}$.
With $\Xi_{+1}=K$ and $\Xi_{-1}=JK$,
\begin{equation}
    H^2=I_d,
    \qquad
    \operatorname{Tr}H=0,
    \qquad
    \Xi_\tau H\Xi_\tau^{-1}=-H.
\end{equation}
Consequently,
\begin{equation}
    f_\tau(\mathcal U_0)=1,
    \qquad
    f_\tau(\mathcal U_1)
    =
    \cos(2\phi)
    =
    1-\epsilon.
\end{equation}

At any shared classical history, let $\rho_Q$ denote the reduced state
of the conditional probe on the query register.
Purifying the probe on an arbitrary reference, the two corresponding
pure outputs have overlap
\begin{equation}
\begin{aligned}
    \left|\operatorname{Tr}(\rho_Q U_1)\right|
    &=
    \left|
        \cos\phi
        -
        i\sin\phi\operatorname{Tr}(\rho_QH)
    \right|\\
    &\geq
    \cos\phi.
    \label{eq:gfp-overlap}
\end{aligned}
\end{equation}
The same conclusion applies to mixed probes by purification.
For pure outputs $\ket a,\ket b$ and any POVM $\{M_y\}$,
\begin{equation}
    |\langle a|b\rangle|
    \leq
    \sum_y|\langle a|M_y|b\rangle|
    \leq
    \sum_y
    \sqrt{
        \langle a|M_y|a\rangle
        \langle b|M_y|b\rangle
    }.
\end{equation}
Thus every conditional pair of classical outcome distributions has
Bhattacharyya coefficient at least $\cos\phi$.

Let $P_b^{(t)}$ be the transcript distribution after $t$ queries under
$\mathcal U_b$, and define
\begin{equation}
    B_t
    =
    \sum_h
    \sqrt{
        P_0^{(t)}(h)P_1^{(t)}(h)
    }.
\end{equation}
If $p_b(y\mid h)$ denotes the distribution of the next outcome conditioned
on a shared history $h$, then
\begin{equation}
\begin{aligned}
    B_{t+1}
    &=
    \sum_h
    \sqrt{
        P_0^{(t)}(h)P_1^{(t)}(h)
    }
    \sum_y
    \sqrt{
        p_0(y\mid h)p_1(y\mid h)
    }\\
    &\geq
    \cos\phi B_t.
\end{aligned}
\end{equation}
The conditional preparation and measurement choices are the same
functions of the shared classical history under both hypotheses, so
this recursion permits arbitrary classical adaptation.
Since $B_0=1$ and
$\cos^2\phi=1-\epsilon/2$,
\begin{equation}
    B_T\geq(\cos\phi)^T,
    \qquad
    d_{\mathrm{TV}}
    \left(P_0^{(T)},P_1^{(T)}\right)
    \leq
    \sqrt{1-B_T^2}
    \leq
    \sqrt{1-(1-\epsilon/2)^T}.
\end{equation}

Any protocol solving the same promise problem with worst-case error
at most $\delta$ must have
\begin{equation}
    d_{\mathrm{TV}}
    \left(P_0^{(T)},P_1^{(T)}\right)
    \geq1-2\delta.
\end{equation}
Hence
\begin{equation}
    T
    \geq
    \frac{
        \log[1/(4\delta(1-\delta))]
    }{
        -\log(1-\epsilon/2)
    }
    =
    \Omega\left(
        \epsilon^{-1}\log(1/\delta)
    \right),
    \label{eq:gfp-accuracy-lower}
\end{equation}
uniformly for
$0<\epsilon\leq1$ and $0<\delta\leq1/4$.
Indeed,
$-\log(1-\epsilon/2)\leq\epsilon$ on this interval, while
$\log[1/(4\delta(1-\delta))]$
is bounded below by a positive constant multiple of
$\log(1/\delta)$.
This proves the theorem.
\end{proof}

We now prove the resource statements following
Theorem~\ref{thm:gfp-testing}.
Completeness of the POVM gives
\begin{equation}
    \sum_y a_y=d,
    \qquad
    \sum_y b_y=1.
\end{equation}
Define
$\chi:=\sum_y|b_y|$,
since
$b_y=\operatorname{Tr}(M_y^{\mathrm T_Q}\rho_{Q,A})$,
the definition of measurement negativity gives
$\chi
\leq
2^{E_N^{\mathrm m}(\mathsf M)}$.
On the other hand, the measurement map associated with the POVM is
trace-norm contractive on Hermitian operators, so
\begin{equation}
    \chi
    \leq
    \|\rho_{Q,A}^{\mathrm T_Q}\|_1
    =
    2^{E_N^{\mathrm s}(\rho_{Q,A})}.
\end{equation}
Consequently,
$\chi
\leq
2^{e_*}$,
where
$e_*=\min\{e_{\mathrm s},e_{\mathrm m}\}$.

Writing
\begin{equation}
    P_b:=\sum_{b_y>0}b_y,
    \qquad
    N_b:=\sum_{b_y<0}|b_y|,
\end{equation}
we have
\begin{equation}
    P_b=\frac{\chi+1}{2},
    \qquad
    N_b=\frac{\chi-1}{2}.
\end{equation}
For $y\in\mathcal F_+$, $b_y=-a_y<0$, whereas for
$y\in\mathcal F_-$, $b_y=a_y>0$.
Therefore
\begin{equation}
    \lambda_+
    \leq
    N_b
    \leq
    \frac{2^{e_*}-1}{2},
    \qquad
    \lambda_-
    \leq
    P_b
    \leq
    \frac{2^{e_*}+1}{2},
\end{equation}
which proves~\eqref{eq:gfp-resource}.

To show attainability, choose an integer $1\leq R\leq d$ and an
auxiliary register $A\simeq\mathbb C^R$.
Let
\begin{equation}
\begin{aligned}
    |\Phi_R\rangle
    &=
    \frac1{\sqrt R}
    \sum_{j=0}^{R-1}|j\rangle_Q|j\rangle_A,
    \qquad
    \rho^{(R)}
    =
    |\Phi_R\rangle\langle\Phi_R|,\\
    P_R
    &=
    \sum_{j=0}^{R-1}|j\rangle\langle j|,\\
    \mathbb{S}_R
    &=
    \sum_{i,j=0}^{R-1}
    |i\rangle\langle j|_Q
    \otimes
    |j\rangle\langle i|_A ,
\end{aligned}
\end{equation}
and define the POVM
\begin{equation}
    M_\pm^{(R)}
    =
    \frac{P_R\otimes I_A\pm \mathbb{S}_R}{2},
    \qquad
    M_{\mathrm{out}}^{(R)}
    =
    (I_Q-P_R)\otimes I_A .
    \label{eq:gfp-rank-construction}
\end{equation}
Zero effects are omitted.
For this pair,
\begin{equation}
    A_\rho
    =
    I_Q\otimes\frac{I_A}{R},
    \qquad
    B_\rho
    =
    \frac{\mathbb{S}_R}{R},
\end{equation}
and hence
\begin{equation}
    a_\pm=\frac{R\pm1}{2},
    \qquad
    b_\pm=\pm a_\pm,
    \qquad
    a_{\mathrm{out}}=d-R,
    \qquad
    b_{\mathrm{out}}=0.
\end{equation}
Thus
\begin{equation}
    \lambda_\tau
    =
    \frac{R-\tau}{2},
    \qquad
    E_N^{\mathrm s}(\rho^{(R)})
    =
    \log_2R.
\end{equation}

It remains to evaluate the measurement negativity.
Since
$\mathbb{S}_R^{\mathrm T_Q}
=
R\rho^{(R)}$,
for any density operator $\sigma$ define
\begin{equation}
    w
    =
    \operatorname{Tr}
    [(P_R\otimes I_A)\sigma],
    \qquad
    z
    =
    \operatorname{Tr}
    (\rho^{(R)}\sigma),
\end{equation}
so that $0\leq z\leq w\leq1$.
Then
\begin{equation}
\begin{aligned}
    \sum_y
    \left|
        \operatorname{Tr}
        [(M_y^{(R)})^{\mathrm T_Q}\sigma]
    \right|=
    \max\{w,Rz\}+1-w\leq R.
\end{aligned}
\end{equation}
Equality is attained by $\sigma=\rho^{(R)}$, proving
\begin{equation}
    E_N^{\mathrm m}(\mathsf M^{(R)})
    =
    \log_2R.
\end{equation}
Therefore the bounds in~\eqref{eq:gfp-resource} are saturated
whenever $R=2^{e_*}$ is an integer.

For arbitrary real budgets, set
\begin{equation}
    \kappa=2^{e_*},
    \qquad
    R=\lfloor\kappa\rfloor.
\end{equation}
If $\kappa\geq2$, then
$R-1\geq\kappa/3$, while for every $\kappa\geq1$,
$R+1\geq\kappa$.
Substituting
$\lambda_\tau=(R-\tau)/2$
into~\eqref{eq:gfp-query} gives
\begin{equation}
    T_\tau
    =
    \mathcal O\left(
        2^{n-e_*}
        \epsilon^{-1}
        \log(1/\delta)
    \right)
\end{equation}
for both signs when $e_*\geq1$, and for $\tau=-1$ also when
$0\leq e_*<1$.
This proves~\eqref{eq:gfp-scaling}.

At fixed accuracy and confidence this upper bound is
$\mathcal O(d/2^{e_*})$.
Theorem~\ref{thm:main-negativity-lower-bound} gives
\begin{equation}
    \Omega\left(
        \min\left\{
            \sqrt d,\frac{d}{2^{e_*}}
        \right\}
    \right)
\end{equation}
queries for the same certification task.
Hence, whenever
$n/2\leq e_*\leq n$,
$d/2^{e_*}\leq\sqrt d$, and the two bounds match:
\begin{equation}
    T^\star
    =
    \Theta(2^{n-e_*})
\end{equation}
at fixed accuracy and confidence.

Finally, large probe and measurement negativities do not by themselves
guarantee $\lambda_\tau>0$ for a specified pair.
To see this, let the auxiliary space decompose as
$A=A_0\oplus A_1$, with
$\dim A_0=\dim A_1=d$.
Place a maximally entangled probe on $Q\otimes A_0$, and define a POVM
that acts as the identity on $Q\otimes A_0$ while implementing the
two SWAP-sector effects on $Q\otimes A_1$, $\mathbb{S}_{Q,A_1}
=
\sum_{i,j=0}^{d-1}
|i\rangle\langle j|_Q\otimes
|j\rangle\langle i|_{A_1}$.
Explicitly,
\begin{equation}
    M_0=I_Q\otimes P_{A_0},
    \qquad
    M_\pm
    =
    \frac{
        I_Q\otimes P_{A_1}
        \pm \mathbb{S}_{Q,A_1}
    }{2}.
\end{equation}
The probe has logarithmic negativity $n$, and the POVM also has
measurement negativity $n$, since its restriction to
$Q\otimes A_1$ contains the full-rank SWAP measurement.
However, on the support of the actual probe,
$a_0=d$,
$b_0=1$,
while
$a_\pm=b_\pm=0$.
Thus
$\lambda_+=\lambda_-=0$.
This proves that $\lambda_\tau>0$ is a genuine compatibility
condition on the specified probe--measurement pair.
In particular, if $e_*=0$,
~\eqref{eq:gfp-resource} forces
$\lambda_+=0$.

\subsection{TRS certification without auxiliary systems by adaptive interference}
\label{app:ancilla-free-protocol}
In this appendix, we provide a protocol for the certification task in~\eqref{eq:main-trs-testing-promise} that uses no auxiliary resources and analyze the query complexity.
\begin{theorem}[TRS certification without auxiliary systems]
\label{thm:main-limited-ancilla-protocol}
Fix either of the two TRS signatures $f_\pm$ in \eqref{eq:main-trs-signatures}.
For every fixed $0<\epsilon\leq1$ and all sufficiently large $n$ depending on $\epsilon$, there exists a classically adaptive protocol using no auxiliary system that,
for any unknown unitary channel $\mathcal U$, distinguishes
\begin{equation}
    f_\pm(\mathcal U)=1
    \qquad\text{from}\qquad
    f_\pm(\mathcal U)\leq1-\epsilon
\end{equation}
with success probability at least $2/3$, using
\begin{equation}
    T=
    \mathcal O \left(
        \frac{2^n\log n}{\epsilon^2 n}
    \right)
\end{equation}
queries.
\end{theorem}

\begin{proof}
Fix $\tau\in\{+1,-1\}$, write $f_{+1}=f_+$ and
$f_{-1}=f_-$, and let
\begin{equation}
    A=
    \begin{cases}
        U,&\tau=+1,\\
        UJ,&\tau=-1,
    \end{cases}
    \qquad
    g
    :=
    \frac1d\operatorname{Tr}(A^\dagger A^{\mathrm T})
    =
    \tau f_\tau(\mathcal U),
    \qquad
    J=i\sigma_y\otimes I_{d/2}.
    \label{eq:af-trs-signed-reduction}
\end{equation}
The identity for $g$ follows directly from the definitions of
$f_\pm$ and $J^{\mathrm T}=J^\dagger=-J$.
Moreover,
\begin{equation}
    \frac{
        \|A^{\mathrm T}-\tau A\|_{\mathrm F}^2
    }{2d}
    =
    1-f_\tau(\mathcal U).
    \label{eq:af-trs-signed-defect}
\end{equation}
Hence $H_0^{(\tau)}$ is equivalent to
$A^{\mathrm T}=\tau A$, whereas
$H_1^{(\tau)}$ implies
$|g-\tau|\geq\epsilon$.

\noindent\textit{Base protocol.}\quad
Choose $W$ Haar-random in $\mathrm U(d)$ and keep it fixed throughout
one execution of the base test.
Define
\begin{equation}
    V=W^{\mathrm T}AW,
    \qquad
    b=
    \left\lfloor
        \frac{\log d}{4\log\log d}
    \right\rfloor,
    \qquad
    m=\left\lceil\frac db\right\rceil,
    \qquad
    r=
    \left\lceil
        \frac{2048d}{\epsilon^2b}
    \right\rceil.
    \label{eq:af-trs-testing-parameters}
\end{equation}
All logarithms are natural, and $d$ is sufficiently large that
$b\geq1$.
Each call to $V$ uses one call to $U$, with preprocessing $W$ for
$\tau=+1$ or $JW$ for $\tau=-1$, and postprocessing $W^{\mathrm T}$.

The base test proceeds as follows.
\begin{enumerate}
    \item
    Prepare $\ket0$, apply $V$, and measure in the computational basis
    for $m$ independent runs.

    \item
    Let $N_j$ denote the resulting outcome counts.
    If no $j\neq0$ satisfies $N_j\geq b$, accept $H_0^{(\tau)}$.
    Otherwise choose such a label $j_\star$ by a fixed deterministic
    rule depending only on the counts.

    \item
    For each of the next $r$ runs, independently sample
    $\theta\in\{0,\pi/2,\pi,3\pi/2\}$ uniformly and define
    \begin{equation}
        \ket{\psi_\theta}
        =
        \frac{
            \ket0+e^{i\theta}\ket{j_\star}
        }{\sqrt2},
        \qquad
        \ket{\chi_{\tau,\theta}}
        =
        \frac{
            \ket0-\tau e^{-i\theta}\ket{j_\star}
        }{\sqrt2}.
        \label{eq:af-trs-interference-states}
    \end{equation}
    Prepare $\ket{\psi_\theta}$, apply $V$, and measure
    $\left\{
        \ketbra{\chi_{\tau,\theta}}{\chi_{\tau,\theta}},
        I-
        \ketbra{\chi_{\tau,\theta}}{\chi_{\tau,\theta}}
    \right\}$.
    Reject $H_0^{(\tau)}$ if the first outcome occurs in any run;
    otherwise accept.
\end{enumerate}

For a selected label $j$, the detection amplitude is
\begin{equation}
    \frac{
        V_{00}
        +
        e^{i\theta}
        (V_{0j}-\tau V_{j0})
        -
        \tau e^{2i\theta}V_{jj}
    }{2}.
\end{equation}
Averaging over the four phases eliminates all cross terms and gives
\begin{equation}
    q_j^{(\tau)}
    :=
    \mathbb E_\theta
    |\bra{\chi_{\tau,\theta}}V\ket{\psi_\theta}|^2
    =
    \frac{
        |V_{00}|^2
        +
        |V_{jj}|^2
        +
        |V_{0j}-\tau V_{j0}|^2
    }{4}.
    \label{eq:af-trs-interference-identity}
\end{equation}

\noindent\textit{Forward-column distribution.}\quad
Write
$W=[w\mid W_\perp]$,
where $w$ is the first column of $W$ and the columns of
$W_\perp$ form an orthonormal basis of $w^\perp$,
and define
\begin{equation}
    \alpha=w^{\mathrm T}Aw=V_{00},
    \qquad
    s=|\alpha|^2,
    \qquad
    \boldsymbol c=(V_{j0})_{j=1}^{d-1},
    \qquad
    \boldsymbol a=(V_{0j})_{j=1}^{d-1}.
    \label{eq:af-trs-column-row-notation}
\end{equation}
Conditional on $w$, the columns of $W_\perp$ form a Haar-random
orthonormal basis of $w^\perp$.
Since
\begin{equation}
    \boldsymbol c=W_\perp^{\mathrm T}Aw,
    \qquad
    \|\boldsymbol c\|^2=1-s,
\end{equation}
its normalized direction is uniform on the complex unit sphere of
dimension $L=d-1$.
Consequently,
\begin{equation}
    |c_j|^2=(1-s)p_j,
    \qquad
    p\sim\operatorname{Dirichlet}(1,\ldots,1),
    \qquad
    p\ \text{independent of }s.
    \label{eq:af-trs-column-dirichlet}
\end{equation}
The exceptional event $s=1$ has Haar probability zero.

For a Haar-random unit vector,
\begin{equation}
    \mathbb E[
        w_iw_j\overline{w_kw_\ell}
    ]
    =
    \frac{
        \delta_{ik}\delta_{j\ell}
        +
        \delta_{i\ell}\delta_{jk}
    }{
        d(d+1)
    }.
    \label{eq:af-trs-haar-vector-moment}
\end{equation}
Contracting this identity with $A$ and $\overline A$ gives
\begin{equation}
    \mathbb E s
    =
    \frac{1+g}{d+1}
    \leq
    \frac2{d+1}.
    \label{eq:af-trs-diagonal-moment}
\end{equation}

\noindent\textit{Selecting a large transition.}\quad
Let the event
$E=\{N_0=0\}$,
then
$\Pr(E^c)
\leq
m\mathbb E s$ with $m$ defined in~\eqref{eq:af-trs-testing-parameters}.
Because
$\Pr(E\mid s,p)=(1-s)^m$,
conditioning on $E$ leaves the prior of $p$ unchanged.
Given $p$ and $E$, the counts are multinomial with $m$ trials and
probabilities $p$.
Integrating against the Dirichlet density yields
\begin{equation}
    \Pr\left(
        (N_j)_{j\neq0}
        =
        (n_j)_{j\neq0}
        \middle|E
    \right)
    =
    \binom{m+L-1}{L-1}^{-1}
\end{equation}
for $L=d-1$ whenever
$\sum_{j\neq0}n_j=m$.

Let
\begin{equation}
    Y
    =
    \sum_{j\neq0}
    \mathbf1\{N_j\geq b\}
\end{equation}
and
\begin{equation}
    h_d
    :=
    \Pr(N_j\geq b\mid E)
    =
    \prod_{t=0}^{b-1}
    \frac{m-t}{m+L-1-t}.
    \label{eq:af-trs-heavy-counts}
\end{equation}
For two distinct labels, the joint probability that both counts exceed
$b$ is the analogous product through $t=2b-1$, which is at most
$h_d^2$.
Therefore
$\operatorname{Var}(Y\mid E)
\leq
Lh_d$
and
\begin{equation}
    \Pr(Y=0\mid E)
    \leq
    \frac1{Lh_d}
    =
    o(1),
    \qquad
    Lh_d=d^{3/4+o(1)}.
    \label{eq:af-trs-heavy-exists}
\end{equation}
Indeed,
$m=\lceil d/b\rceil$ and $b^2/m\to0$ imply
\begin{equation}
    \log h_d
    =
    -b\log b+\mathcal O(1)
    =
    -\frac14\log d+o(\log d).
\end{equation}

Conditional on $E$ and the count vector,
\begin{equation}
    p\mid(E,N)
    \sim
    \operatorname{Dirichlet}
    ((N_j+1)_{j\neq0}).
\end{equation}
For any selected label $j_\star$, its posterior mean is
$\mu_\star
=
(N_{j_\star}+1)/(L+m)$
and the variance is
$\mu_\star(1-\mu_\star)/(L+m+1)$.
Hence
\begin{equation}
    \Pr\left(
        p_{j_\star}<\frac{\mu_\star}{2}
        \middle|E,N
    \right)
    \leq
    \frac4{N_{j_\star}+1}
    \leq
    \frac4{b+1}.
    \label{eq:af-trs-posterior-tail}
\end{equation}
For sufficiently large $d$,
$L+m\leq2d$ and
$\Pr(s>1/2)\leq2\mathbb E s$.
Combining these estimates,
\begin{equation}
    \Pr(\mathsf G)
    \geq
    1-\zeta_d,
\end{equation}
where the ``good'' event is
\begin{equation}
    \mathsf G
    :=
    \left\{
        j_\star\text{ is selected and }
        |c_{j_\star}|^2\geq\frac b{8d}
    \right\},
\end{equation}
and
\begin{equation}
    \zeta_d
    :=
    \frac{2(m+2)}{d+1}
    +
    \frac1{Lh_d}
    +
    \frac4{b+1}
    =
    o(1).
    \label{eq:af-trs-heavy-transition}
\end{equation}
This estimate is uniform in the unknown unitary and in $\tau$.

\noindent\textit{Completeness of the base test.}\quad
Under $H_0^{(\tau)}$,
$V^{\mathrm T}
=
\tau V$.
For $\tau=-1$, the diagonal of $V$ vanishes, so~\eqref{eq:af-trs-interference-identity} gives
$q_j^{(-1)}=0$.
Thus the base test accepts with probability one.
For $\tau=+1$, $V$ is a symmetric unitary whose distribution is
invariant under unitary congruences.
Conditional on its first column
$v=(\alpha,\boldsymbol c)^{\mathrm T}$, write
\begin{equation}
    V=
    \begin{bmatrix}
        \alpha&\boldsymbol c^{\mathrm T}\\
        \boldsymbol c&B
    \end{bmatrix},
    \qquad
    \boldsymbol u
    =
    \frac{\boldsymbol c}{\sqrt{1-s}}.
    \label{eq:af-trs-null-block}
\end{equation}
Let
$Q:\mathbb C^{d-2}\to\boldsymbol u^\perp$
be an isometry.
Unitarity and $B^{\mathrm T}=B$ imply
$B
=
-\bar\alpha
\boldsymbol u\boldsymbol u^{\mathrm T}
+
QCQ^{\mathrm T}$,
where, conditional on $v$,
$C$ is a symmetric unitary invariant under all
$\mathrm U(d-2)$ congruences.
Consequently,
\begin{equation}
    \mathbb E[
        C_{ab}\overline{C_{k\ell}}
        \mid v
    ]
    =
    \frac{
        \delta_{ak}\delta_{b\ell}
        +
        \delta_{a\ell}\delta_{bk}
    }{
        d-1
    }.
\end{equation}
Contracting with the $j$th row of $Q$ yields
\begin{equation}
    \mathbb E[
        |V_{jj}|^2
        \mid v
    ]
    \leq
    s+\frac2{d-1}.
    \label{eq:af-trs-selected-diagonal}
\end{equation}
The forward-count likelihood depends only on $v$, so the same estimate
holds after conditioning on the forward record and selecting
$j_\star$.
Using $V_{0j}=V_{j0}$ and
$\mathbb E s=2/(d+1)$,
\begin{equation}
    \Pr(
        \mathrm{reject}
        \mid
        H_0^{(+1)}
    )
    \leq
    r
    \left(
        \frac1{d+1}
        +
        \frac1{2(d-1)}
    \right)
    =
    \mathcal O\left(
        \frac1{\epsilon^2b}
    \right)
    =
    o_\epsilon(1).
    \label{eq:af-trs-completeness}
\end{equation}

\noindent\textit{Soundness of the base test.}\quad
Suppose $H_1^{(\tau)}$ holds and define
$z
=
w^\dagger
A^\dagger A^{\mathrm T}
w$.
Since $A^\dagger A^{\mathrm T}$ is unitary with normalized trace $g$,
~\eqref{eq:af-trs-haar-vector-moment} gives
\begin{equation}
    \mathbb E z=g,
    \qquad
    \mathbb E|z-g|^2
    =
    \frac{1-g^2}{d+1}
    \leq
    \frac1{d+1}.
    \label{eq:af-trs-overlap-concentration}
\end{equation}
Moreover, for
$\boldsymbol a
=
W_\perp^{\mathrm T}A^{\mathrm T}w$ defined in~\eqref{eq:af-trs-column-row-notation}, we have
\begin{equation}
    \|\boldsymbol a\|^2
    =
    \|\boldsymbol c\|^2
    =
    1-s,
    \qquad
    \boldsymbol c^\dagger\boldsymbol a
    =
    z-s.
    \label{eq:af-trs-row-column-gram}
\end{equation}
Conditional on $w$ and $\boldsymbol c$, we decompose $\boldsymbol a
=
\gamma\boldsymbol c+\boldsymbol\xi$ with
\begin{equation}
    \gamma
    =
    \frac{z-s}{1-s},
    \qquad
    \boldsymbol\xi\perp\boldsymbol c,
    \qquad
    \|\boldsymbol\xi\|^2
    =
    (1-s)(1-|\gamma|^2).
\end{equation}
The direction of $\boldsymbol\xi$ is uniform in
$\boldsymbol c^\perp$ whenever it is nonzero, and therefore
\begin{equation}
    \mathbb E[
        |\xi_j|^2
        \mid
        w,\boldsymbol c
    ]
    \leq
    \frac1{d-2}.
    \label{eq:af-trs-residual-moment}
\end{equation}
Conditioning further on the forward counts preserves this estimate.

By Markov's and Chebyshev's inequalities, the event
$\left(|z-g|\leq\epsilon/4\right)\cap\left(s\leq\epsilon/8\right)$
has probability at least
$1-
32/((d+1)\epsilon^2)$.
On this event, using $|g-\tau|\geq\epsilon$,
\begin{equation}
    |\gamma-\tau|
    =
    \frac{
        |z-\tau+(\tau-1)s|
    }{
        1-s
    }
    \geq
    \frac\epsilon2.
    \label{eq:af-trs-regression-gap}
\end{equation}
On event $\mathsf G$, conditional Markov applied to~\eqref{eq:af-trs-residual-moment} gives
\begin{equation}
    \Pr\left(
        |\xi_{j_\star}|
        >
        \frac\epsilon4
        |c_{j_\star}|
        \middle|
        w,\boldsymbol c,N
    \right)
    \leq
    \frac{
        128d
    }{
        (d-2)\epsilon^2b
    }.
    \label{eq:af-trs-selected-residual}
\end{equation}
Whenever all these ``good'' events hold,
\begin{equation}
\begin{aligned}
    |V_{0j_\star}-\tau V_{j_\star0}|^2
    =
    |(\gamma-\tau)c_{j_\star}
      +\xi_{j_\star}|^2\geq
    \frac{\epsilon^2}{16}
    |c_{j_\star}|^2\geq
    \frac{\epsilon^2b}{128d}.
\end{aligned}
\end{equation}
Thus
\begin{equation}
    q_{j_\star}^{(\tau)}
    \geq
    \frac{\epsilon^2b}{512d}.
\end{equation}
Conditional on the fixed $W$ and the forward record, the interference
runs are independent, so
$(1-q_{j_\star}^{(\tau)})^r
\leq
e^{-4}$.
Combining all failure events,
\begin{equation}
\begin{aligned}
    \Pr(
        \mathrm{accept}
        \mid
        H_1^{(\tau)}
    )
    \leq
    e^{-4}
    +
    \zeta_d
    +
    \frac{32}{
        (d+1)\epsilon^2
    }+
    \frac{
        128d
    }{
        (d-2)\epsilon^2b
    }
    =
    e^{-4}+o_\epsilon(1).
\end{aligned}
\label{eq:af-trs-soundness}
\end{equation}

For every fixed $0<\epsilon\leq1$, there therefore exists
$d_0(\epsilon)$ such that, for all $d\geq d_0(\epsilon)$, both the
false-rejection and false-acceptance probabilities of the base test
are at most $1/3$, uniformly over all channels satisfying the
corresponding promise.
The base test uses at most
\begin{equation}
    T_0
    =
    m+r
    =
    \mathcal O\left(
        \frac{
            d\log\log d
        }{
            \epsilon^2\log d
        }
    \right)
\end{equation}
queries.

\noindent\textit{Confidence amplification.}\quad
To obtain an arbitrary failure probability
$0<\delta\leq1/3$, independently repeat the entire base test $K$ times.
Each repetition uses a fresh Haar-random unitary $W$ and fresh
measurement randomness, while the same unknown channel
$\mathcal U$ is queried throughout.
Output the majority decision.

For a fixed promised channel, let
$Z_k\in\{0,1\}$ indicate whether repetition $k$ gives the wrong
decision.
Conditional on the fixed channel, the variables $Z_k$ are independent
and satisfy
$\mathbb E Z_k
\leq
1/3.$
For odd $K$, a wrong majority implies
$\sum_{k=1}^K Z_k\geq K/2$.
Hoeffding's inequality therefore gives
\begin{equation}
\begin{aligned}
    \Pr(\text{majority decision is wrong})
    \leq
    \exp\left[
        -2K
        \left(
            \frac12-\frac13
        \right)^2
    \right]=
    e^{-K/18}.
\end{aligned}
\end{equation}
Choosing, for example,
\begin{equation}
    K
    =
    2
    \left\lceil
        9\log\frac1\delta
    \right\rceil
    +1
\end{equation}
makes the worst-case failure probability at most $\delta$.

The amplified protocol therefore uses
\begin{equation}
\begin{aligned}
    T
    =
    KT_0=
    \mathcal O\left(
        \frac{
            d\log\log d
        }{
            \epsilon^2\log d
        }
        \log\frac1\delta
    \right)=
    \mathcal O\left(
        \frac{
            2^n\log n
        }{
            \epsilon^2 n
        }
        \log\frac1\delta
    \right)
\end{aligned}
\end{equation}
queries.
Every preparation, known control, and measurement acts only on the
query register, and all quantum systems are discarded after every
channel use.
Thus the protocol uses no auxiliary system and retains only classical
memory between channel uses.
This proves
Theorem~\ref{thm:main-limited-ancilla-protocol}.
\end{proof}

\end{document}